\documentclass[lettersize,journal]{IEEEtran}
\usepackage{amsmath,amsfonts}
\usepackage{algorithmic}
\usepackage{algorithm}
\usepackage{array}
\usepackage[caption=false,font=normalsize,labelfont={footnotesize,rm},textfont={footnotesize,rm}]{subfig}
\usepackage{textcomp}
\usepackage{stfloats}
\usepackage{url}
\usepackage{verbatim}
\usepackage{bm,graphicx}
\usepackage{cite}
\usepackage{outlines}
\begin{document}
\title{Efficient Modeling of Depolarizing Mueller BRDFs}

\author{Quinn Jarecki and Meredith Kupinski
\thanks{}
}



\maketitle

\begin{abstract}
Light-matter interactions within indoor environments are significantly depolarizing. Nonetheless, the relatively small polarization attributes are informative. To make use of this information, polarized-BRDF (pBRDF) models for common indoor materials are sought. Fresnel reflection and diffuse partial polarization are popular terms in pBRDF models, but the relative contribution of each is highly material-dependent and changes based on scattering geometry and albedo. An efficient pBRDF would describe these dependencies with as few parameters as possible while retaining physical significance and task-relevant information. 
This work compares a triply-degenerate (TD)-Mueller matrix (MM) model to measurements of 3D printed objects. 
In this TD-MM model, the radiometric, polarimetric, and depolarization attributes are decoupled to reduce the number of parameters. The depolarization is quantified by a single geometry-dependent parameter, four geometry-independent material constants describe the polarization properties, and our TD-MM model is normalized to unit radiance so that the BRDF is decoupled. To test an application of the TD-MM model the material constants are assumed and the geometry-dependent depolarization parameter for a red 3D printed sphere is estimated from linear Stokes images. The geometry-averaged error of the depolarization parameter is 4.2\% at 662 nm (high albedo) and 11.7\% at 451 nm (low albedo). Since the error is inversely proportional to albedo and depolarization, the TD-MM model is referred to as appropriate for depolarization-dominant materials. The robustness of the TD-MM model is also tested by comparing ground-truth Mueller images to extrapolations of a red 3D printed Stanford bunny under arbitrary polarized illumination. 

\end{abstract}


\section{Introduction}
As light propagates through a given path connecting the source and observer, each light-matter interaction can change the orientation of electric field oscillations. The polarization changes between an incident and exitant ray direction after interaction are quantified by a polarized bidirectional reflectance distribution function (pBRDF) \cite{Boerner_1981,Hopcraft_Smith}. The unpolarized, or radiometric, scattering behavior of a material is characterized by the scalar-valued BRDF. A BRDF is the ratio of the differential reflected radiance to the differential input irradiance and, in general, is a function of the ray directions \cite{Bartell_BRDF_theory}. The pBRDF is a Mueller matrix (MM)-valued function of the ray directions. A MM is a 4$\times$4 real-valued matrix that describes the linear transformation of incident to exitant Stokes parameters \cite{:48,Parke_1949,chipman}. 

In high-precision metrology applications, such as defect detection for silicon wafer manufacturing or ellipsometry, the pBRDF is modeled by Fresnel interactions at multiple boundaries, e.g. thin film stacks \cite{scatmech_proc}. Some work has even been done in efficiently parameterizing polarized scattering from rough surfaces using Fresnel reflections, though depolarization is not accounted for ignored\cite{Renhorn:15}.
However, in the regime of everyday materials found within common indoor environments, the polarization attributes are dominated by depolarization. In these scenes, the polarization attributes are relatively small, but nonetheless informative and have demonstrated utility. 

Interest in polarization for computer graphics and vision algorithms has grown since commercial off-the-shelf (COTS) polarization cameras became available in 2018. The ability to track polarization in physics-based rendering (PBR) engines increased in relevance as polarimetric measurements became more accessible \cite{PolCourse,PolCourse23,10.5555/3056989.3056994,Berger2019IntroductionTP,s23104592}. Very recently, data-driven models successfully used COTS polarimeters for shape from polarization (SfP) and material texture discrimination \cite{Baek2018,ECCV_DeepSfP_Ba,dave_pandora_2022,lei2021shape,Ellipsometry:SIG:2022}. These studies confirm that polarimetry offers unique information about a scene; however, analytic pBRDF models for common indoor materials remain a barrier to exploring new applications using physics-based approaches. 

Both BRDF and pBRDF models typically contain a specular component that describes light scattered from the surface of a material and a diffuse component attributed to light scattered from within the material \cite{Wolff_diffuse}. 
The polarization of the specular component, more accurately called a first-surface reflection component, is commonly modeled as Fresnel reflection from hypothetical sub-resolution microfacets\cite{Torrance:67,priest_germer}. 
Early work on polarized light scattering assumed that the diffuse component becomes completely depolarized such that diffuse contributions to reflection can be eliminated by using a polarizer \cite{Wolff_reflection_components}. Although first-surface reflection tends to produce greater polarimetric modulation, it is generally incorrect to assume that diffuse reflection is completely unpolarized \cite{Atkinson_Hancock_2006}. Many SfP approaches consider both first-surface and diffuse polarized light scattering but use either a purely specular or purely diffuse pBRDF for a given material\cite{Atkinson_Hancock_2006,cui_sfp,kadambi_sfp}. Data driven models have had success with a decomposition into diffuse and specular terms as input into neural networks \cite{s23104592,ECCV_DeepSfP_Ba}. 

More rigorous pBRDF models characterize polarimetric light scattering in terms of a combination of first-surface and diffuse polarized components. The pBRDF model introduced by Baek et al. consists of a first-surface microfacet term and a polarized diffuse term\cite{Baek_model}. The diffuse term describes Fresnel transmission into, depolarizing scattering within, and Fresnel transmission out of the material. In this model, depolarization arises from the diffuse reflection term and from the opposing polarizance orientations between first-surface microfacet and diffuse terms. Kondo et al. extended this model by including an ideal depolarizer as an independent third term \cite{kondo2020accurate}.

An empirical BRDF is generated by measuring the reflected radiance distribution over the hemisphere of output directions for every point in the hemisphere of illumination directions. A popular way to expedite this process is to image a homogeneous spherical object with multiple illumination geometries\cite{Marschner_Westin_Lafortune_Torrance_2000,Matusik_Pfister_Brand_McMillan_2003}. Imaging a sphere captures many scattering geometries simultaneously, and intermediate scattering geometries can be interpolated. A pBRDF can be obtained in a similar way by performing MM imaging of homogeneous spheres as in work by Baek et al.\cite{Baek_database,Baek_model}. 

Complete MM imaging requires at least sixteen polarimetric linearly-independent measurements to constrain the sixteen degrees of freedom of a MM\cite{MMA,chipman}. For measuring general MMs, it is an established practice to use more than sixteen measurements because the pseudoinverse solution becomes more robust to noise as additional linearly-dependent measurements are performed\cite{Hagen_Oka_Dereniak_2007,Dubreuil:07,Lemaillet:08,Alenen_Tyo_2012}. Dual rotating retarder (DRR) polarimeters are commonly used to perform MM imaging and a DRR polarimeter is used in this work as a baseline for polarimetric accuracy\cite{Azzam_1978,Goldstein_1992,LopezTellez_rgb950}. The design of MM imaging measurement protocols are well-studied \cite{Hagen_Oka_Dereniak_2007,Dubreuil:07,Lemaillet:08,Alenen_Tyo_2012}.
DRR polarimeters (and similar polarimeters which modulate the polarized signal with respect to time) can be considered ill-suited for measuring objects with polarization that changes in time and applications where a large number of objects or geometries must be measured. In these cases, partial polarimetry can be employed with the goal of maximizing polarimetric information in a reduced period of time\cite{Kupinski:18}. 

Partial MM polarimeters perform fewer than sixteen linearly independent measurements and therefore cannot reconstruct the full MM. Such polarimeters and techniques to analyze their data have been developed for use in fields such as biomedicine and computer vision. Qi et al. demonstrated the use of a linear partial MM polarimeter for endoscopy\cite{Qi:13}.
For polarimetric imaging of biological tissue near normal incidence, Novikova et al. showed the analysis of partial MM data can yield comparable results to complete MM data\cite{Novikova:22}. Gonzalez et al. introduced a decomposition technique for MMs lacking the fourth row\cite{Gonzalez_2021}.
Kupinski et al. optimized the polarization state generator (PSG) and polarization state analyzer (PSA) states to perform a binary classification task\cite{Kupinski:17}.


This work makes use of the triple-degenerate (TD) MM model introduced by Li and Kupinski\cite{Li_singleParam}. The TD-MM model decouples depolarization from the largest contributing MJM, given the symbol $\widehat{\mathbf{m}}_0$, in a Cloude decomposition\cite{Cloude1986GroupTA,cloude_entropy}. Brosseau showed a similar decomposition of a MM into a MJM and an ideal depolarizer which performs an equivalent transformation of an input Stokes vector to an output Stokes vector as a depolarizing MM, where the MJM component was dependent on the input Stokes vector\cite{Brosseau_1998}. In a TD-MM model, $\widehat{\mathbf{m}}_0$ does not depend on the incident Stokes vector. Ossikovski and Arteaga defined the integral decomposition of a MM, where a depolarizing MM is written as the sum of the largest contributing MJM and a residual depolarizing MM\cite{Ossikovski:15}. The residual term is the sum of the three remaining MJMs, using their corresponding coherency eigenvalues as weights. The integral decomposition is similar to the TD model, but the TD model is the special case where the three smaller eigenvalues are equal. Since these three eigenvalues are equal, a fraction of the contribution of $\widehat{\mathbf{m}}_0$ is combined with the remaining three MJMs to yield $\mathbf{m}_{ID}$ as shown in Fig.~\ref{fig:TD}, simplifying the description of depolarization. 

Prior work has demonstrated the broad applicability of this TD-MM model and the use of the largest eigenvalue, $\xi_0$, as an important attribute that depends on a material's texture and albedo, and is a function of scattering geometry. Omer and Kupinski demonstrated the efficiency of a TD representation by truncating measured MM data to compress tabulated pBRDFs by 50\%, with the additional benefit of enforcing physicality when interpolating to unmeasured geometries\cite{KhalidRendering}. Jarecki and Kupinski demonstrated using an assumption of a TD coherency eigenspectrum and modeling $\widehat{\mathbf{m}}_0$ as Fresnel reflection to extrapolate MM images from linear Stokes measurements\cite{Jarecki_proc,jarecki2022}. They performed MM extrapolations on a set of two colors of plastic LEGO bricks treated to have varying surface roughness. This dataset represented an ensemble of many albedo and texture cases for a single material, and the observed results obeyed Umov's effect (\emph{i.e.} that depolarization magnitude should trend positively with albedo) and that depolarization should also depend on texture and geometry\cite{umow1905chromatische,AoLP_Kupinski}. 
They used a microfacet-inspired Fresnel reflection model for $\widehat{\mathbf{m}}_0$ and compared DRR polarimeter measurements to extrapolations from a COTS linear Stokes camera. However, the range of measured scattering geometries was small due to imaging (macroscopically) flat samples, so  purely first-surface reflection model based on Fresnel reflection was sufficient to describe polarization scattering behavior for that ensemble of materials. The material studied in this work is better described using a model which combines Fresnel reflection and diffuse polarization.
The objective of this work is to introduce a TD-MM model that efficiently describes measurements of spherical objects in a DRR Mueller polarimeter. The accuracy of this model is assessed as performance in the task of estimating $\xi_0$ from an assumed form of $\widehat{\mathbf{m}}_0$ and partial polarimetric measurements. 

The details of the model are described in Sect.~\ref{sect:model}, the method for estimating $\xi_0$ is described in Sect.~\ref{sect:method}, and quantitative results for a 3D printed red sphere and qualitative results for a 3D printed Stanford bunny of the same material are described in Sect.~\ref{sect:results}

\section{A Triple Degenerate pBRDF for Depolarization-Dominant Materials}

\begin{figure}[h]
    \centering
        \subfloat[Cloude decomposition]{\includegraphics[trim= 0 10 10 10, clip,width=.33\columnwidth]{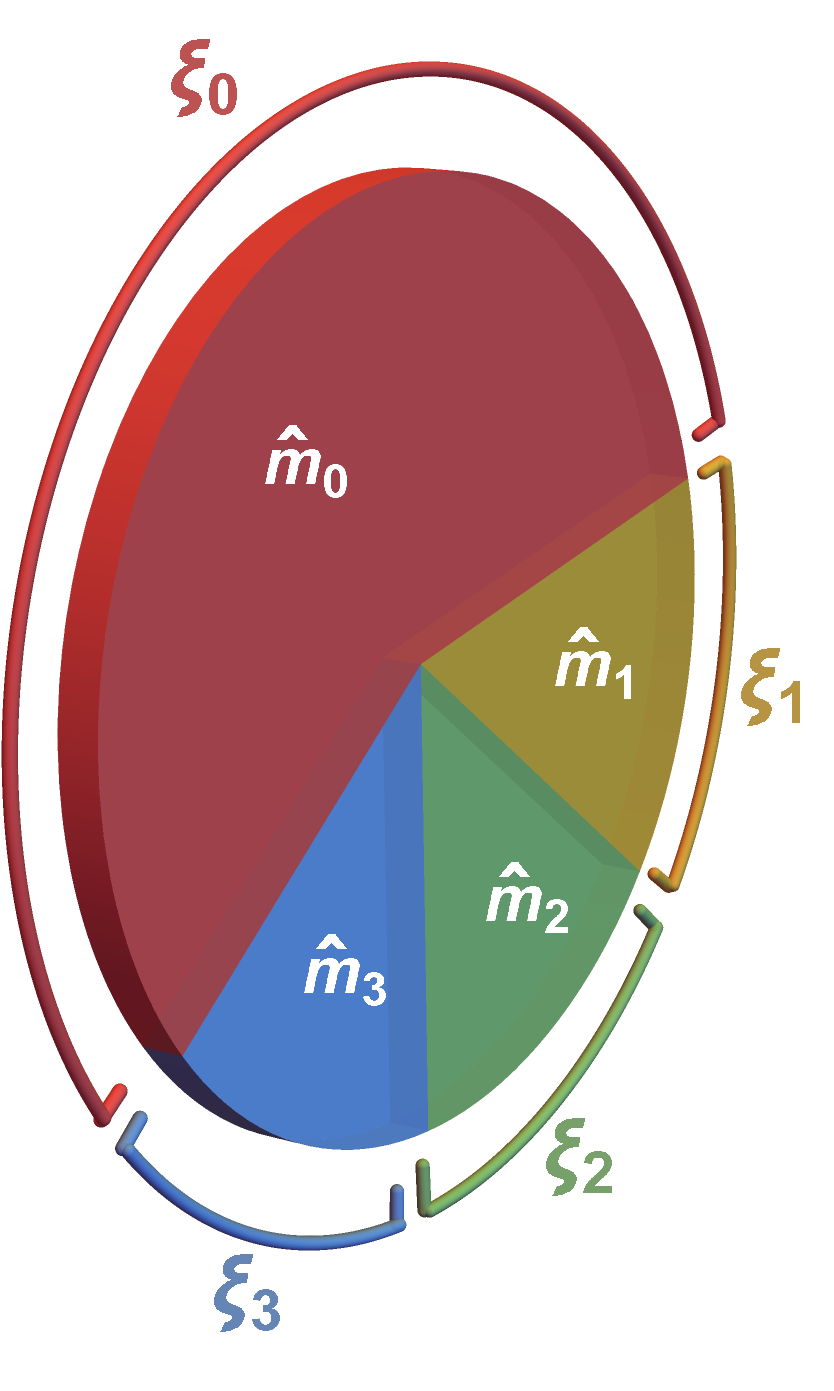}}\label{fig:TD1}
        \subfloat[Equal Weights]{\includegraphics[trim= 40 20 10 50, clip,width=.32\columnwidth]{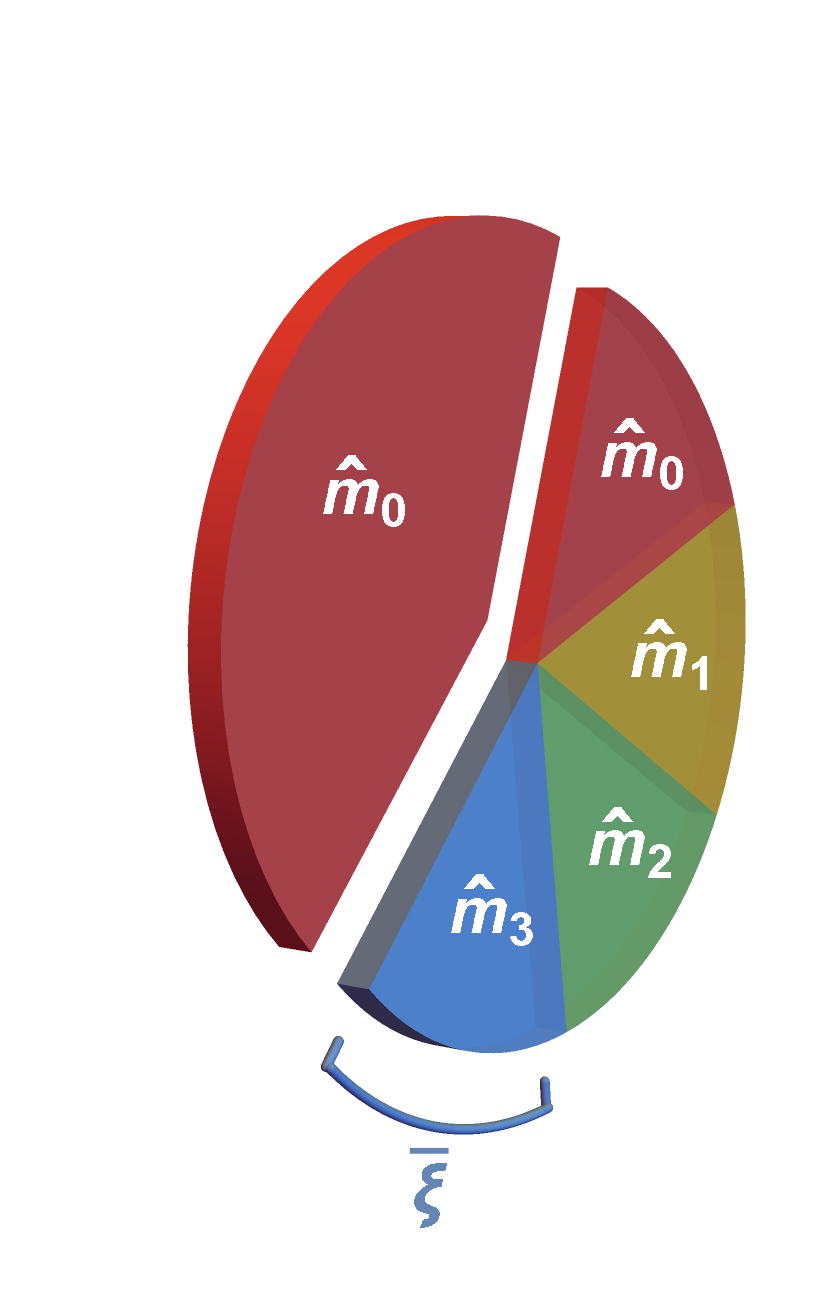}}
        \subfloat[TD Model]{\includegraphics[trim= 40 20 20 70, clip,width=.3\columnwidth]{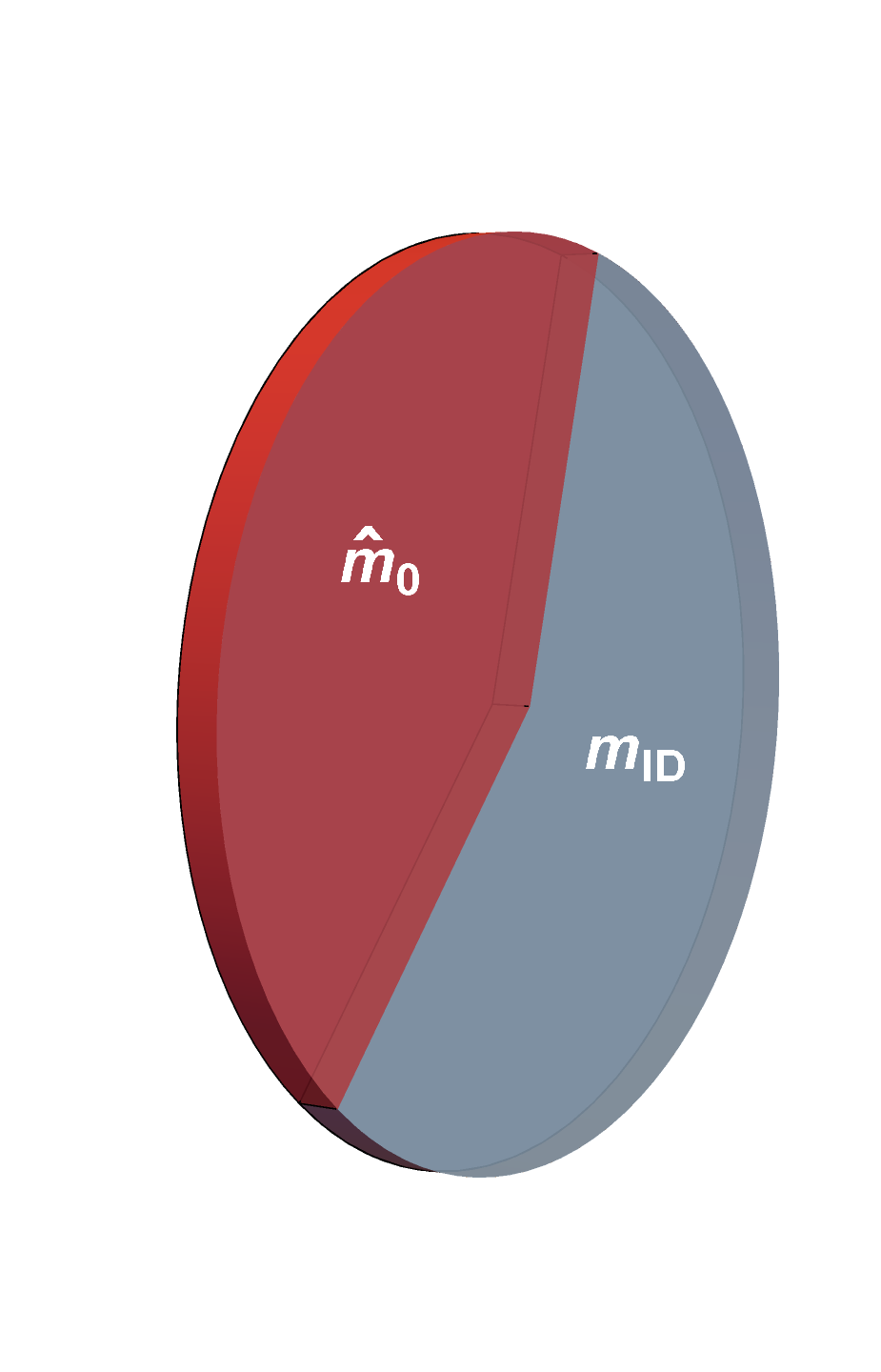}}
            \caption{Visualization of Cloude decomposition where $\xi_1=\xi_2=\xi_3=\overline{\xi}$ is denoted a triple degenerate (TD) model. As shown in Eq.~\ref{eq:1par_a}, the relative weight between $\widehat{\mathbf{m}}_0$ and $\mathbf{m}_{ID}$ is controlled by the largest coherency eigenvalue $\xi_0$, making it a single-parameter depolarization model. In (a), the Cloude decomposition of a depolarizing MM into a convex sum of four MJMs where the fractional size of each wedge corresponds to the coherency eigenvalues and the weight of each MJM. In (b), the TD weights allow the four MJMs to be separated in equal contribution. In (c), the equal contributions combine to form an ideal depolarizer $\mathbf{m}_{ID}$. }\label{fig:TD}
\end{figure}
A MM maps an input polarization state to an output polarization state\cite{chipman,MMA}. A MM has sixteen degrees of freedom: one for average throughput, three for diattenuation, three for retardance, and nine for depolarization\cite{chipman}. Diattenuation is the polarization-dependent reflectance which is encoded in the top row of the MM and denoted by the vector$\mathbf{D}$. Retardance is the polarization-dependent phase. Depolarization is the randomization of polarization with respect to position, angle, time, and/or wavelength faster than the particular detector in use can resolve\cite{Billings_monochromatic}. Depolarization is of particular interest in this work because it is the dominant polarization property in most every-day environments. Several different metrics exist for depolarization such as the depolarization index and the coherency eigenvalues\cite{gil_depIndex,Ossikovski_eigenvalue-based}.

The eigenvectors of a MM do not strictly correspond to physical Stokes parameters. Instead, eigenanalysis is performed on the linearly related coherency matrix\cite{Cloude1986GroupTA,cloude_entropy}. The eigenvectors of the coherency matrix can be used to decompose a depolarizing MM into a convex sum of up to four non-depolarizing MMs $\mathbf{m}=\sum_{n=0}^3\xi_n\widehat{\mathbf{m}}_n$. 
Each of these non-depolarizing MM has an equivalent Jones matrix representation $\mathbf{J}$ and therefore is also called a Mueller-Jones matrix (MJM), and is notated with a hat $\widehat{\cdot}$.
The $\xi_n$ in the convex sum are called the coherency matrix eigenvalues, and are normalized such that $\sum_{n=0}^3\xi_n=1$. With this normalization, each eigenvalue $\xi_n$ represents the fraction of light which is transformed by the MJM $\widehat{\mathbf{m}}_n$ as shown in Fig.~\ref{fig:TD}~(a). The set of four MJM in this decomposition are orthonormal in the sense that $\frac{1}{2}\mathrm{tr}(\mathbf{J}_n^\dagger\mathbf{J}_m)=\delta_{nm}$ and $\frac{1}{4}\sum_{n=0}^3\widehat{\mathbf{m}}_n=\mathbf{m}_{ID}$.

A useful special case is a triply-degenerate (TD) eigenspectrum ($\xi_1=\xi_2=\xi_3=\overline{\xi}$), where $\overline{\xi}=(1-\xi_0)/3$ as in Fig.~\ref{fig:TD}~(b). In the TD case, the depolarizing MM can be written as the convex sum of a MJM and an ideal depolarizer
\begin{equation}
    \mathbf{M}=\frac{4M_{00}}{3}\left[\left(\xi_0-\frac{1}{4}\right)\widehat{\mathbf{m}}_0+\left(1-\xi_0\right)\mathbf{m}_{ID}\right],
    \label{eq:1par_a}
\end{equation}
where $\xi_0$ is the largest eigenvalue, $\widehat{\mathbf{m}}_0$ is the dominant MJM, $\mathbf{m}_{ID}$ which is zero everywhere except the top-left element equals one, and $M_{00}$ is the throughput for unpolarized light. The model in Eq.~\ref{eq:1par_a} is represented by the pie chart in Fig.~\ref{fig:TD}~(c).

In a TD-MM model, the degrees of freedom are reduced from 16 to eight: one for throughput, one for depolarization $\xi_0$, and six for the dominant MJM $\widehat{\mathbf{m}}_0$. The single degree of freedom for depolarization is $\xi_0$ which controls the relative weight between an ideal depolarizer and a MJM. The largest eigenvalue is bounded in the range $0.25\leq\xi_0\leq 1.0$, where $\xi_0=0.25$ means the MM is the ideal depolarizer and $\xi_0=1.0$ means the MM is a MJM. In a TD-MM model, the depolarization index $DI$ is monotonically related to $\xi_0$ by the equation $DI=\frac{1}{3}\sqrt{16\xi_0^2-8\xi_0+1}$. Diattenuation and retardance orientations match those of $\widehat{\mathbf{m}}_0$ and are invariant to $\xi_0$. The maximum diattenuation and retardance magnitudes in a TD-MM match those of $\widehat{\mathbf{m}}_0$ when $\xi_0=1$, but are reduced as $\xi_0$ approaches $0.25$ where $\mathbf{m}_{ID}$ dominates the sum.

\begin{figure}[!hb]
        \centering
    \makebox[5pt]{{\includegraphics[trim={0 0 0 0},clip,width=0.35\columnwidth]{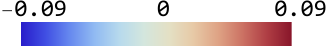}}}
    \hspace{3.7cm}
    \makebox[5pt]{{\includegraphics[trim={0 0 0 0},clip,width=0.35\columnwidth]{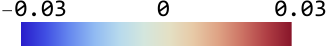}}}
    \\
        \subfloat[$\xi_1-\overline{\xi}$ 451 nm]{\includegraphics[width=.35\columnwidth]{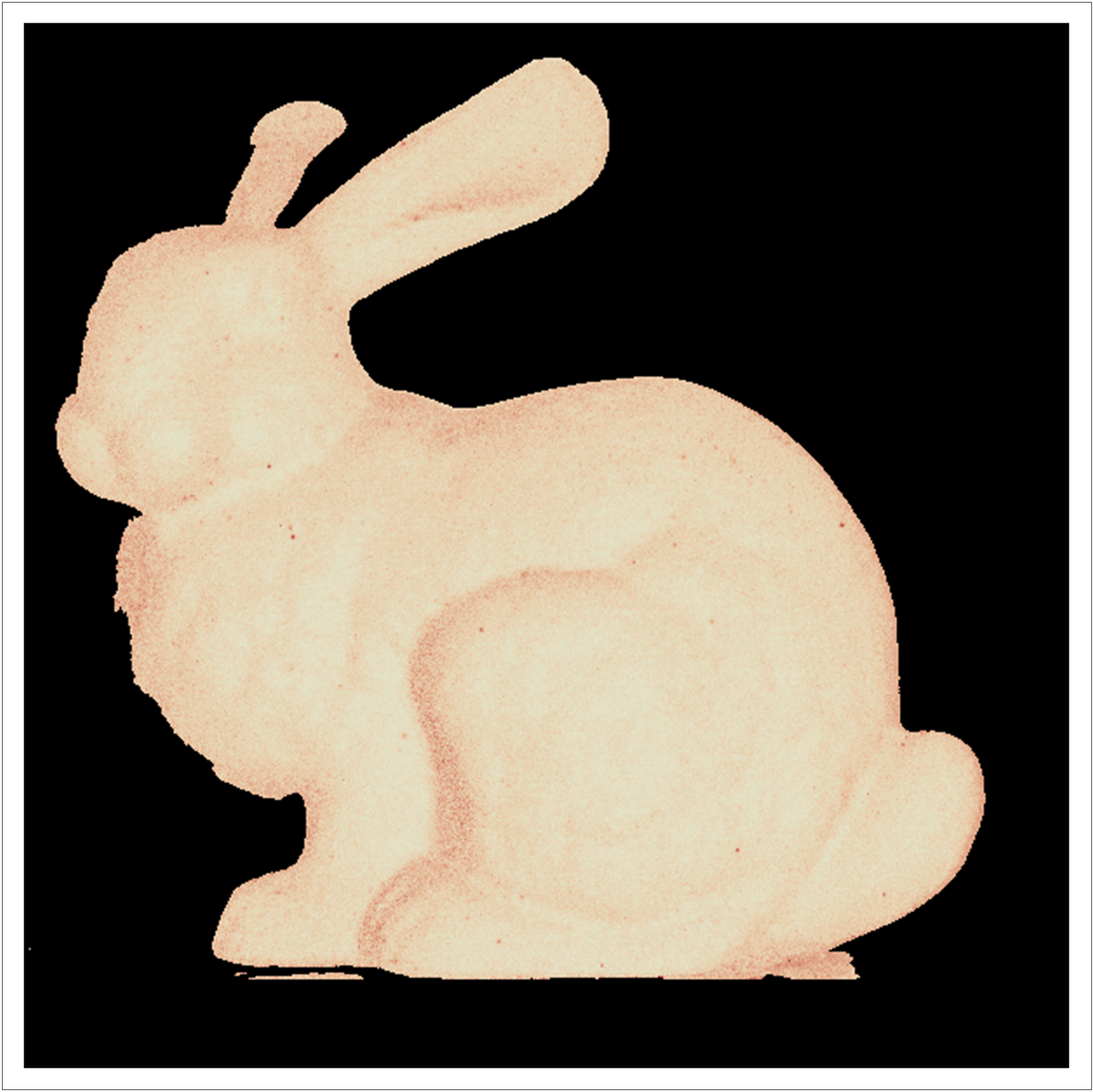}}\label{fig:xi1_451}
        \hspace{.7cm}
        \subfloat[$\xi_1-\overline{\xi}$ 662 nm]{\includegraphics[width=.35\columnwidth]{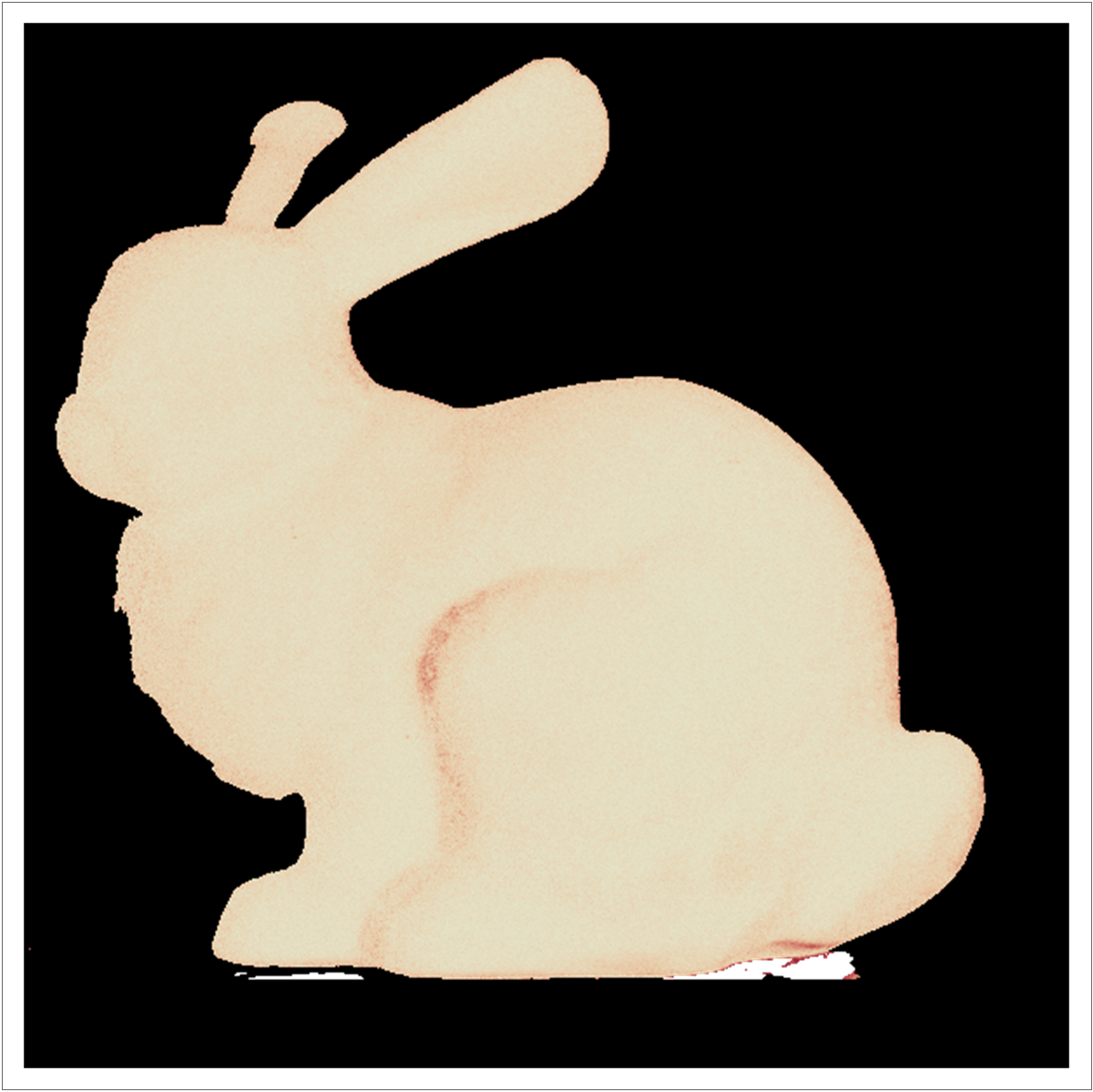}}\label{fig:xi1_662}\\    
        \subfloat[$\xi_2-\overline{\xi}$ 451 nm]{\includegraphics[width=.35\columnwidth]{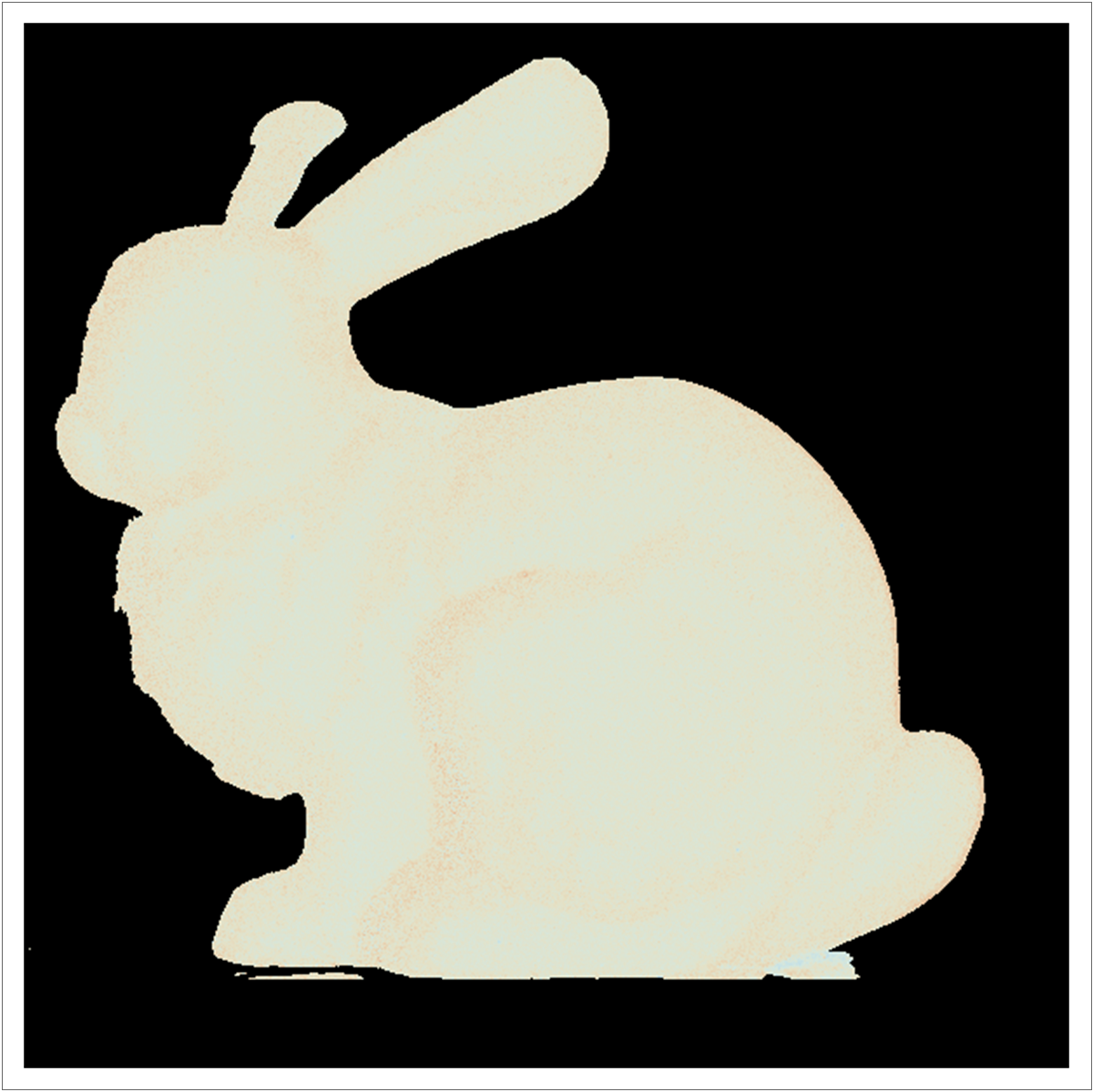}}\label{fig:xi2_451}
        \hspace{.7cm}
        \subfloat[$\xi_2-\overline{\xi}$ 662 nm]{\includegraphics[width=.35\columnwidth]{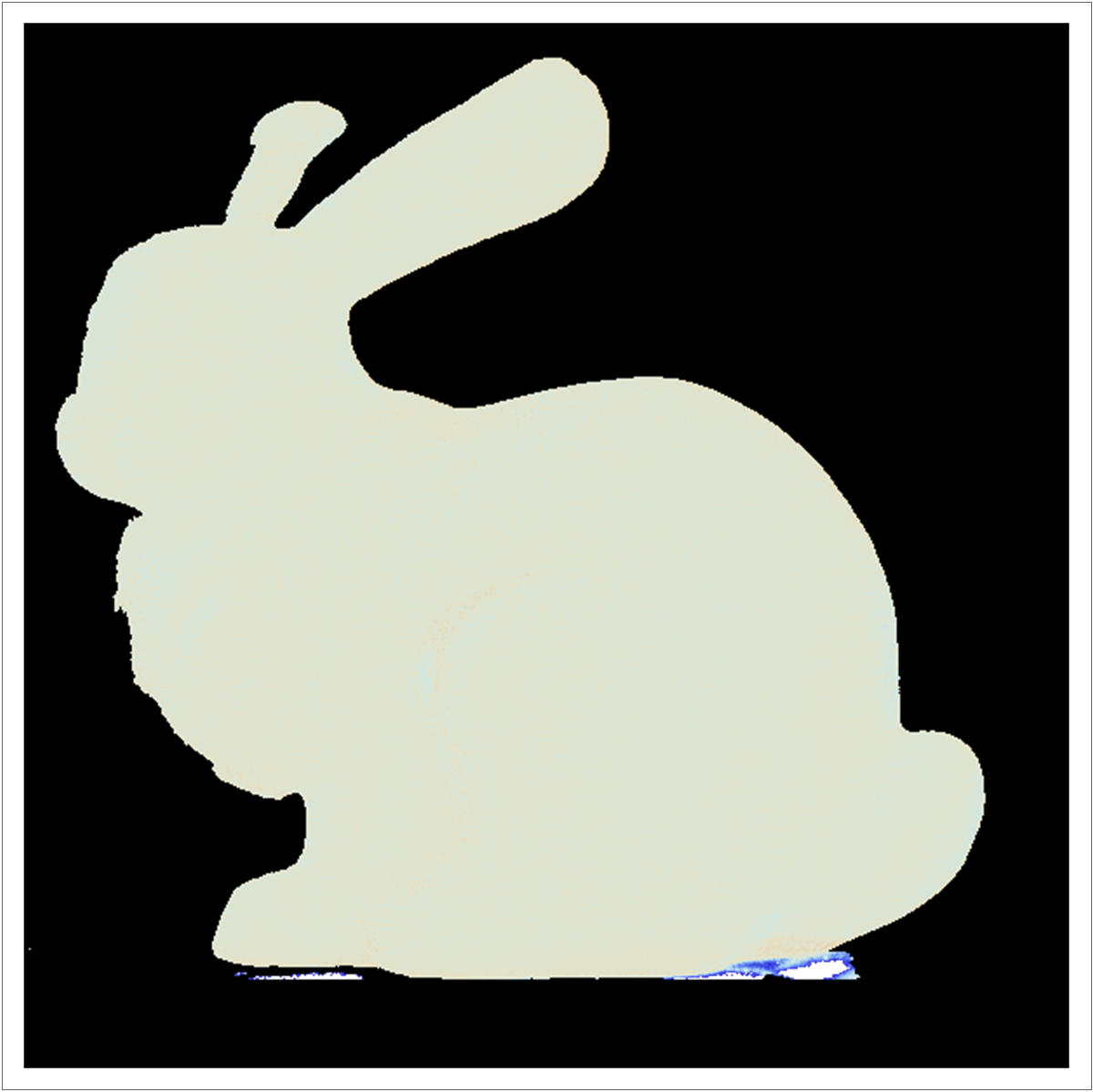}}\label{fig:xi2_662}\\
        \subfloat[ $\xi_3-\overline{\xi}$ 451 nm]{\includegraphics[width=.35\columnwidth]{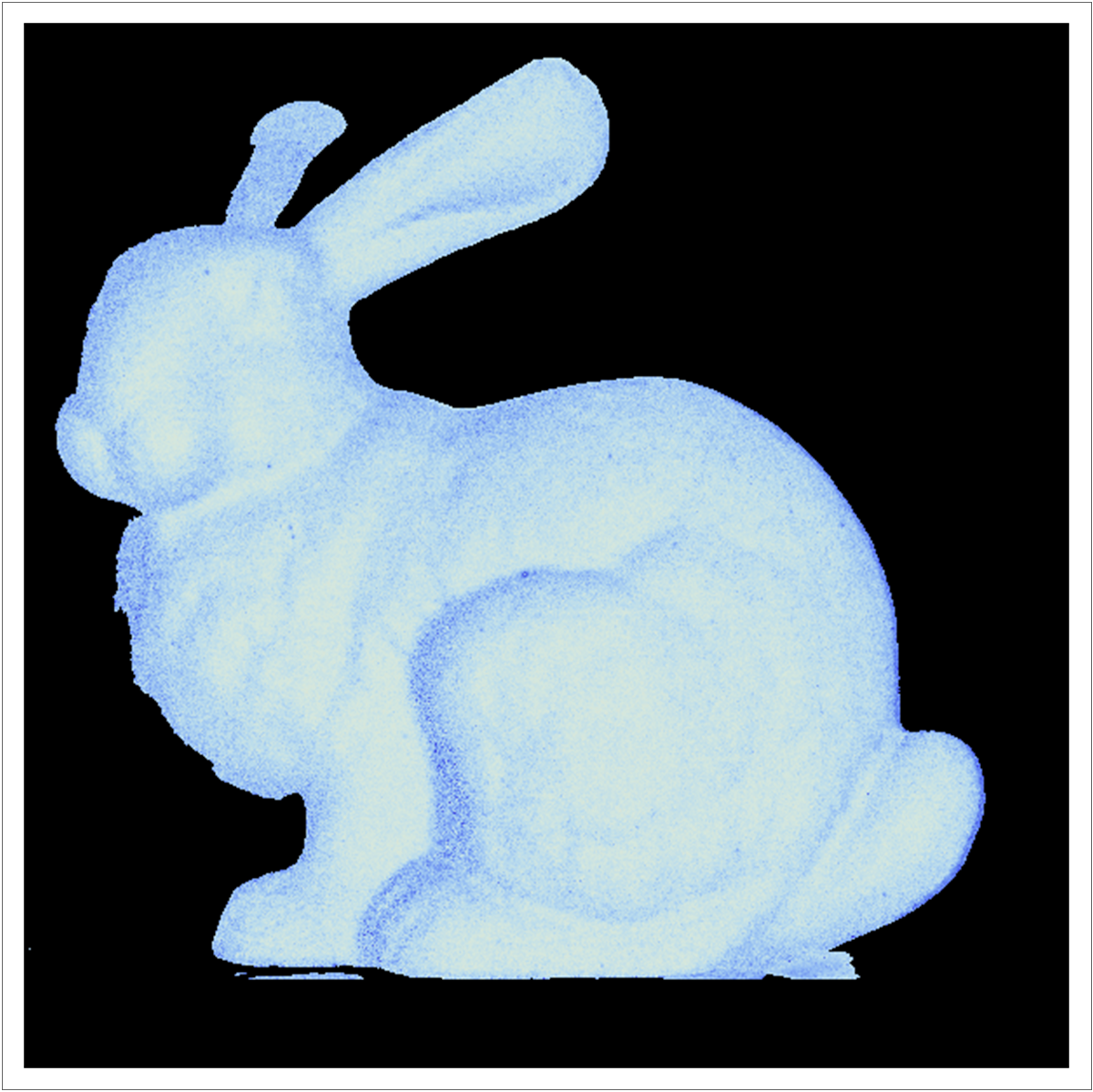}}\label{fig:xi3_451}  
        \hspace{.7cm}
        \subfloat[ $\xi_3-\overline{\xi}$ 662 nm]{\includegraphics[width=.35\columnwidth]{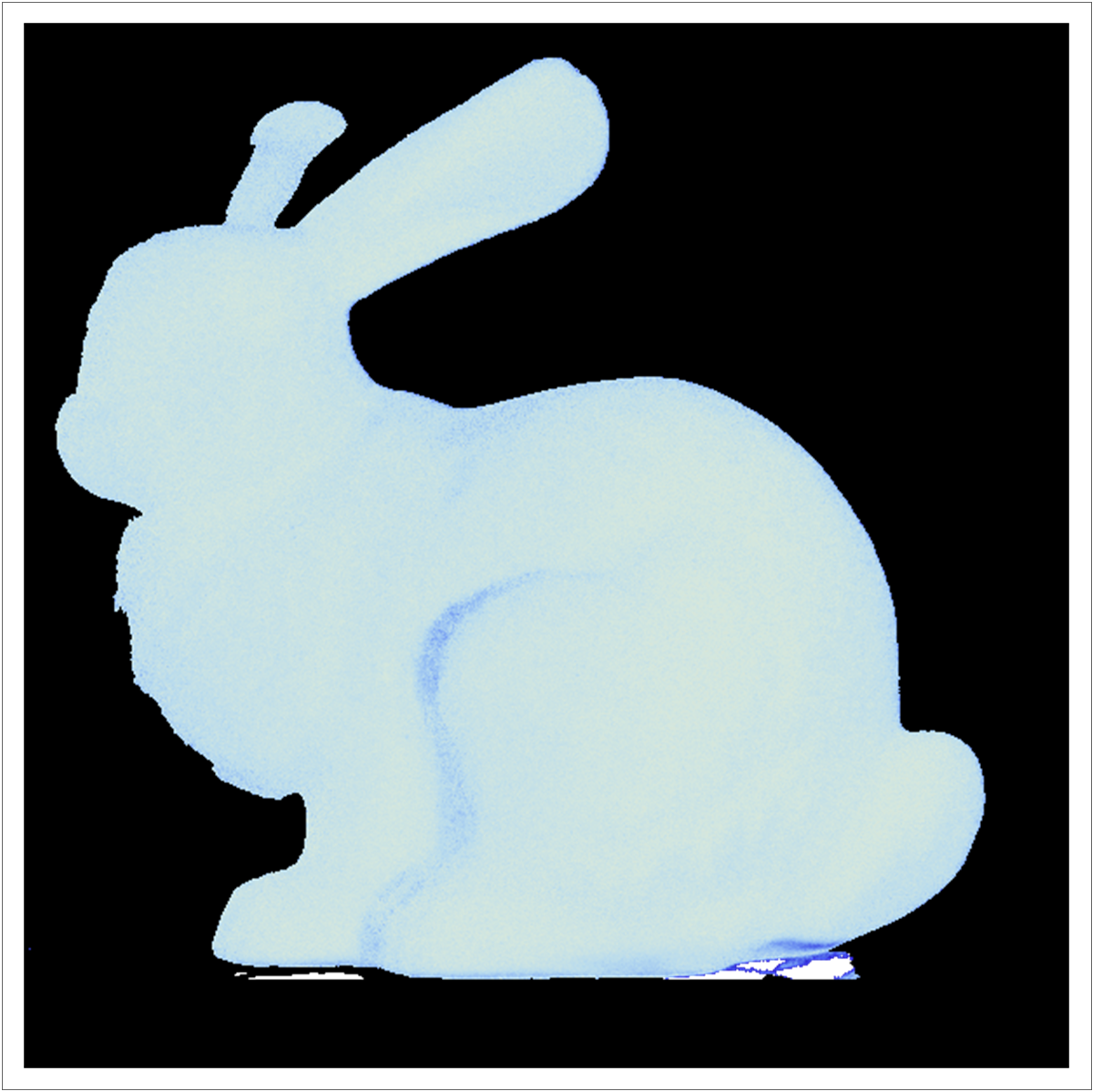}}\label{fig:xi3_662}
        \caption{Deviation of each of the smaller three coherency eigenvalues $\xi_1$, $\xi_2$, and $\xi_3$ from $\overline{\xi}=(1-\xi_0)/3$ for a red 3D printed bunny. Values shown in (a), (c), (e) are at 451 nm and values shown in (b), (d), (f) are at 662 nm. Each column shares a colorbar. When $\xi_1$, $\xi_2$, and $\xi_3$ are all equal to $\overline{\xi}$, then the coherency eigenspectrum is TD and the depolarizing MM can be exactly described as a sum of a dominant MJM and an ideal depolarizer (see Fig.~\ref{fig:TD}). 
        The eigenspectrum for the red 3D printed bunny under 451 nm illumination (low albedo, low depolarization per Umov's effect) has larger deviation from TD than 662 nm (high albedo, high depolarization).
        This trend with wavelength suggests that even if the dominant MJM model $\widehat{\mathbf{m}}_0$ is accurate, the error in assuming the MM fits Eq.~\ref{eq:1par_a} would be greater at 451 nm than 662 nm. In particular, regions of high curvature deviate more from TD as shown on the left side of the bunny's hind leg.}\label{fig:bunny_xi}
\end{figure}

To demonstrate the appropriateness of a TD assumption for the red 3D printing material used in this work, Fig.~\ref{fig:bunny_xi} shows the deviation of the coherency eigenspectrum from TD of a printed Stanford bunny at two wavelengths. Since the eigenvalues are normalized to sum to unity, the difference of a given eigenvalue from $\overline{\xi}$ can be thought of as the fraction of light that is not described by a TD model.  The low albedo case of 451 nm illumination deviates from TD for as much as 9\% of the light while the high albedo case of 662 nm illumination only deviates as much as 3\%. For both wavelengths, the deviation from TD is larger in regions of high curvature. This could be due to increased likelihood of multiple ray bounces resulting in more complex polarized light scattering behavior. In regions with high deviation from a TD eigenspectrum, tasks such as estimation of $\xi_0$ will be less accurate even when $\widehat{\mathbf{m}}_0$ is perfectly known.

\section{Mueller-Jones Model of First-Surface and Diffuse Polarization Attributes}\label{sect:model}
\subsection{Rusinkiewicz Coordinate System}\label{sect:rusink}

\begin{figure}[!h]
    \centering
        \subfloat[Measurement configuration]{\fbox{\includegraphics[trim={73 55 87 50},clip, width=.45\columnwidth]{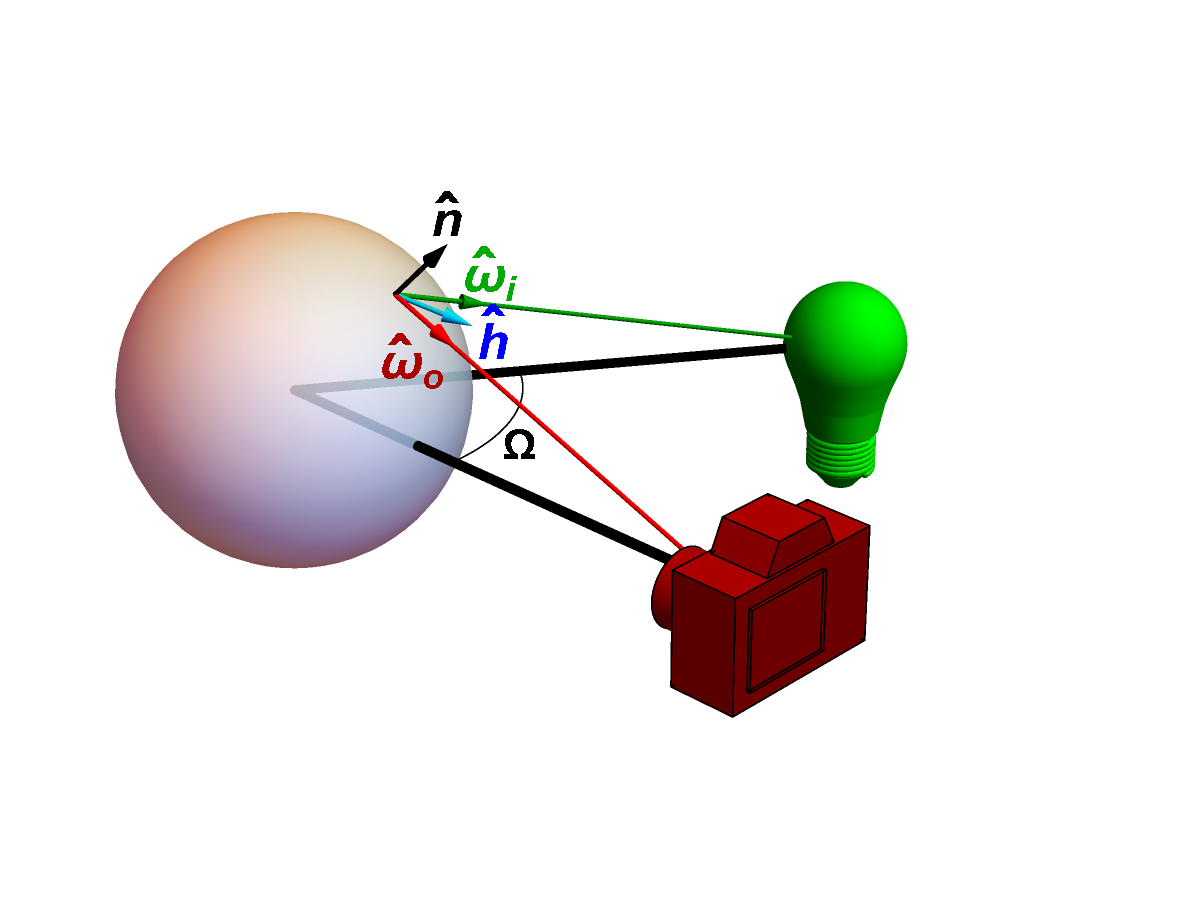}}}\label{fig:sphere}
        \subfloat[Rusinkiewicz coordinate system]{\fbox{\includegraphics[trim={0 12 0 18},clip,width=.45\columnwidth]{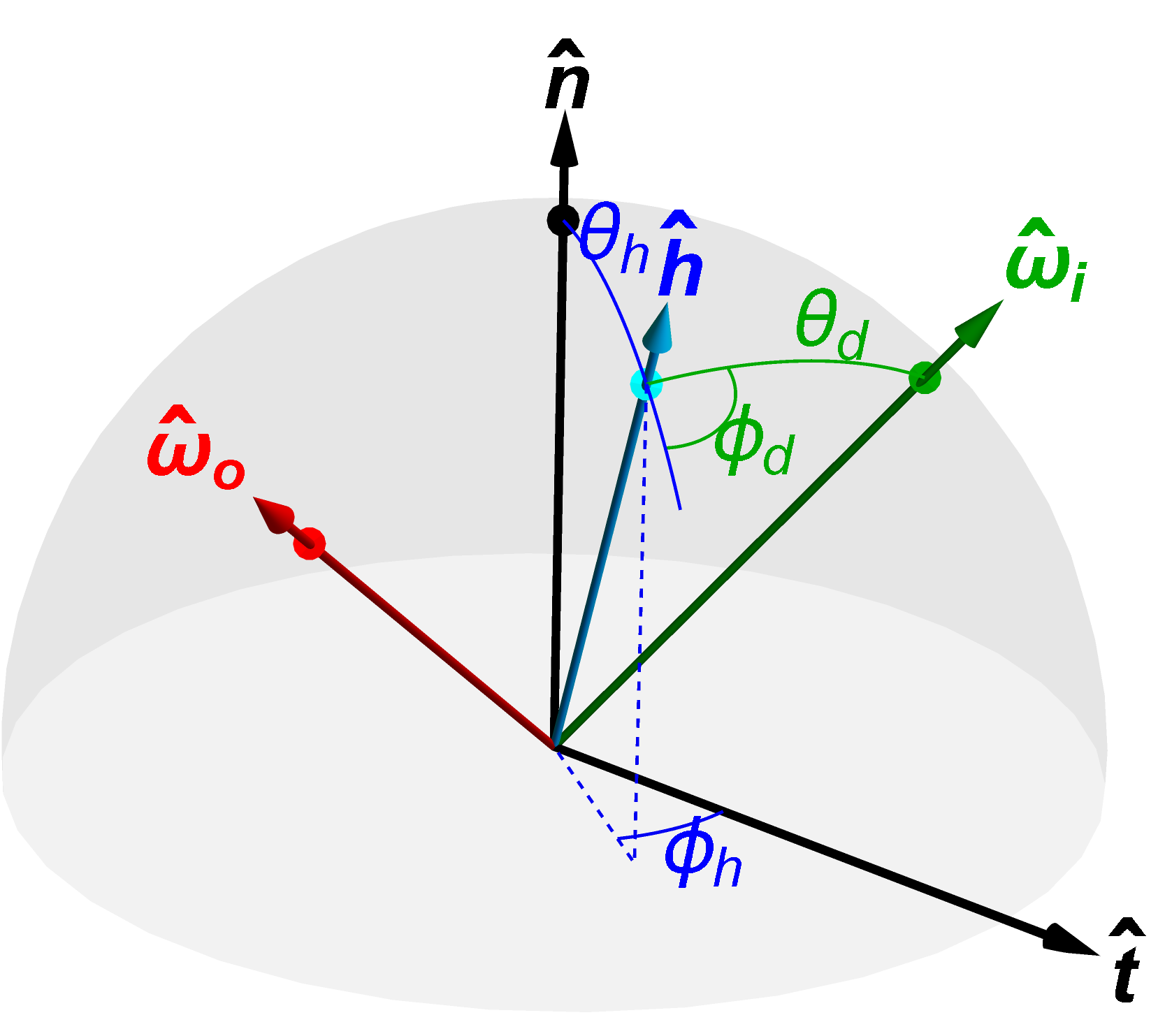}}}\label{fig:rusink}
            \caption{In (a), a spherical object by a camera and light source separated by an angle $\Omega$. The point on the sphere imaged by a given pixel is observed along the red vector $\widehat{\pmb{\omega}}_o$ and is illuminated along the green vector $\widehat{\pmb{\omega}}_i$. This point on the sphere has a surface normal along the black vector $\widehat{\mathbf{n}}$. The blue halfway vector $\widehat{\mathbf{h}}$ bisects $\widehat{\pmb{\omega}}_o$ and $\widehat{\pmb{\omega}}_i$ and is used in (b) to parameterize the scattering geometry in terms of Rusinkiewicz coordinates. This work considers isotropic surfaces where the pBRDF is invariant to $\phi_h$ since $\widehat{\mathbf{t}}$ is arbitrary. }\label{fig:rusinkCoords}
\end{figure}

The measurement configuration for a spherical object and the parameters used to define the scattering geometry are shown in Figure~\ref{fig:rusinkCoords}~(a). The green light bulb indicates the source which is connected to a point on the object by the vector $\widehat{\pmb{\omega}}_i$. The red vector from this object point to a given camera element is $\widehat{\pmb{\omega}}_o$. The angle between the source and the camera is denoted $\Omega$. Figure~\ref{fig:rusinkCoords}~(b) shows a zoomed in view where these two vectors and the object surface normal are parameterized in a Rusinkiewicz coordinate system.

Rusinkiewicz coordinates are an established parameterization that reduces dimensionality for isotropic surfaces and recasts the scattering geometry in terms of variables with useful physical interpretations\cite{Rusinkiewicz}. The parameterization defines a halfway vector $\widehat{\mathbf{h}}$ (so named for bisecting $\widehat{\pmb{\omega}}_i$ and $\widehat{\pmb{\omega}}_o$) in terms of spherical coordinates $\theta_h$ and $\phi_h$ about the surface normal $\widehat{\mathbf{n}}$. The incoming and outgoing ray directions, $\widehat{\pmb{\omega}}_i$ and $\widehat{\pmb{\omega}}_o$, are then defined with spherical coordinates $\theta_d$ and $\phi_d$ about $\widehat{\mathbf{h}}$. The angle between the surface normal and the halfway vector, $\theta_h$, describes how a particular scattering geometry deviates from specular reflection. $\phi_h$ orients the scattering geometry relative to an axis of anisotropy $\widehat{\mathbf{t}}$. For isotropic surfaces the BRDF/pBRDF is invariant to $\phi_h$, reducing the functional dependence on four variables to three. $\theta_d$ can be thought of as the angle of incidence onto a hypothetical microfacet as further described below. Last, $\phi_d$ describes the deviation of a scattering geometry from being ``in-plane.'' ``Out-of-plane'' scattering produces changes in polarization due to geometric effects rather than polarization-dependent reflectance or optical path length. 

\begin{figure}[!h]
\centering
    \makebox[5pt]{{\includegraphics[trim={0 0 0 0},clip,width=0.3\textwidth]{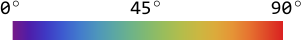}}}\vspace{-.2cm}\\
    \centering
        \subfloat[Sphere $\theta_h$]{\includegraphics[width=.35\columnwidth]{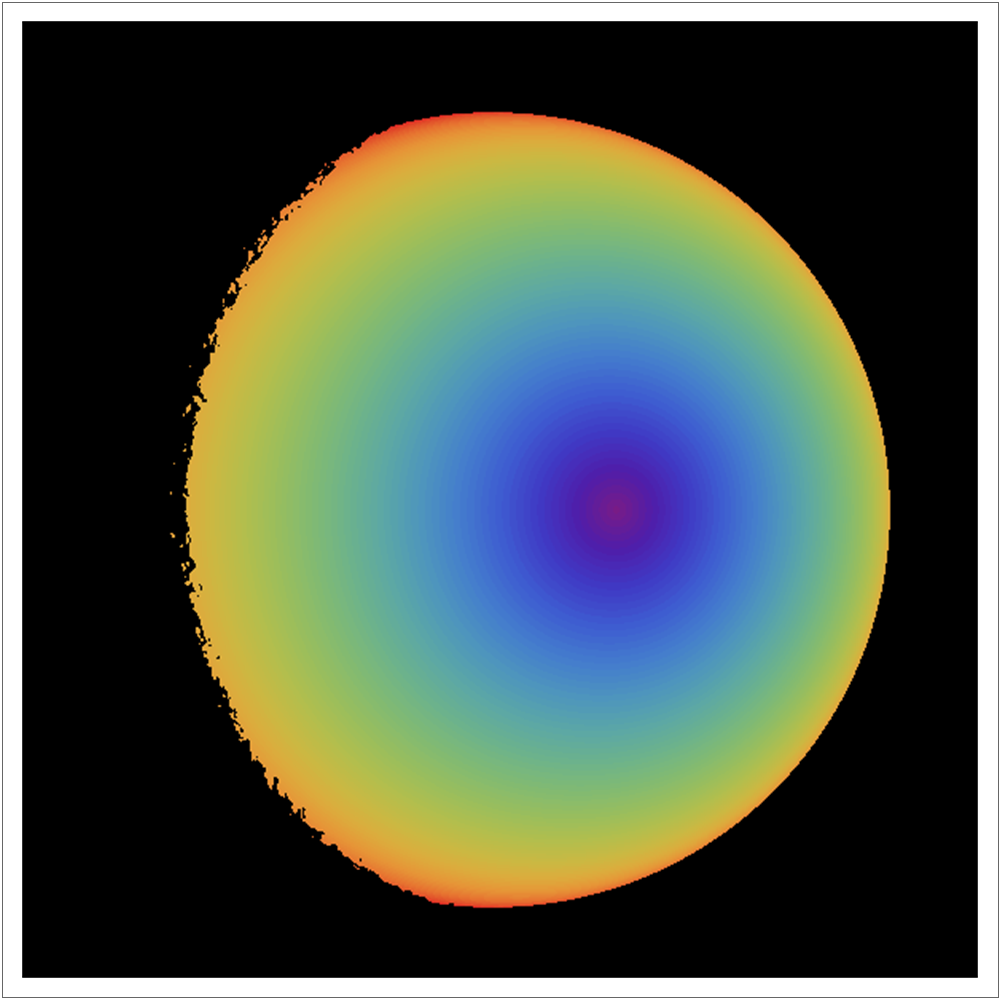}}\label{fig:theta_h}
        \hspace{.7cm}
        \subfloat[Bunny $\theta_h$]{\includegraphics[width=.35\columnwidth]{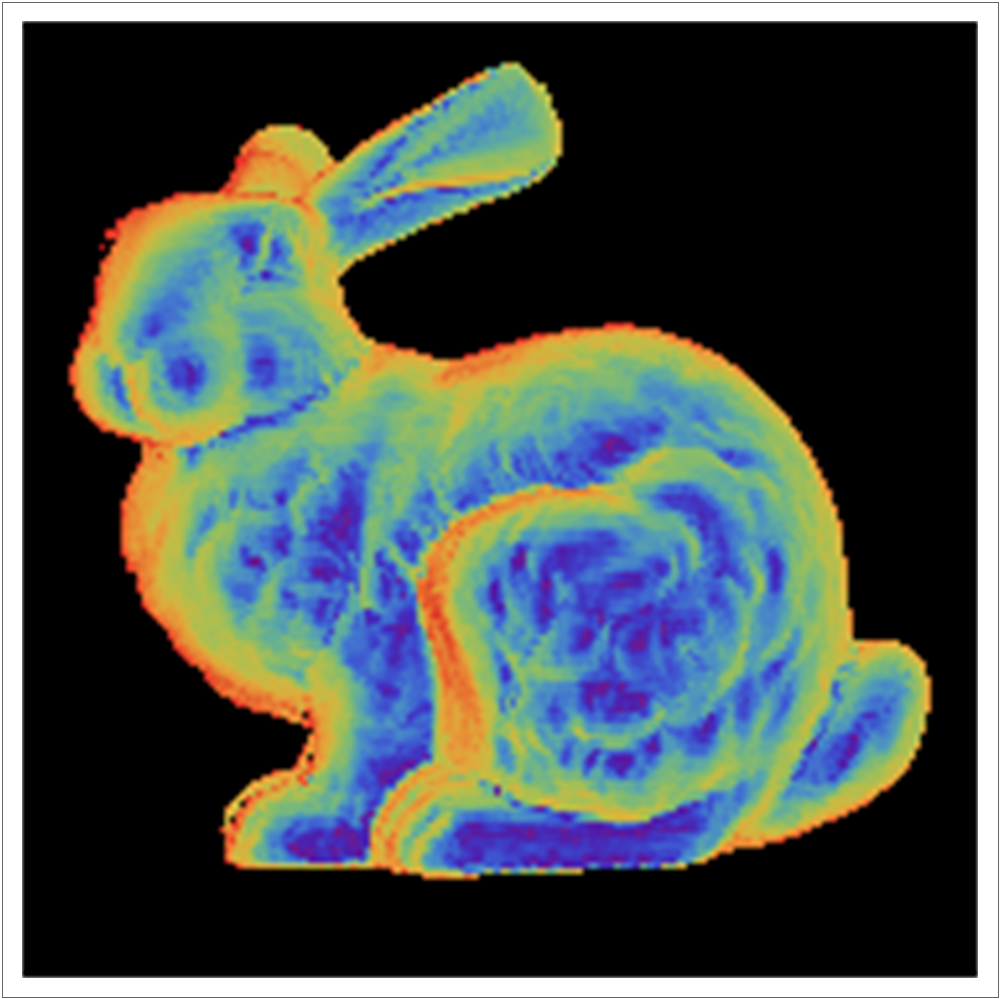}}\label{fig:bun_theta_h}
        \\
        \centering
    \makebox[5pt]{{\includegraphics[trim={0 0 0 -5},clip,width=0.3\textwidth]{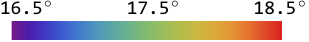}}}\vspace{-.2cm}\\
    \centering
        \subfloat[Sphere $\theta_d$]{\includegraphics[width=.35\columnwidth]{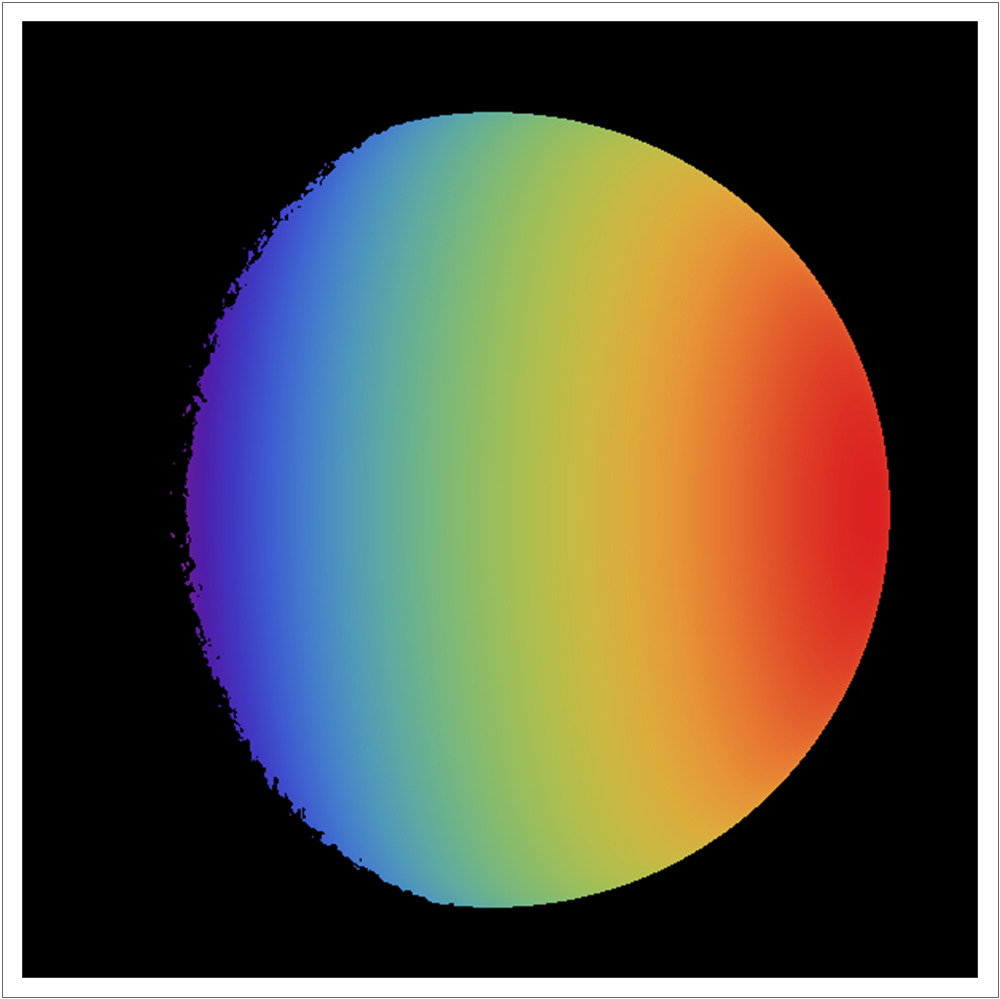}}\label{fig:theta_d}
        \hspace{.7cm}
        \subfloat[Bunny $\theta_d$]{\includegraphics[width=.35\columnwidth]{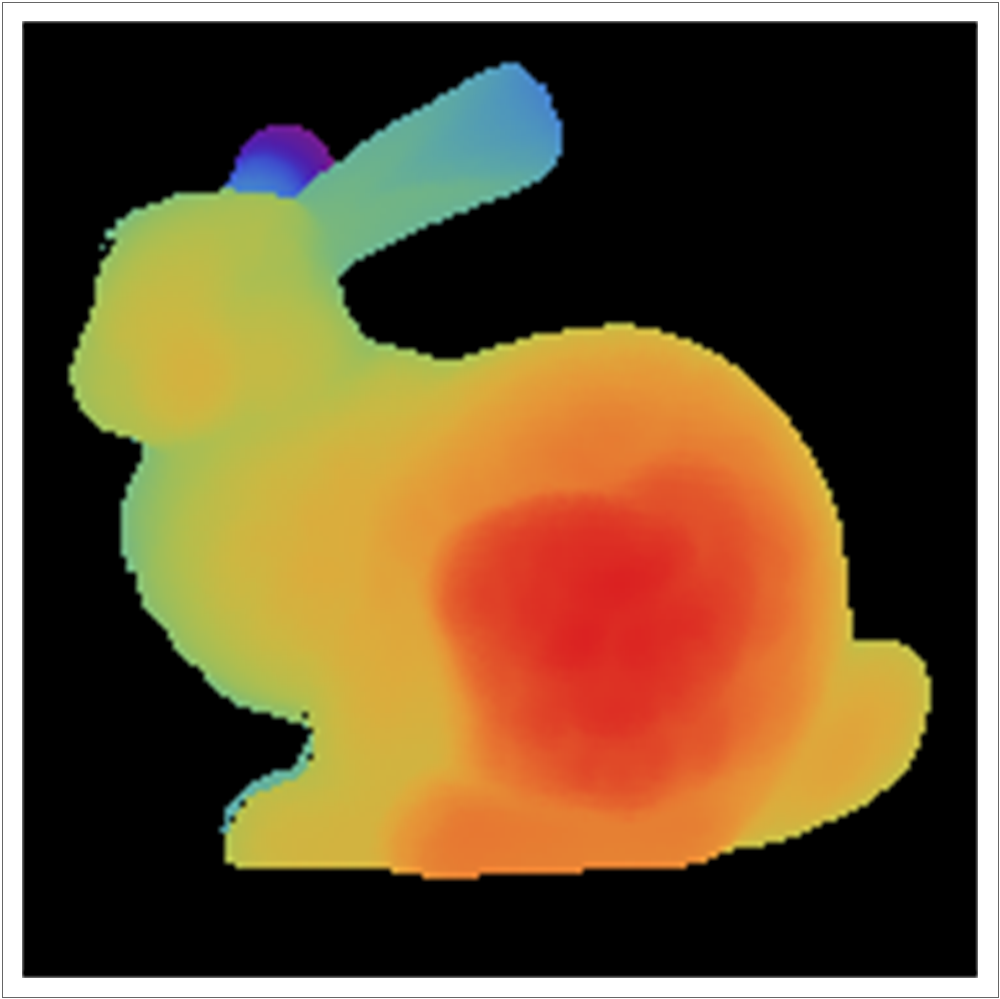}}\label{fig:bun_theta_d}
        \\
        \centering
        \makebox[5pt]{{\includegraphics[trim={0 0 0 -5},clip,width=0.3\textwidth]{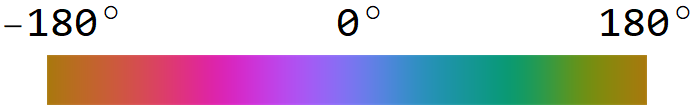}}}\vspace{-.2cm}\\
        \centering
        \subfloat[Sphere $\phi_d$]{\includegraphics[width=.35\columnwidth]{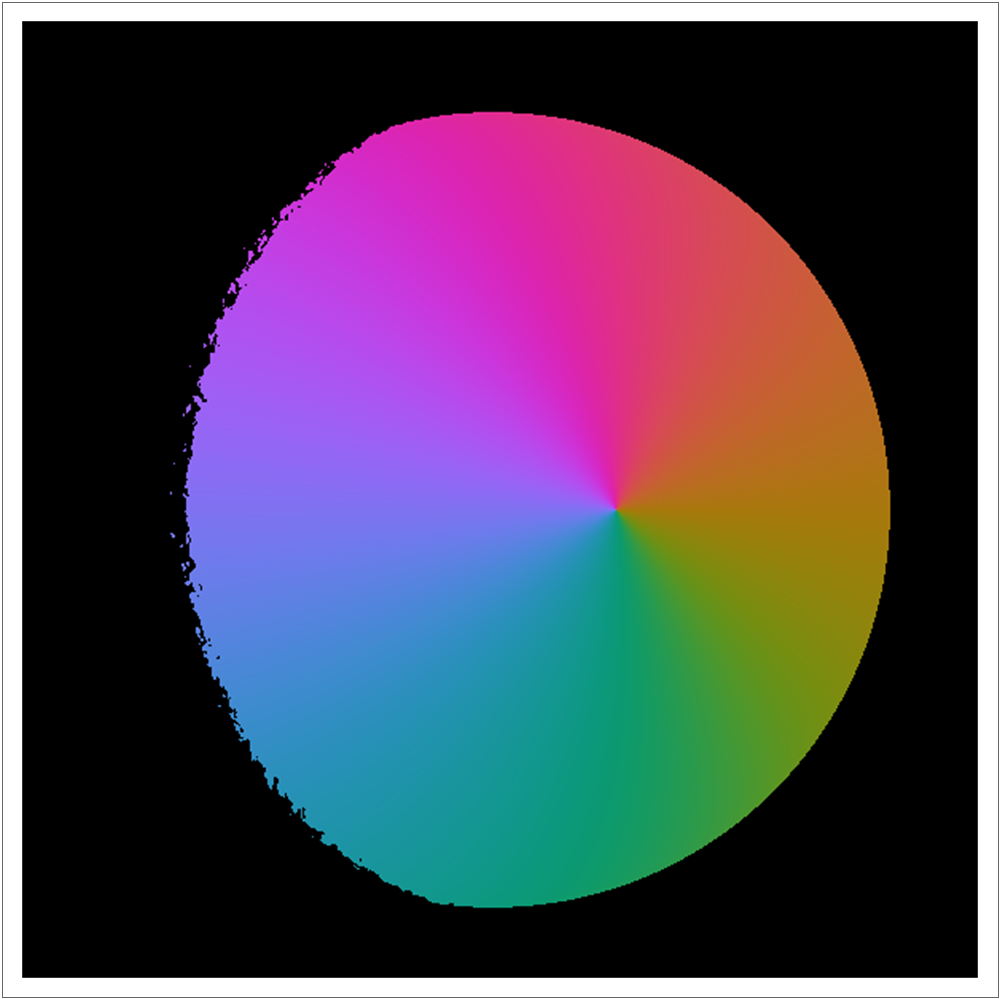}}\label{fig:phi_d}
        \hspace{.7cm}
        \subfloat[Bunny $\phi_d$]{\includegraphics[width=.35\columnwidth]{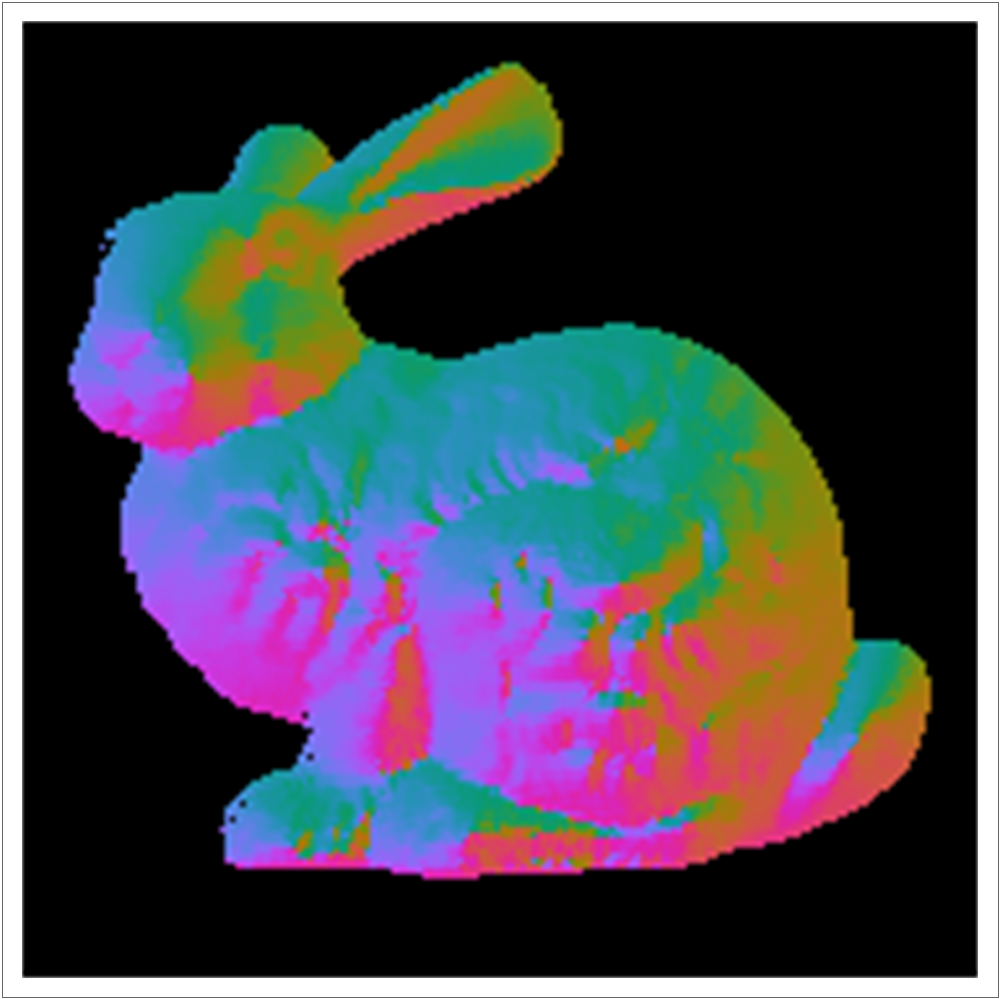}}\label{fig:bun_phi_d}
        \centering
            \caption{Rusinkiewicz coordinates $\theta_h$, $\theta_d$, and $\phi_d$ are shown in (a), (c), and (e), respectively, for a spherical object where $\Omega=35^\circ$ and in (b), (d), and (f), respectively, a Stanford bunny. 
            The Rusinkiewicz definitions are shown in Fig.~\ref{fig:rusinkCoords}. The regions in (a) and (b) where $\theta_h$ is small correspond to near-specular geometries where Fresnel reflection dominates. }\label{fig:measAngles}
\end{figure}

\subsection{First-Surface Reflection Component}\label{sect:specular}
The first-surface reflection, frequently referred to as specular reflection, is commonly modeled as Fresnel reflection from a hypothetical sub-resolution feature called a microfacet (the term first-surface reflection will be used going forward so that specular may be reserved to describe scattering configurations where $\theta_h=0^\circ$)\cite{priest_germer}. The microfacet is oriented such that it satisfies the Law of Reflection for a given pair of input and output ray directions. 

\begin{figure}[!h]
\centering
    \makebox[5pt]{{\includegraphics[trim={0 0 -10 0},clip,width=0.3\textwidth]{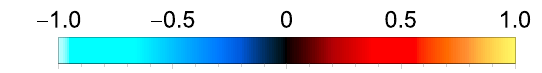}}}\vspace{-.2cm}\\
    \centering
        \subfloat[First surface reflection MJM]{\includegraphics[width=.23\textwidth]{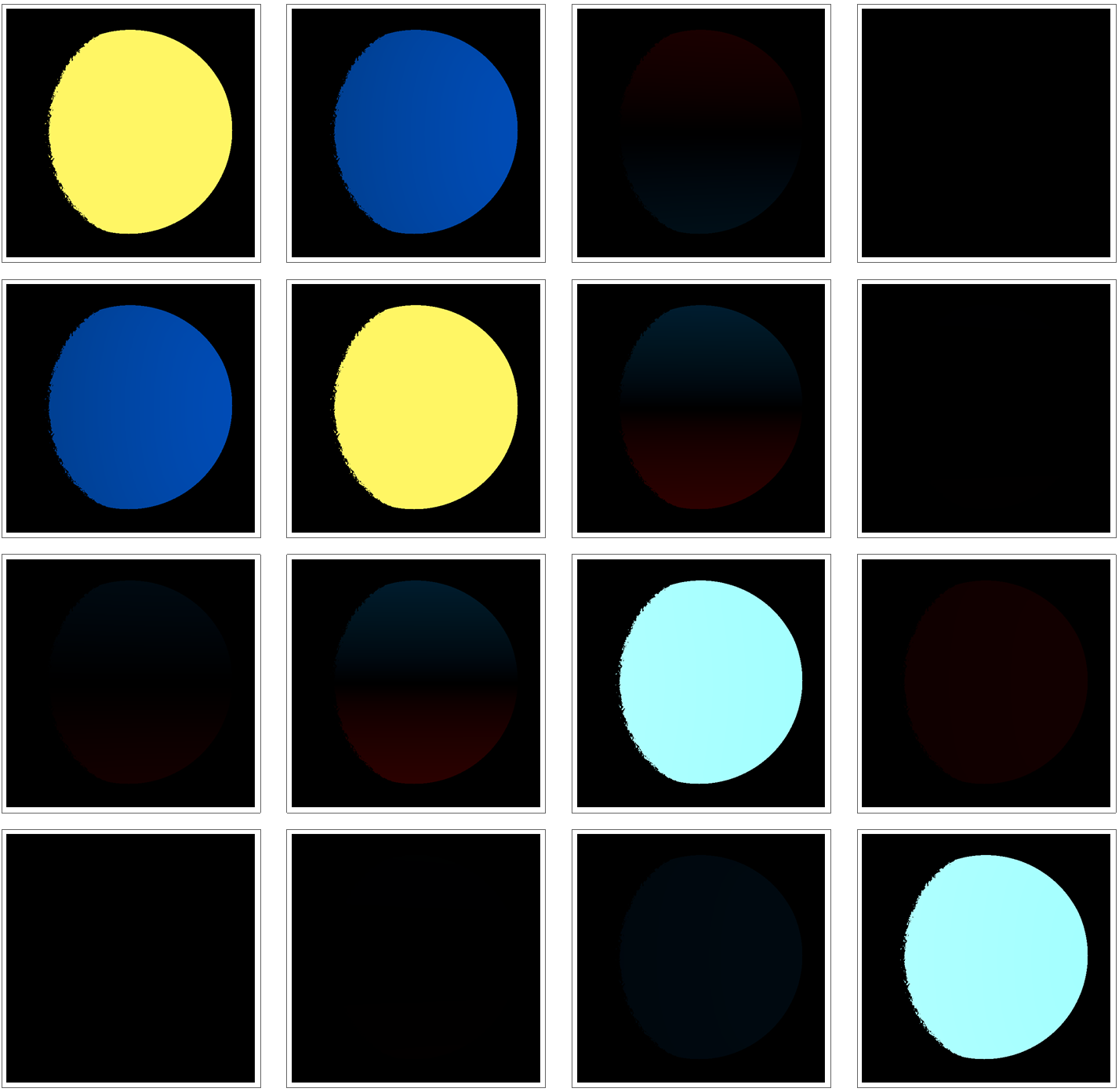}}\label{fig:specular}
        \hspace{.25cm}
        \subfloat[Diffuse polarization MJM]{\includegraphics[width=.23\textwidth]{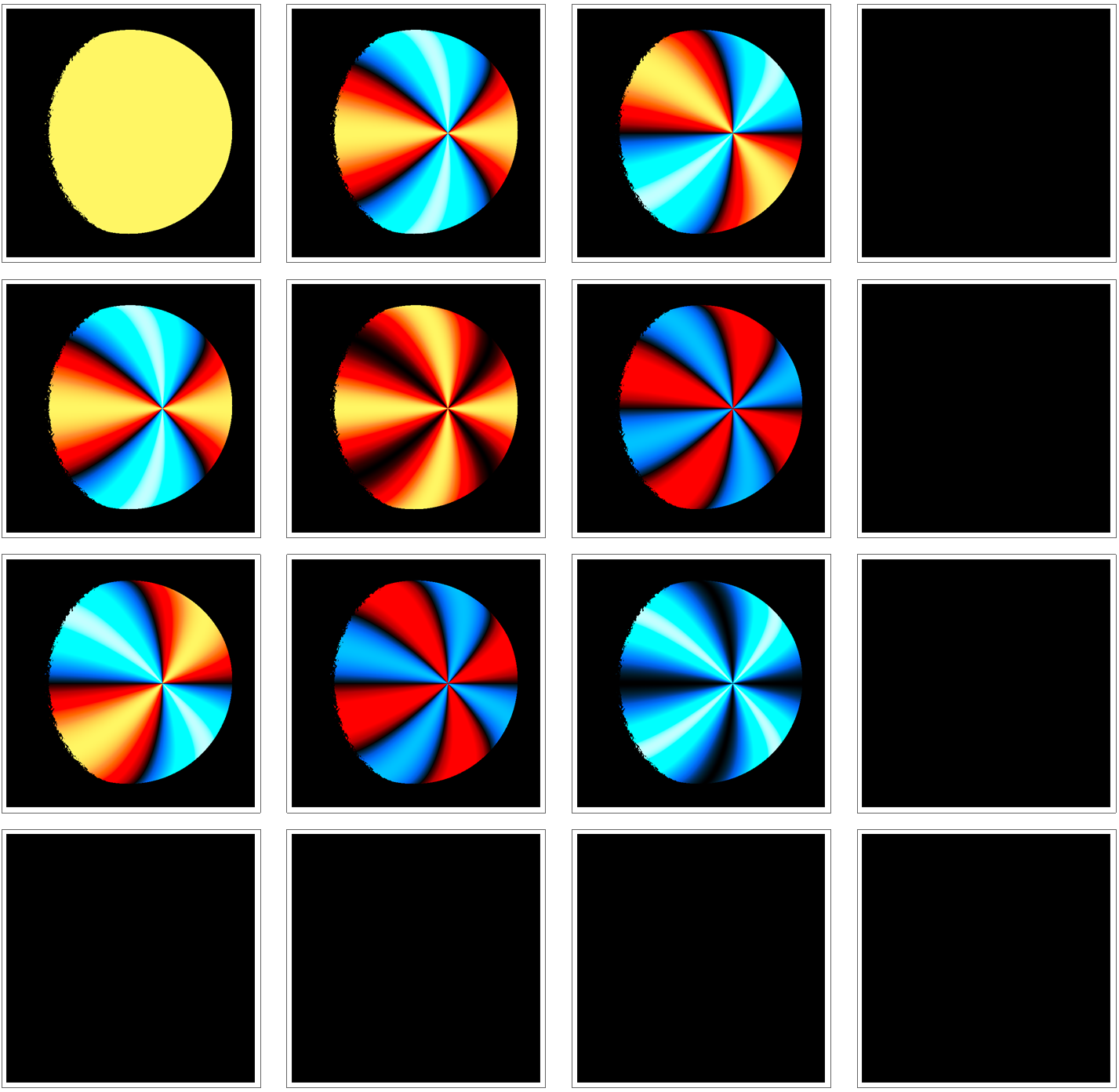}}
        \label{fig:diffuse}
            \caption{MM images of (a) the first-surface reflection model in Eq.~\ref{eq:spec} and (b) the diffuse model in Eq.~\ref{eq:diff} evaluated over a sphere in the same scattering configuration as in Fig.~\ref{fig:measAngles}. The first-surface reflection component is modeled as Fresnel reflection from microfacets, so the polarized scattering behavior does not depend directly on the surface normal. For this reason, the diattenuation and polarizance orientations do not follow the surface of the sphere. The non-zero values present in the $m_{1,2}$ and $m_{2,1}$ elements of (a) and (b) represent changes in the polarization state due to geometric effects rather than retardance.}\label{fig:modelMMs}
\end{figure}

This model is conceptually simple, however, implementation is complicated by the need to consider rotations in the plane transverse to the direction of propagation. The orientations of $s$- and $p$-polarizations in space vary based on the scattering geometry, and the orientations of horizontal and vertical for both the input and output coordinate systems are implicit parameters of the Jones matrix (JM) or MM. The JM for Fresnel reflection from a microfacet $\mathbf{F}$ is 
\begin{equation}
    \!\!\!\!\!\!\mathbf{F}_{n_{\lambda}}\!(\widehat{\pmb{\omega}}_i,\widehat{\pmb{\omega}}_o)\!=\!\!\begin{bmatrix}\widehat{\mathbf{x}}_{o}\\ \widehat{\mathbf{y}}_{o}\end{bmatrix} \!\!\!\begin{bmatrix}\widehat{\mathbf{s}}_o \\ \widehat{\mathbf{p}}_o\end{bmatrix}^{\mathrm{T}}\!\!\! \begin{bmatrix}r_s(n_\lambda,\theta_d) & 0\\0 & r_p(n_\lambda,\theta_d)\end{bmatrix}\!\!\!
    \begin{bmatrix}\widehat{\mathbf{s}}_i\\ \widehat{\mathbf{p}}_i\end{bmatrix}\!\!\! \begin{bmatrix}\widehat{\mathbf{x}}_{i} \\ \widehat{\mathbf{y}}_{i}\end{bmatrix}^{\mathrm{T}}\!\!\!\!\!.\label{eq:spec}
\end{equation}
Here $\widehat{\pmb{\omega}}_i$ and $\widehat{\pmb{\omega}}_o$ are the input (i) and output (o) ray directions, respectively; $\widehat{\mathbf{x}}_{i/o}$, $\widehat{\mathbf{y}}_{i/o}$ are the horizontal and vertical directions for the input and output coordinate system, respectively; $\widehat{\mathbf{s}}_{i/o}$ and $\widehat{\mathbf{p}}_{i/o}$ are the directions of $s$- and $p$-polarization for the incoming and outgoing rays with respect to the microfacet, respectively. 
The directions of $x$- and $y$-polarization are found $\widehat{\mathbf{x}}_{i/o}=\widehat{\mathbf{n}}\times\widehat{\pmb{\omega}}_{i/o}$ and $\widehat{\mathbf{y}}_{i/o}=\widehat{\pmb{\omega}}_{i/o}\times\widehat{\mathbf{x}}_{i/o}$.
The directions of $s$- and $p$-polarization are found $\widehat{\mathbf{s}}_{i}=\widehat{\mathbf{s}}_o=\widehat{\mathbf{h}}\times\widehat{\pmb{\omega}}_i$ and $\widehat{\mathbf{p}}_{i/o}=\widehat{\pmb{\omega}}_{i/o}\times\widehat{\mathbf{s}}_{i/o}$. The Fresnel coefficients $r_s$ and $r_p$ depend on the refractive index ratio $n_\lambda$ and the angle of incidence onto the microfacet $\theta_d$\cite{priest_germer}. The values used for $n_\lambda$ are presented in Tab.~\ref{tab:matProps}. 

Eq.~\ref{eq:spec} is invariant to the surface normal since Fresnel reflection is only defined for a normal vector that is halfway between the illumination and view vectors (i.e.~a microfacet). Instead, the Fresnel term only depends on the incident and scattered ray directions $\widehat{\pmb{\omega}}_i$ and $\widehat{\pmb{\omega}}_o$. The diattenuation magnitude depends only on $\theta_d$, which is half of the angle between $\widehat{\pmb{\omega}}_i$ and $\widehat{\pmb{\omega}}_o$. $\theta_d$ varies slowly over the sphere as shown in the example geometry in Fig.~\ref{fig:measAngles}, so the diattenuation magnitude also varies slowly. Variation in the diattenuation orientation is caused by variation in the orientation of $s$- and $p$-polarizations relative to the input and output coordinate systems. Both the $s$- and $p$-directions and the input and output coordinate systems vary over the field of view. These coordinate transforms lead to coupling between linear polarization states which can be observed faintly in the center $2\times2$ elements of the MM image in Fig.~\ref{fig:MMs}~(a). This should not be confused with true circular retardance, as it is the result of geometric effects rather than a relative phase difference between polarization states. When the microfacet surface normal $\widehat{\mathbf{h}}$ and the macroscopic surface normal $\widehat{\mathbf{n}}$ are coplanar (and therefore also coplanar with $\widehat{\pmb{\omega}}_i$ and $\widehat{\pmb{\omega}}_o$), there is no coupling between linear states caused by geometric effects.
This can be noted in the MM image of a purely Fresnel reflection $\widehat{\mathbf{m}}_0$ is shown in Fig.~\ref{fig:modelMMs}~(a), where there is a dark band through the center of the $m_{1,2}$ and $m_{2,1}$ elements. The geometries in this region are referred to as ``in-plane" scattering.

It is more correct to think of $\mathbf{J}_s(\widehat{\pmb{\omega}}_i,\widehat{\pmb{\omega}}_o)$ as a first-surface reflection model rather than a microfacet model. This is because the distribution of microfacet normals and adjacency effects are not explicitly included. Conventional microfacet models use distributions on the orientations of the microfacets to model surface roughness \cite{Walter2007,Breon,ashikmin2000microfacet,breon2017brdf}. Adjacency effects, specifically shadowing and masking, and the microfacet distribution can be interpreted as dictating the likelihood of an incident ray undergoing first-surface reflection toward the detector\cite{SmithShadow}. This likelihood corresponds to the relative contribution of first-surface reflection to other terms in the pBRDF model. In this work, the adjacency effects and microfacet distribution are absorbed into the dominant eigenvalue $\xi_0(\widehat{\pmb{\omega}}_i,\widehat{\pmb{\omega}}_o)$ from Eq.~\ref{eq:1par_a} which controls the weight between and the first-surface to diffuse weighting function $f_\lambda(\theta_h)$.

\subsection{Diffuse Reflection Component}
\begin{figure*}[!b]
\centering
    \makebox[5pt]{{\includegraphics[trim={0 0 -18 0},clip,width=0.4\textwidth]{figures/MM_colorbar_ticksontop.pdf}}}\\
    \centering
        \subfloat[Measured $\widehat{\mathbf{m}}_0$ 451 nm]{\includegraphics[width=.23\textwidth]{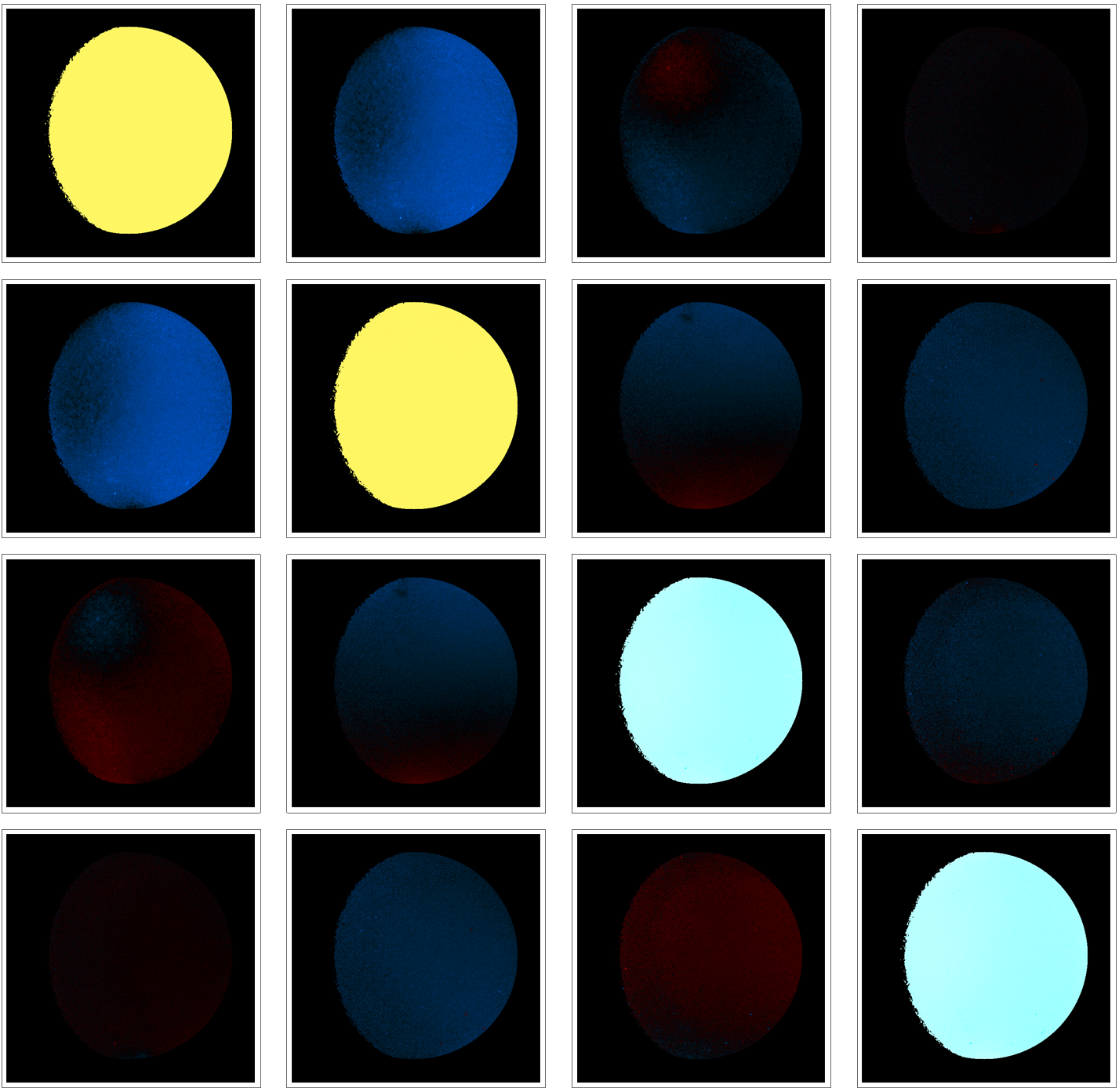}}\label{fig:meas451sphere}
        \hspace{.25cm}
        \subfloat[Modeled $\widehat{\mathbf{m}}_0$ 451 nm]{\includegraphics[width=.23\textwidth]{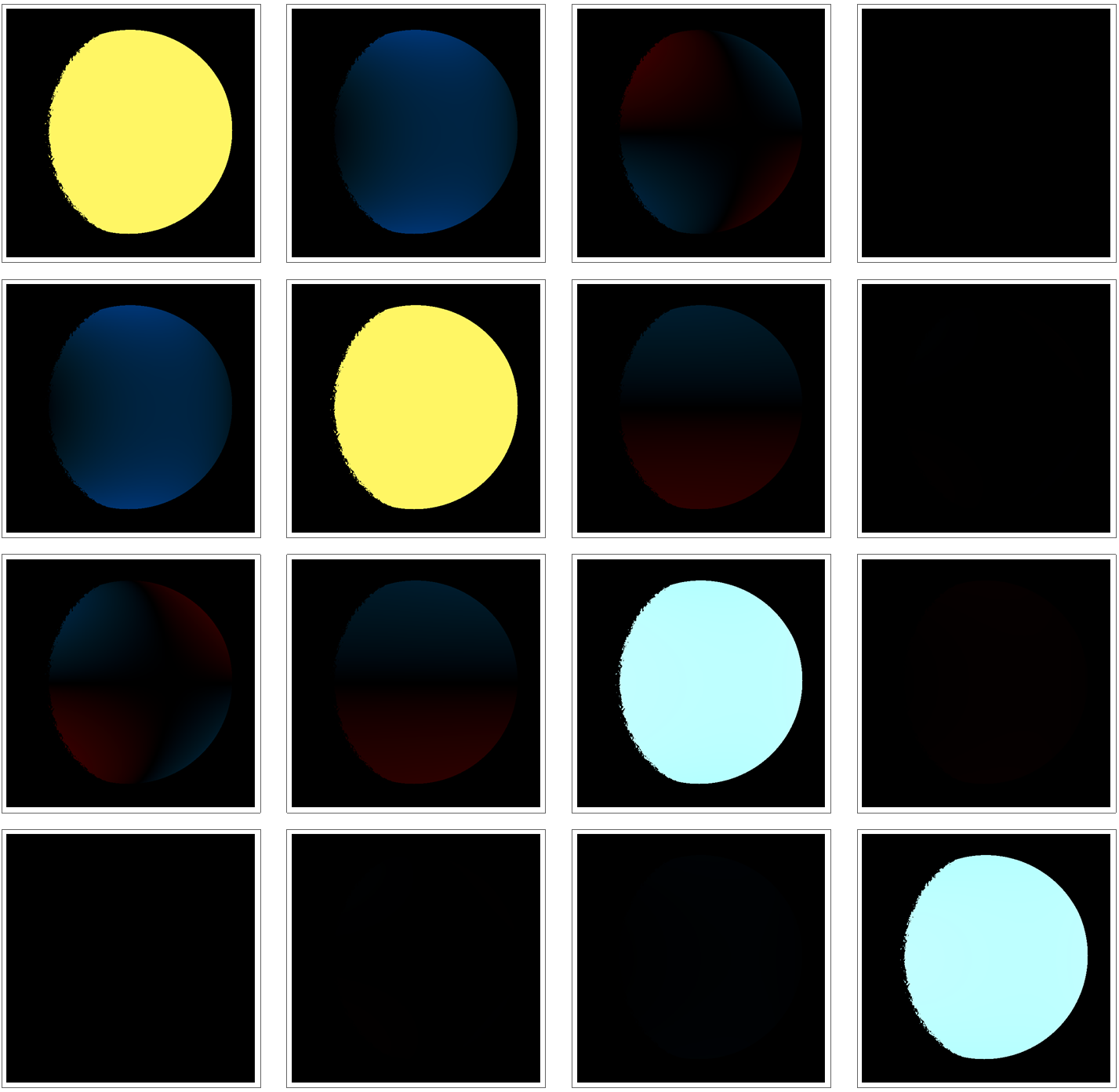}}\label{fig:meas662sphere}
        \hspace{.25cm}
        \subfloat[Measured $\widehat{\mathbf{m}}_0$ 662 nm]{\includegraphics[width=.23\textwidth]{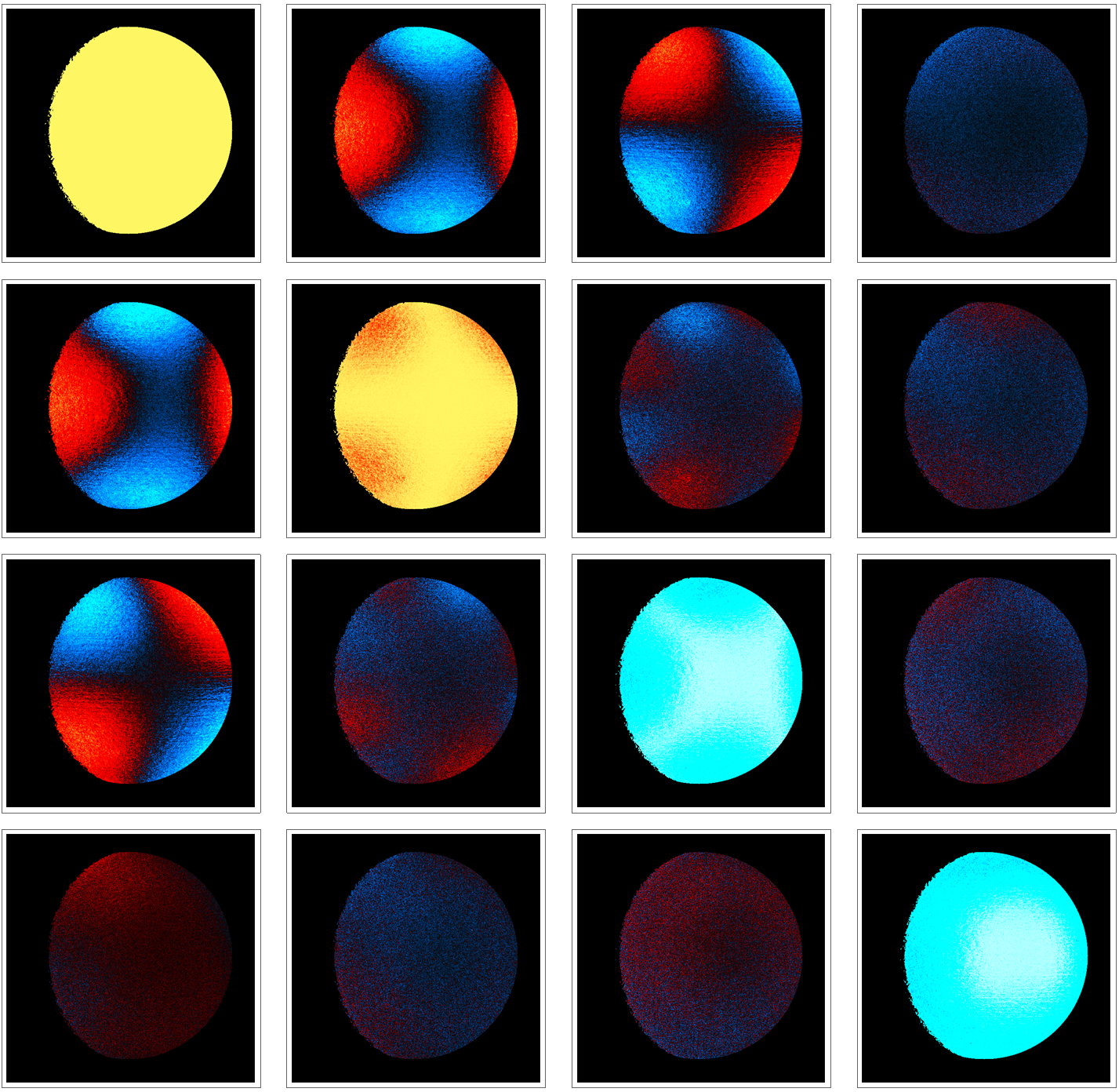}}\label{fig:model451sphere}
        \hspace{.25cm}
        \subfloat[Modeled $\widehat{\mathbf{m}}_0$ 662 nm]{\includegraphics[width=.23\textwidth]{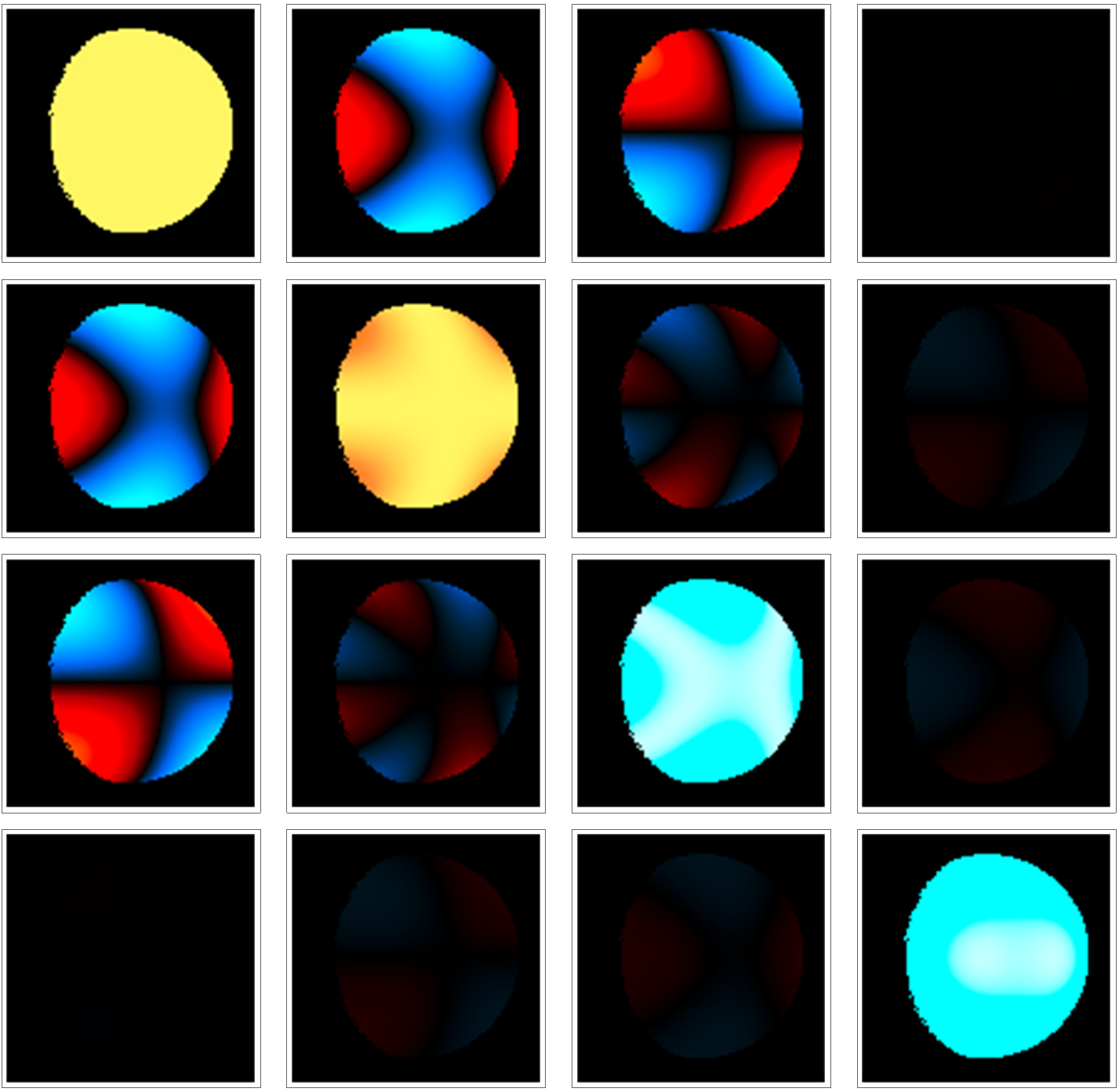}}\label{fig:model662sphere}\\
        \subfloat[Measured $\widehat{\mathbf{m}}_0$ 451 nm]{\includegraphics[width=.23\textwidth]{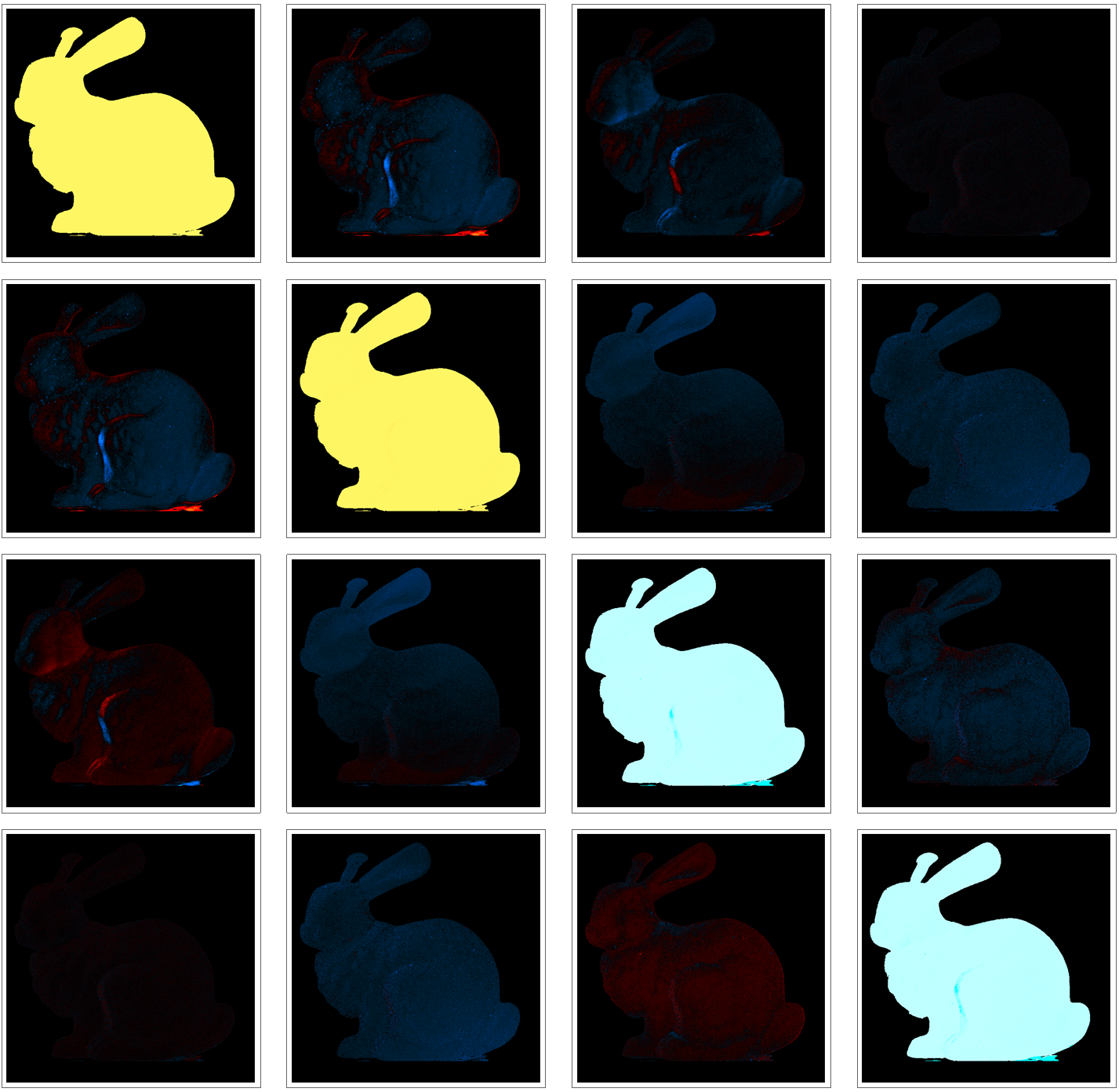}}\label{fig:meas451bunny}
        \hspace{.25cm}
        \subfloat[Modeled $\widehat{\mathbf{m}}_0$ 451 nm]{\includegraphics[width=.23\textwidth]{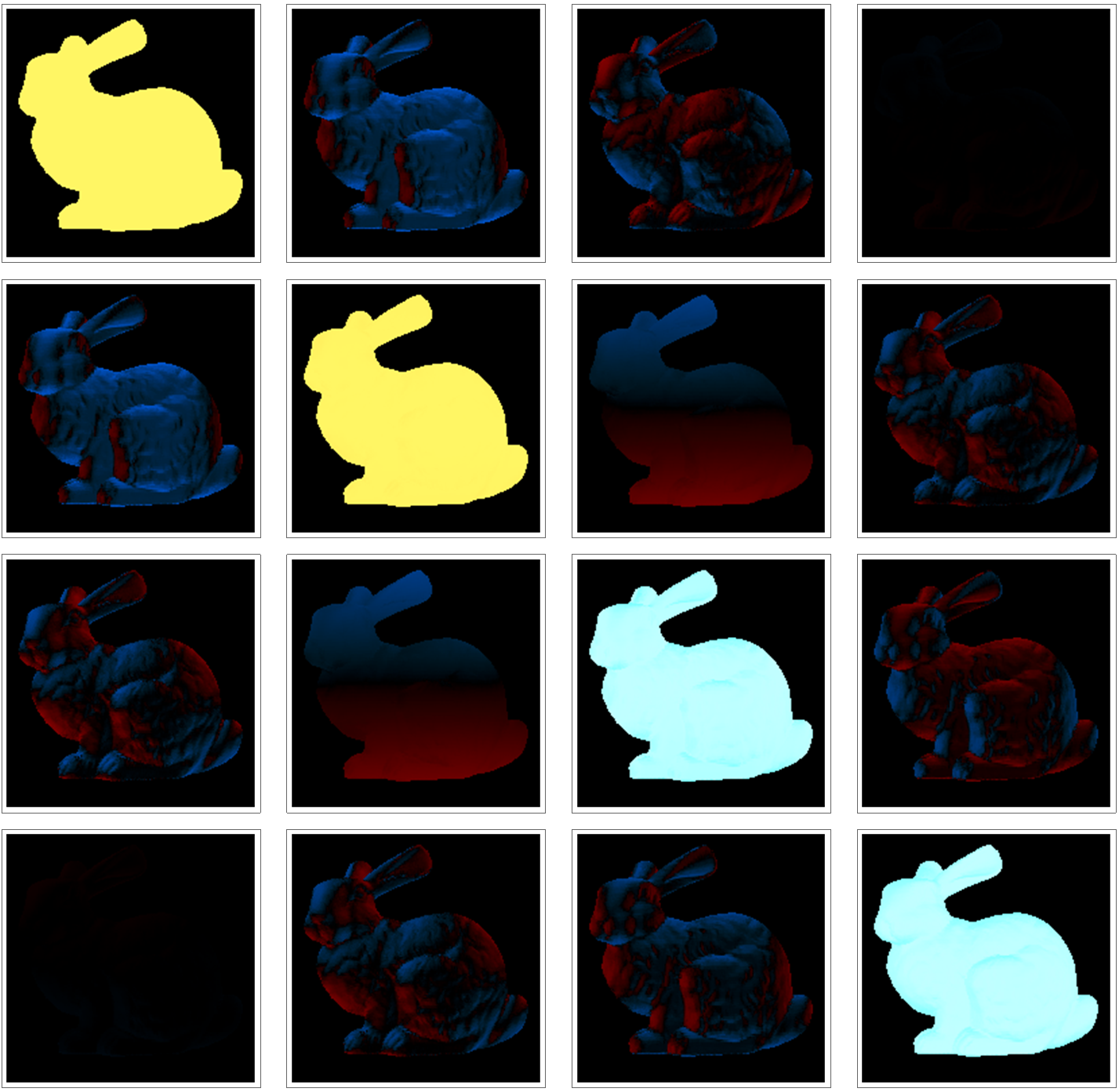}}\label{fig:meas662bunny}
        \hspace{.25cm}
        \subfloat[Measured $\widehat{\mathbf{m}}_0$ 662 nm]{\includegraphics[width=.23\textwidth]{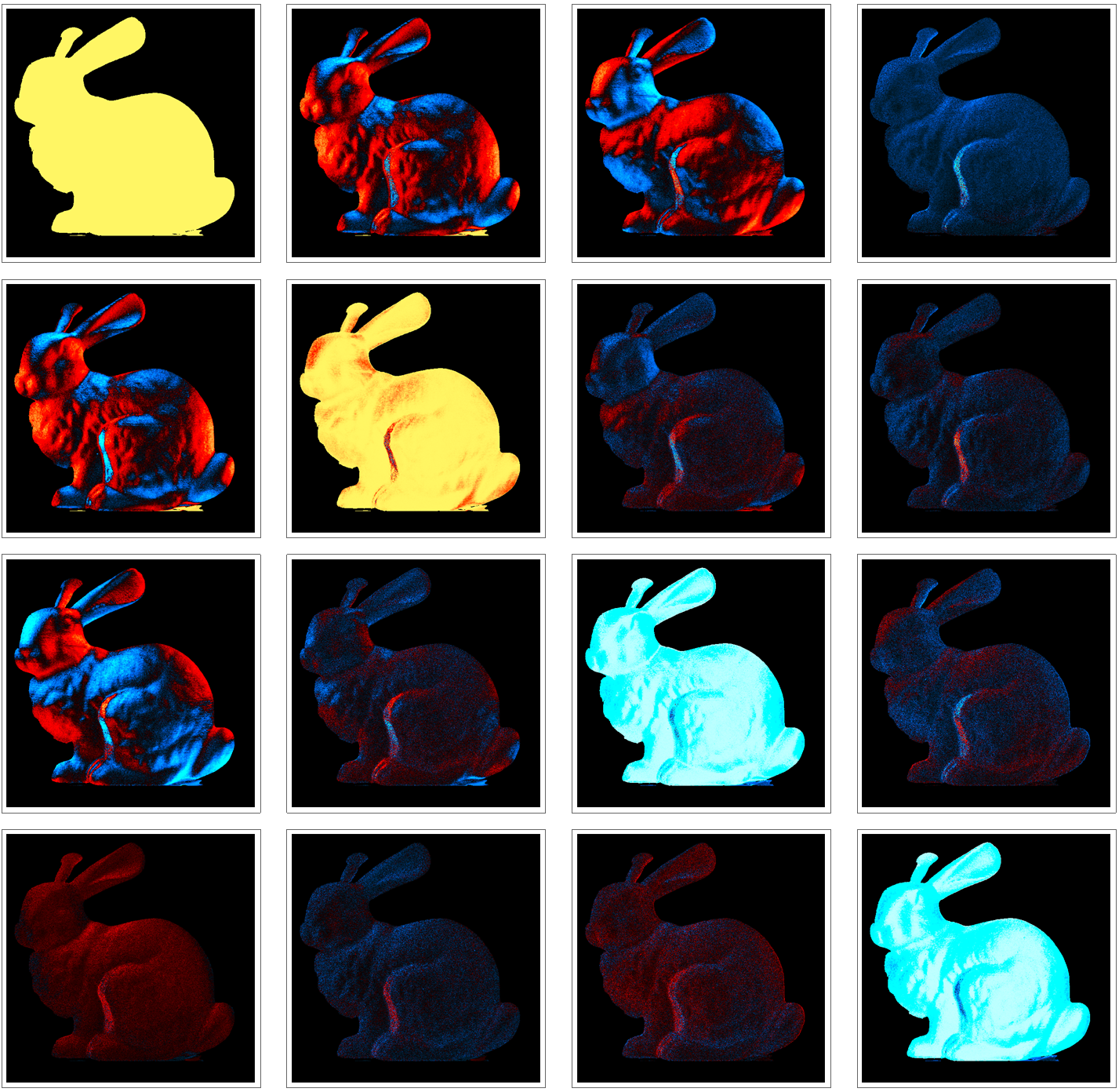}}\label{fig:model451bunny}
        \hspace{.25cm}
        \subfloat[Modeled $\widehat{\mathbf{m}}_0$ 662 nm]{\includegraphics[width=.23\textwidth]{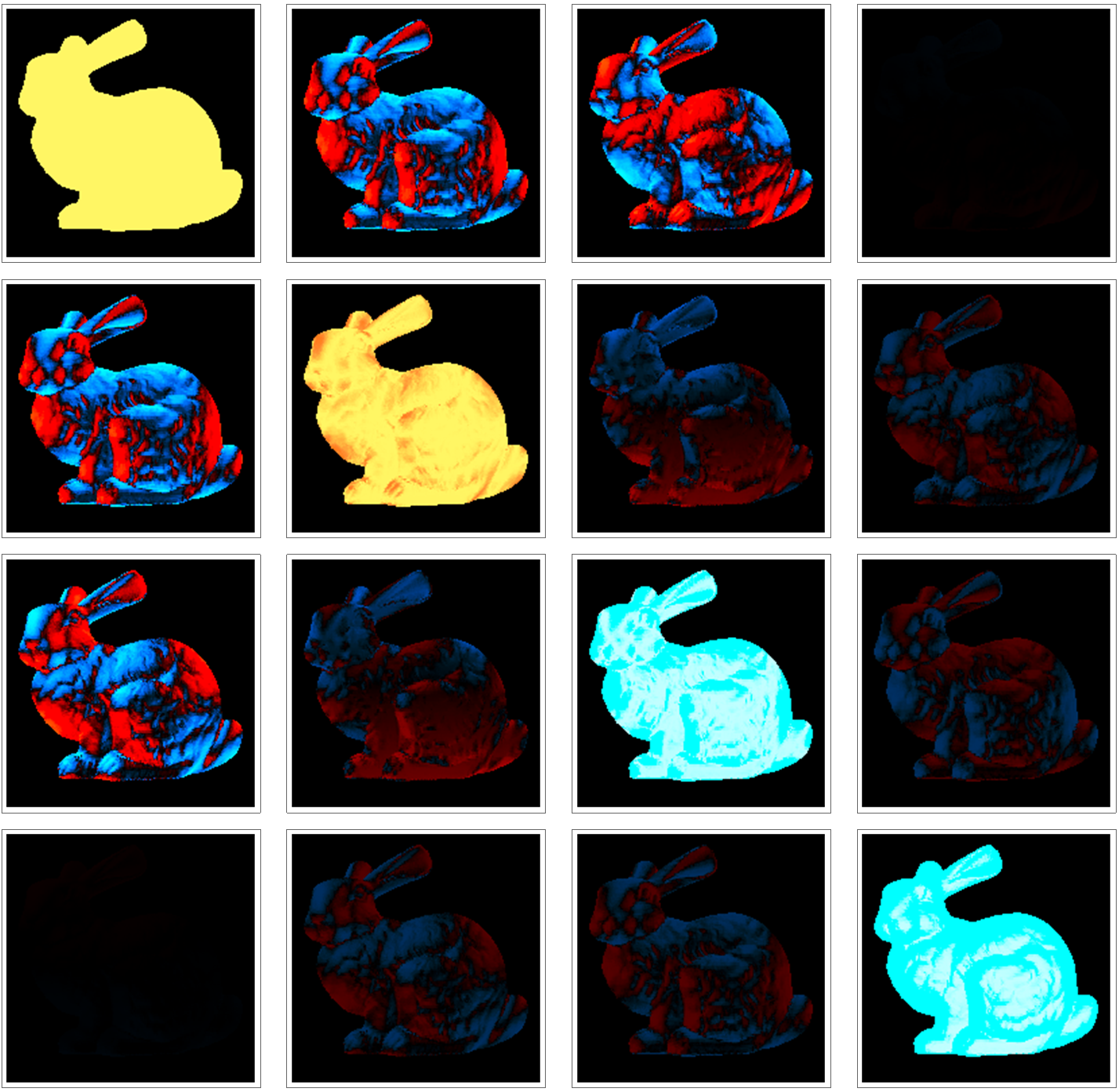}}\label{fig:model662bunny}
            \caption{Comparisons of dominant MJM from DRR measurements at $\Omega=35^\circ$ and from the models (see Eq.~\ref{eq:mixed}) for a red sphere (top row) and a red bunny (bottom row) of the same material. At 451 nm (left side) the low albedo creates lower diffuse scattering compared to 662 nm (right side). Properties of the capture system are not included in the MJM images from the models, specifically the finite spatial and polarimetric resolutions. Additionally, the model bunny MJM images are based on the geometry captured for the MM extrapolation from Stokes data experiment detailed in Sect.~\ref{sect:results}~C. There is therefore some inherent visual disagreement due to the geometry.  }\label{fig:MMs}
\end{figure*}

The diffuse reflection component is modeled as
\begin{align}
    \mathbf{S}(\widehat{\pmb{\omega}}_i,\widehat{\pmb{\omega}}_o,\widehat{\mathbf{n}})
    &=\begin{bmatrix}1&0\\0&-1\end{bmatrix}\mathbf{R}(\phi_d)\begin{bmatrix}1&0\\0&0\end{bmatrix}\mathbf{R}(-\phi_d),\label{eq:diff}
\end{align}
where the rightmost three JMs represent a polarizer with unit diattenuation and a transmission axis oriented at $\phi_d$. This matrix has a diattenuation orientation that is in the plane spanned by the surface normal $\widehat{\mathbf{n}}$ and $\widehat{\pmb{\omega}}_o$, a pattern which is centered about the point on the sphere where $\theta_h=0^\circ$ as shown in Fig.~\ref{fig:measAngles}~(a). When left multiplied by the reflection matrix, the $45^\circ$ and $135^\circ$ polarizations are reversed. This pattern being centered about $\theta_h=0^\circ$ is consistent with the $\widehat{\mathbf{m}}_0$ images observed from DRR measurements of spheres (see Fig.~\ref{fig:MMs}~(c)), however, it does mark a departure from many other diffuse polarization models. In models such as the one proposed by Atkinson and Hancock, the orientation of the diffuse term is centered about the central camera axis\cite{Atkinson_Hancock_2006}. 

\subsection{Mixed Dominant Mueller-Jones Matrix Model}
Previous work demonstrated extrapolating MM images using a purely Fresnel reflection model for the dominant MJM, but this work was performed on flat objects which have a smaller range of scattering geometries\cite{Jarecki_proc,jarecki2022}. The objects and scattering geometries described in this work require models for $\widehat{\mathbf{m}}_0$ that introduce a polarized diffuse reflection term. This term is characterized by a diattenuation orientation that depends on the surface normal. However, the dominant MJM necessarily must be non-depolarizing, and simply summing different MMs, in general, introduces depolarization.

To create a mixed polarization model without introducing depolarization, the individual first-surface and diffuse components are combined as Jones matrices as in

\begin{equation}
    \mathbf{J}(\widehat{\pmb{\omega}}_i,\widehat{\pmb{\omega}}_o,\widehat{\mathbf{n}}|n_\lambda,a_\lambda,b_\lambda) = \mathbf{F}_{n_\lambda}+a_\lambda \mathrm{sin}^{b_{\lambda}}(\theta_h)\mathbf{S}(\widehat{\mathbf{n}}).\label{eq:mixed}
\end{equation}
Here, the functional dependencies of $\mathbf{J}$ are separated into the scattering geometry and material parameters, with the separation notated by the vertical bar $|$. There are four material-dependent parameters which are constant for all scattering geometries: the real and imaginary components of the refractive index $n_\lambda$, and two parameters $a_\lambda$ and $b_\lambda$ which control how the weight of the diffuse polarization term depends on scattering geometry. The four material parameters efficiently reduce the original six degrees of freedom for the diattenuation and retardance of the MJM. Both matrices on the right-hand-side depend on $\widehat{\pmb{\omega}}_i$ and $\widehat{\pmb{\omega}}_o$, but this has been dropped for brevity.

The surface normal of the material $\widehat{\mathbf{n}}$ (see Fig.~\ref{fig:rusinkCoords}) affects the diffuse but not Fresnel term. $\mathbf{F}(\widehat{\pmb{\omega}}_i,\widehat{\pmb{\omega}}_o,n_\lambda)$ is the first-surface reflection component modeled as Fresnel reflection based on refractive index ratio $n_\lambda$ at wavelength $\lambda$, $\mathbf{S}(\widehat{\pmb{\omega}}_i,\widehat{\pmb{\omega}}_o,\widehat{\mathbf{n}})$ is the diffuse polarization term and is defined with unit throughput, so its relative weight as a function of scattering geometry is given by the sine function. 
$\mathbf{J}$ is converted to a MJM, normalized, and then used as $\widehat{\mathbf{m}}_0$ for a TD-MM model as in Eq.~\ref{eq:1par_a}. 

\begin{table}
\begin{center}
\caption{Material constants determined \emph{ad hoc} to match the models to the observed MM measurements. These parameters are constant with respect to scattering geometry.} 
\label{tab:matProps}
\begin{tabular}{| c| c |c| }
\hline
 Material constants & 451 nm & 662 nm \\ 
 \hline\hline
 $n_\lambda$ & $1.20+0.25i$ & $1.30+0.08i$ \\  
 \hline
 $a_\lambda$ & 0.03 & 0.17 \\ 
 \hline
 $b_\lambda$ & 2.5 & 2.0\\
 \hline
\end{tabular}
\end{center}
\end{table}

The values for the material properties are found in Tab.~\ref{tab:matProps}. These values, as well as the choice to use a sine function for the geometric dependence of the weight, were determined \emph{ad hoc} to match the measurements. 
The weight function is modeled as only depending on $\theta_h$ because $\theta_h$ is a measure of deviation from a specular scattering configuration. The sine function was chosen for the weight functions based on the observation that Fresnel reflection-like behavior dominates for near-specular scattering geometries (where $\theta_h$ is small) and diffuse polarization behavior tends to become dominant away from specular geometries. The coefficients and powers in these equations were empirically chosen to match the diattenuation trends observed in measurements.

\section{Eigenvalue Estimation Method}\label{sect:method}
 A collection of noise-free polarimetric measurements $\mathbf{P}$ of a general MM is modeled as $\mathbf{P}=\mathbf{W}_N\mathbf{M}$, where $\mathbf{W}_N$ is the polarimetric measurement matrix with $N$ number of measurements and the MM is expressed as a 16 $\times$ 1 vector $\mathbf{M}$. If $\mathbf{M}$ is accurately described by a TD-MM model as in Eq.~\ref{eq:1par_a}, then the noise-free polarimetric measurements can be modeled as $\mathbf{P}=\boldsymbol{\Phi}\pmb{\alpha}.$
Here $\boldsymbol{\Phi}$ is an $N\times2$ matrix
\begin{equation}
    \boldsymbol{\Phi}=
    \begin{bmatrix}
        \mathbf{p}_0 & 
        \mathbf{p}_{ID}
    \end{bmatrix}=\mathbf{W}_N
    \begin{bmatrix}
        \widehat{\mathbf{m}}_0 & 
        \mathbf{m}_{ID}
    \end{bmatrix},  
    \label{eq:basis}
\end{equation}
where $\pmb{\alpha}$ is a vector whose elements are the coefficients on each MM in the TD model
\begin{equation}
    \pmb{\alpha}=
    \begin{bmatrix}
        \alpha_0\\ \alpha_{ID}
    \end{bmatrix}=\frac{4M_{00}}{3}
    \begin{bmatrix}
        \xi_0-\frac{1}{4}\\1-\xi_0
    \end{bmatrix}.
    \label{eq:alpha}
\end{equation}

The estimate for $\widetilde{\pmb{\alpha}}$ is given by applying the pseudoinverse of $\boldsymbol{\Phi}$ to noisy measurements $\widetilde{\mathbf{P}}$ as in $\widetilde{\pmb{\alpha}}=\boldsymbol{\Phi}^{+}\widetilde{\mathbf{P}}$.
The estimates for the dominant eigenvalue $\widetilde{\xi}_0$ and for the average reflectance $\widetilde{M}_{00}$ are calculated from the elements of $\widetilde{\pmb{\alpha}}$
\begin{equation}
    \widetilde{\xi}_0=\frac{\frac{1}{4}+\widetilde{\alpha}_0/\widetilde{\alpha}_{ID}}{1+\widetilde{\alpha}_0/\widetilde{\alpha}_{ID}},\label{eq:xi0_est}
\end{equation}
and
\begin{equation}
    \widetilde{M}_{00}=\frac{3\widetilde{\alpha}_0}{4\widetilde{\xi}_0-1}.
    \label{eq:M00_est}
\end{equation}
In this work, the estimated average reflectance $\widetilde{M}_{00}$ is not considered because the polarimeters used are not calibrated to produce results that are scaled as fractions of the illumination radiance. By normalizing the results, or setting $M_{00}=1$, the polarimetric properties can be directly compared.

For a general MM, $\mathbf{W}$ must be rank sixteen to perform a complete reconstruction. In the case of a TD-MM when $\widehat{\mathbf{m}}_0$ is known \emph{a priori} or provided by a model, the two unknown degrees of freedom can be estimated with a rank two measurement matrix. The complete DRR Mueller imaging polarimeter used in this work performs 40 polarimetric measurements, $\mathbf{W}_{40}$, making it an overdetermined system \cite{LopezTellez_rgb950}. The partial polarimeter used in this work is a Sony Triton 5.0MP Polarization Camera. This is a division of the focal plane linear Stokes camera. The light source from the DRR polarimeter is used to illuminate the object with horizontally polarized light. The measurement matrix for the linear Stokes camera, $\mathbf{W}_{4}$, has four rows but is rank three. This is underdetermined for reconstructing a general MM, but is overdetermined for estimating the two remaining degrees of freedom in Eq.~\ref{eq:alpha}.

\section{Results}\label{sect:results}
\subsection{Dominant MJM Model Comparison}
\begin{figure}[!h]
\centering
    \makebox[5pt]{{\includegraphics[trim={0 0 0 0},clip,width=0.3\textwidth]{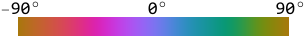}}}\vspace{-.2cm}\\
    \centering
        \subfloat[Measured $\psi$ 451 nm]{\includegraphics[width=.35\columnwidth]{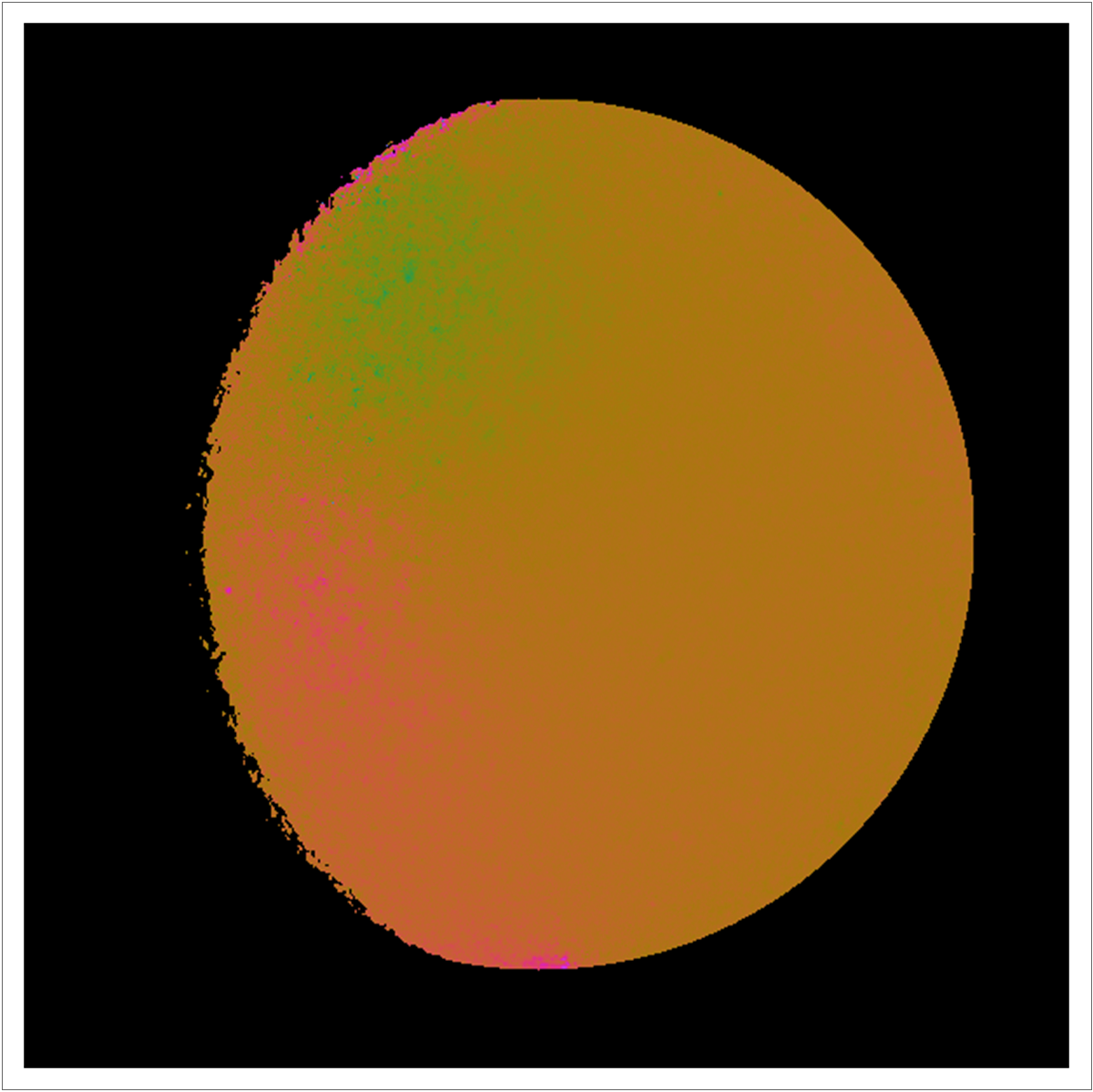}}\label{fig:RGB950_aolp_451}
        \hspace{.7cm}
        \subfloat[Measured $\psi$ 662 nm]{\includegraphics[width=.35\columnwidth]{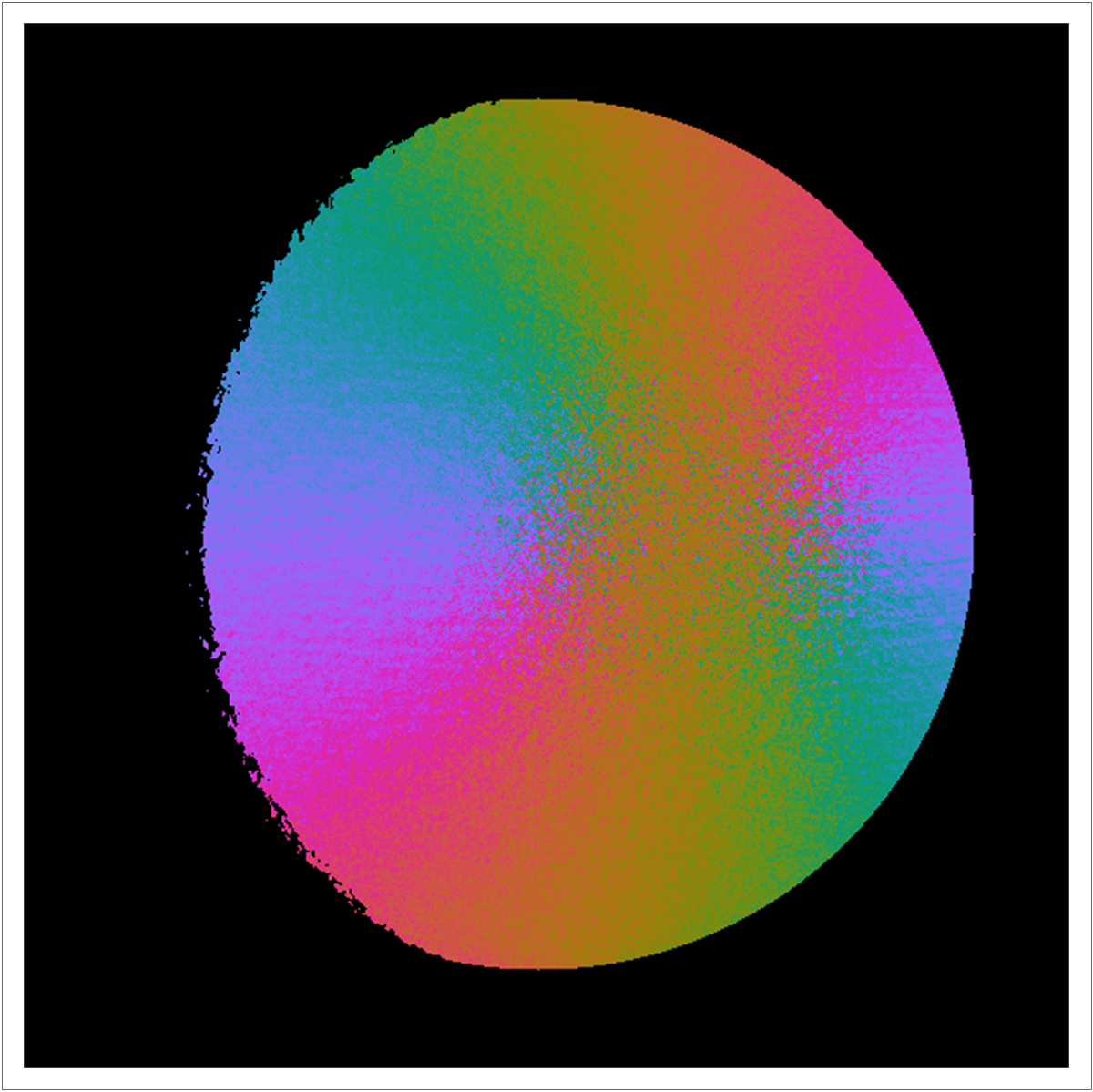}}\label{fig:RGB950_aolp_662}
        \\
        \subfloat[Modeled $\widetilde{\psi}$ 451 nm]{\includegraphics[width=.35\columnwidth]{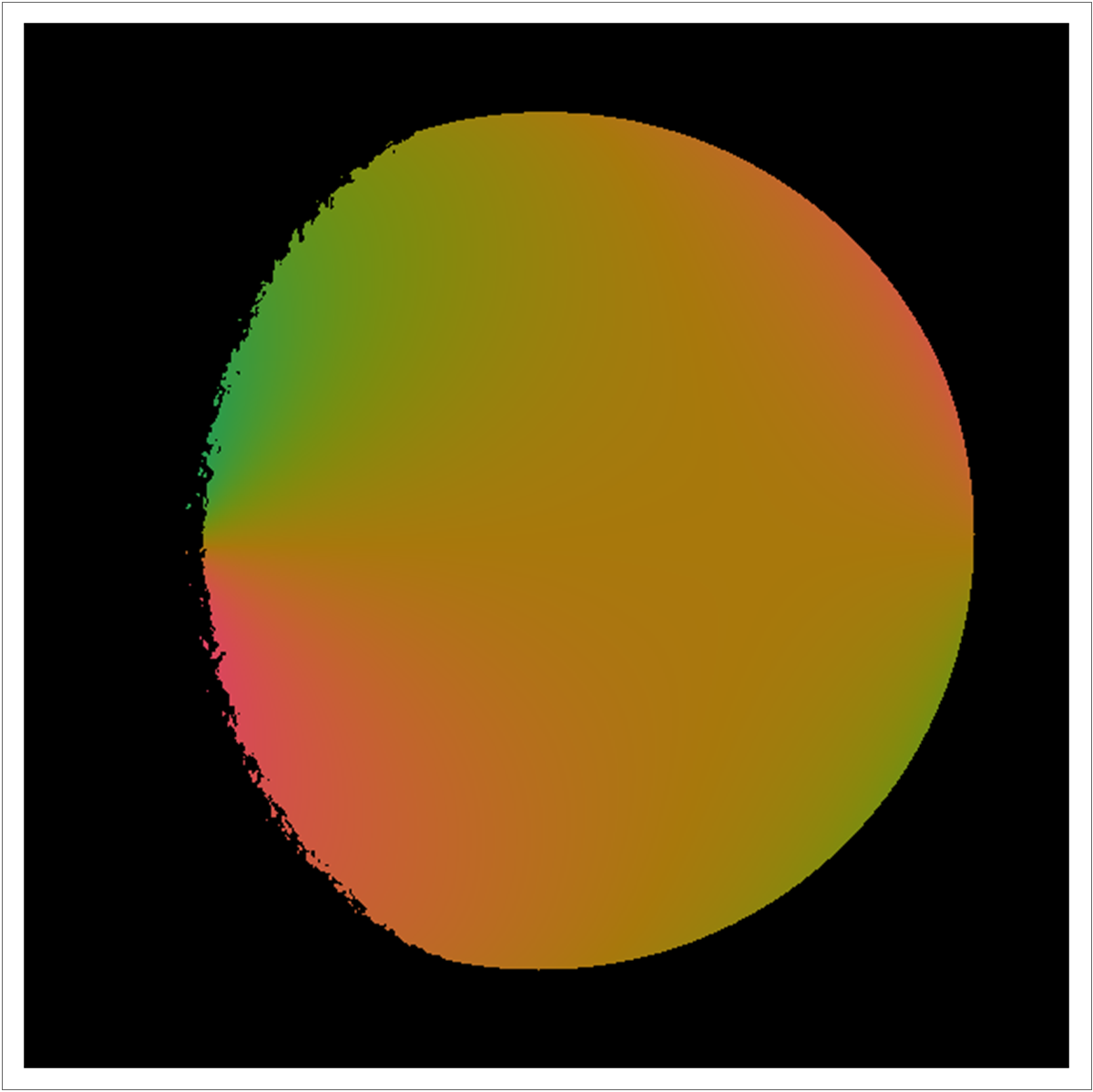}}\label{fig:model_aolp_451}
        \hspace{.7cm}
        \subfloat[Modeled $\widetilde{\psi}$ 662 nm]{\includegraphics[width=.35\columnwidth]{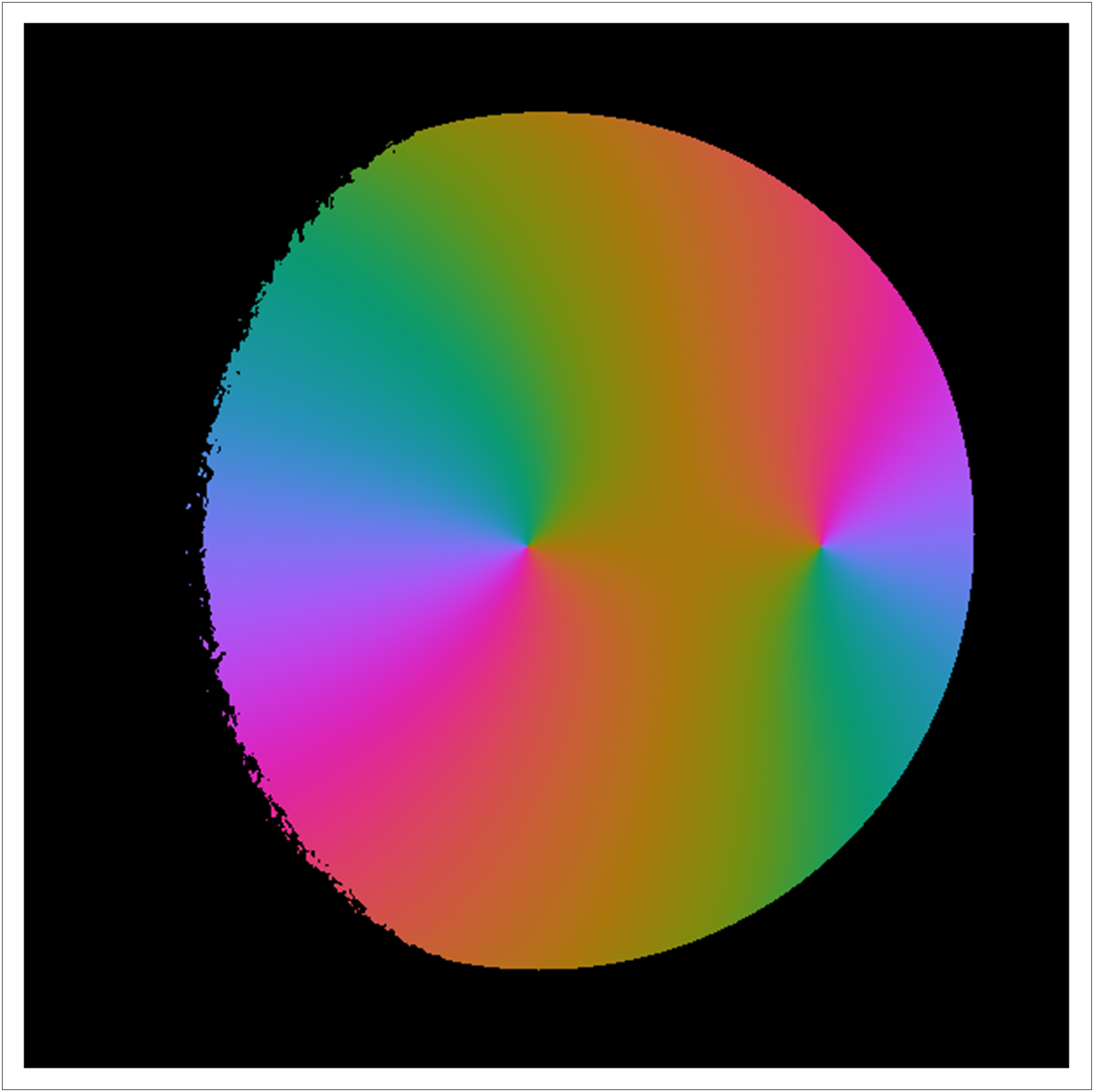}}\label{fig:model_aolp_662}\\
        \centering
    \makebox[5pt]{{\includegraphics[trim={0 0 0 -10},clip,width=0.3\textwidth]{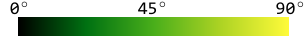}}}\vspace{-.2cm}\\
        \subfloat[$\psi$ deviation 451 nm]{\includegraphics[width=.35\columnwidth]{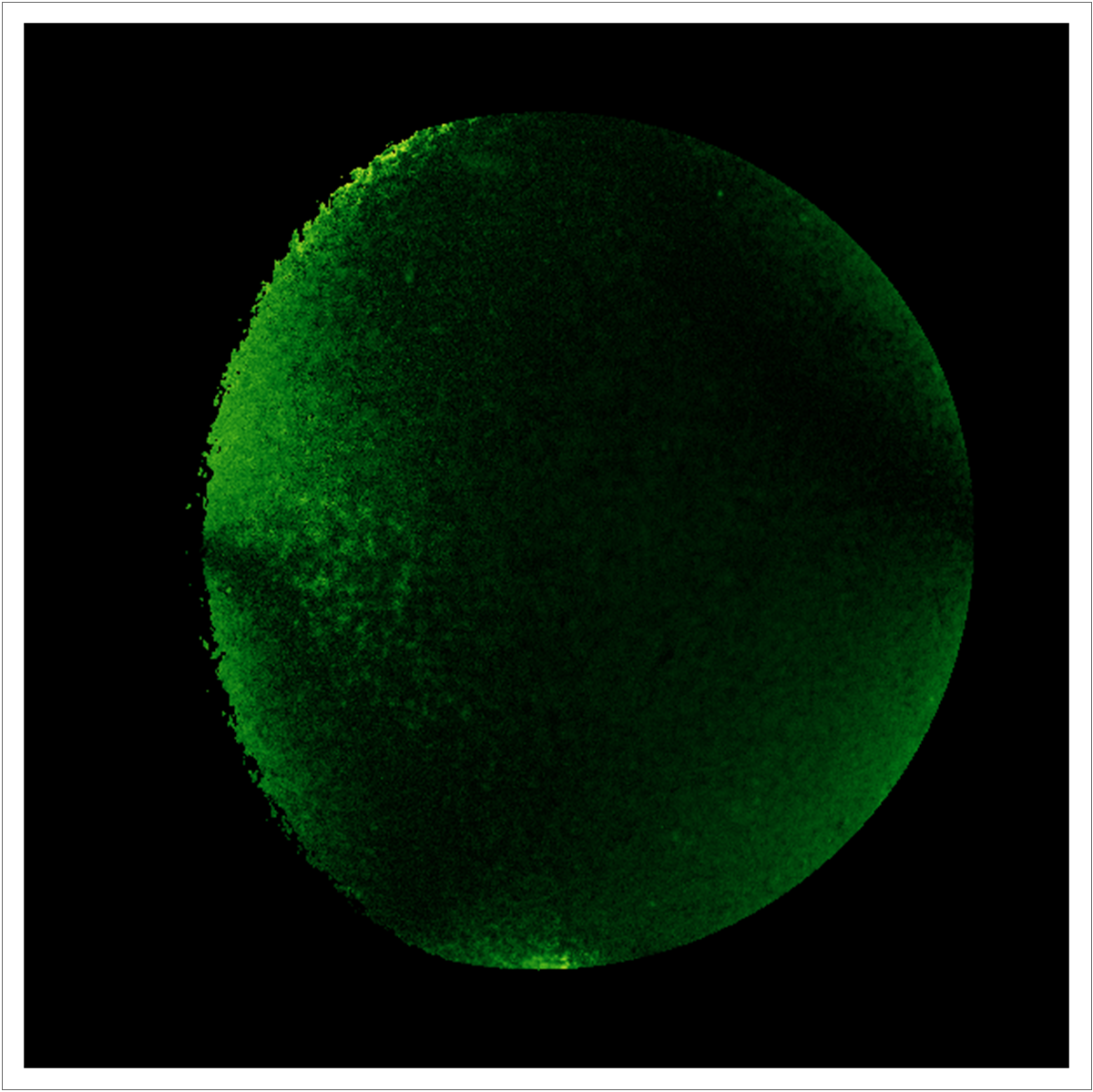}}\label{fig:error_aolp_451}
        \hspace{.7cm}
        \subfloat[$\psi$ deviation 662 nm]{\includegraphics[width=.35\columnwidth]{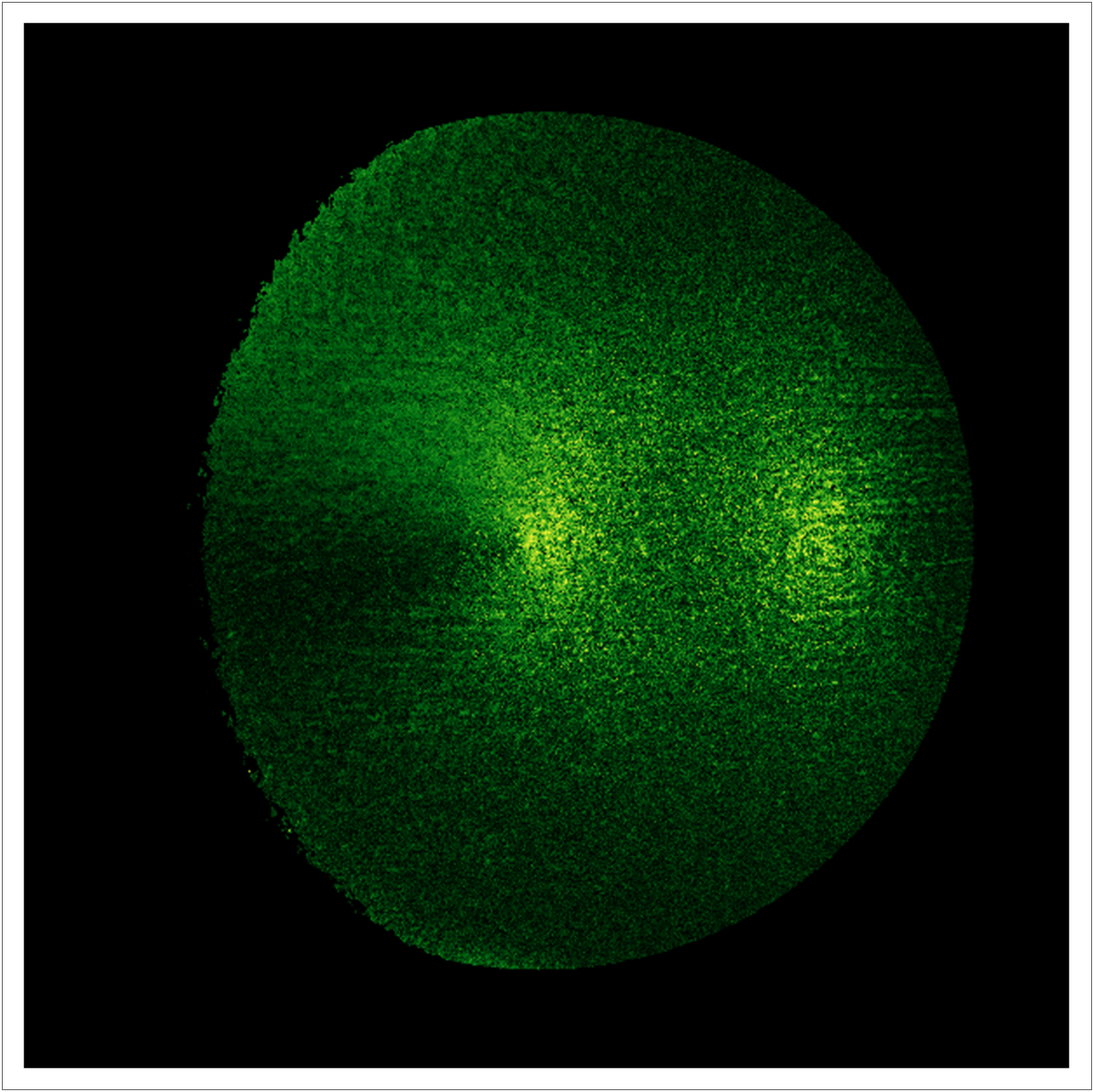}}
        \centering
            \caption{Diattenuation orientation $\psi$ images of the $\widehat{\mathbf{m}}_0$ models at $\Omega=35^\circ$ used in the eigenvalue estimation at (a) 451 nm and at (b) 662 nm, and of the $\widehat{\mathbf{m}}_0$ calculated from DRR MM polarimeter data at (c) 451 nm and at (d) 662 nm. The angle between the measured and modeled diattenuation orientations is shown in (e) and (f) for 451 nm and 662 nm respectively. The diattenuation orientation in (a) is primarily vertically oriented which is consistent with Fresnel reflection-dominated scattering as expected from a low albedo case. The orientation in (b) processes smoothly around the edge of the sphere which is consistent with diffuse polarization. This procession is radially oriented but with the 45$^\circ$/135$^\circ$ components flipped due to the measurement being a reflection configuration. The center of the pattern shows vertically oriented diattenuation which suggests that Fresnel reflection dominates in that region. The $\widetilde{\psi}$ values in the extrapolations (c) and (d) are invariant to the estimation of $\widetilde{\xi}_0$ and only depend on the model $\widehat{\mathbf{m}}_0$. Agreement between the measurements and extrapolations is summarized according to Eq.~\ref{eq:diatOriErr} in Fig.~\ref{fig:diat_error}~(a). }\label{fig:aolp}
\end{figure}

\begin{figure}[!ht]
    \centering
    \makebox[5pt]{{\includegraphics[trim={0 0 0 0},clip,width=0.35\columnwidth]{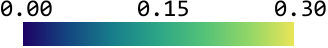}}}
    \hspace{3.5cm}
    \makebox[5pt]{{\includegraphics[trim={0 0 0 0},clip,width=0.35\columnwidth]{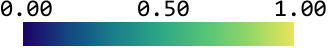}}}\vspace{-.2cm}
    \\
    \centering
        \subfloat[Model $|\mathbf{D}|$ 451 nm]{\includegraphics[trim={0 0 0 50},clip,width=.34\columnwidth]{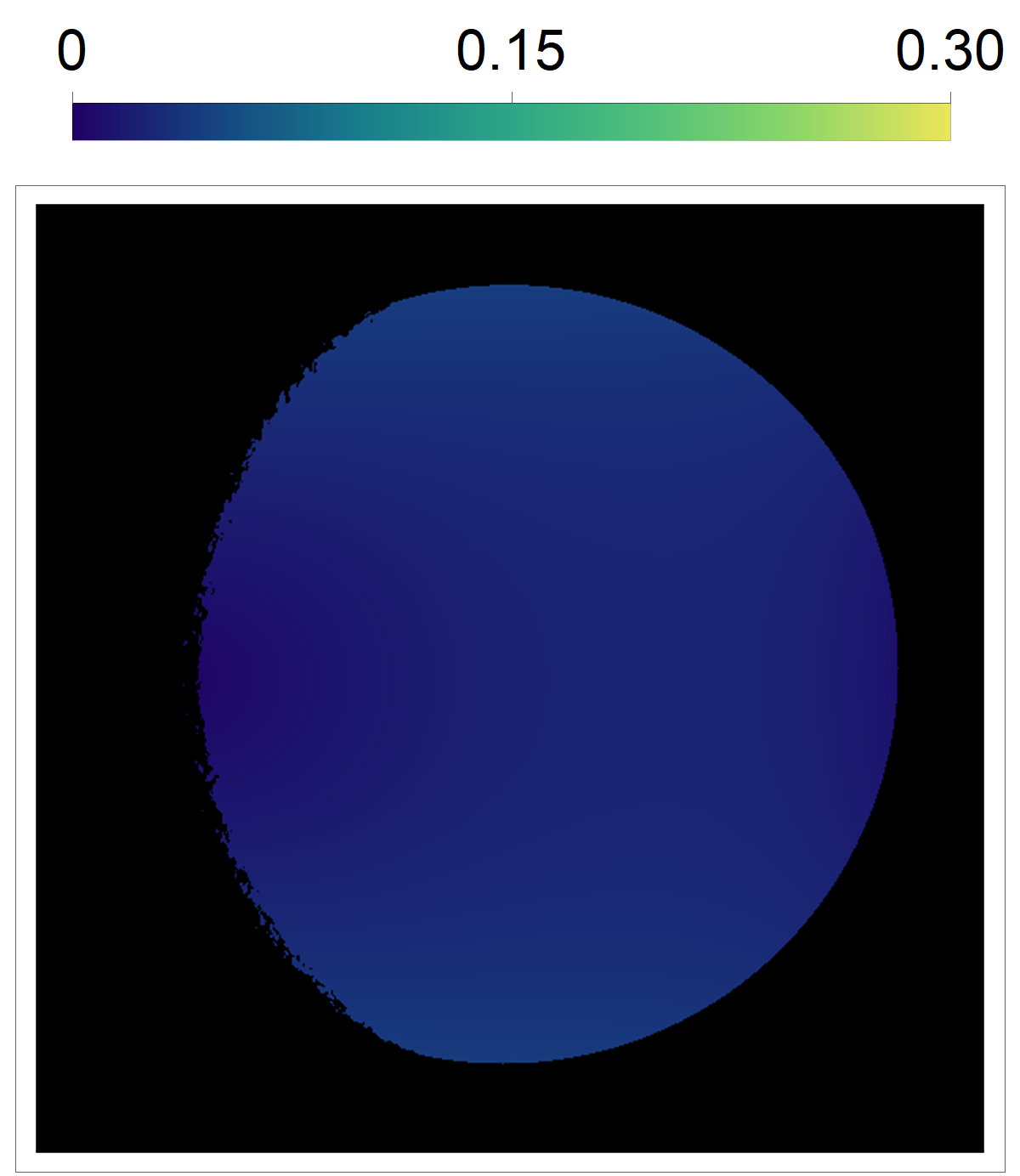}}\label{fig:lucid_diatMag_451}
        \hspace{.7cm}
        \subfloat[Model $|\widetilde{\mathbf{D}}|$ 662 nm]{\includegraphics[trim={0 0 0 50},clip,width=.34\columnwidth]{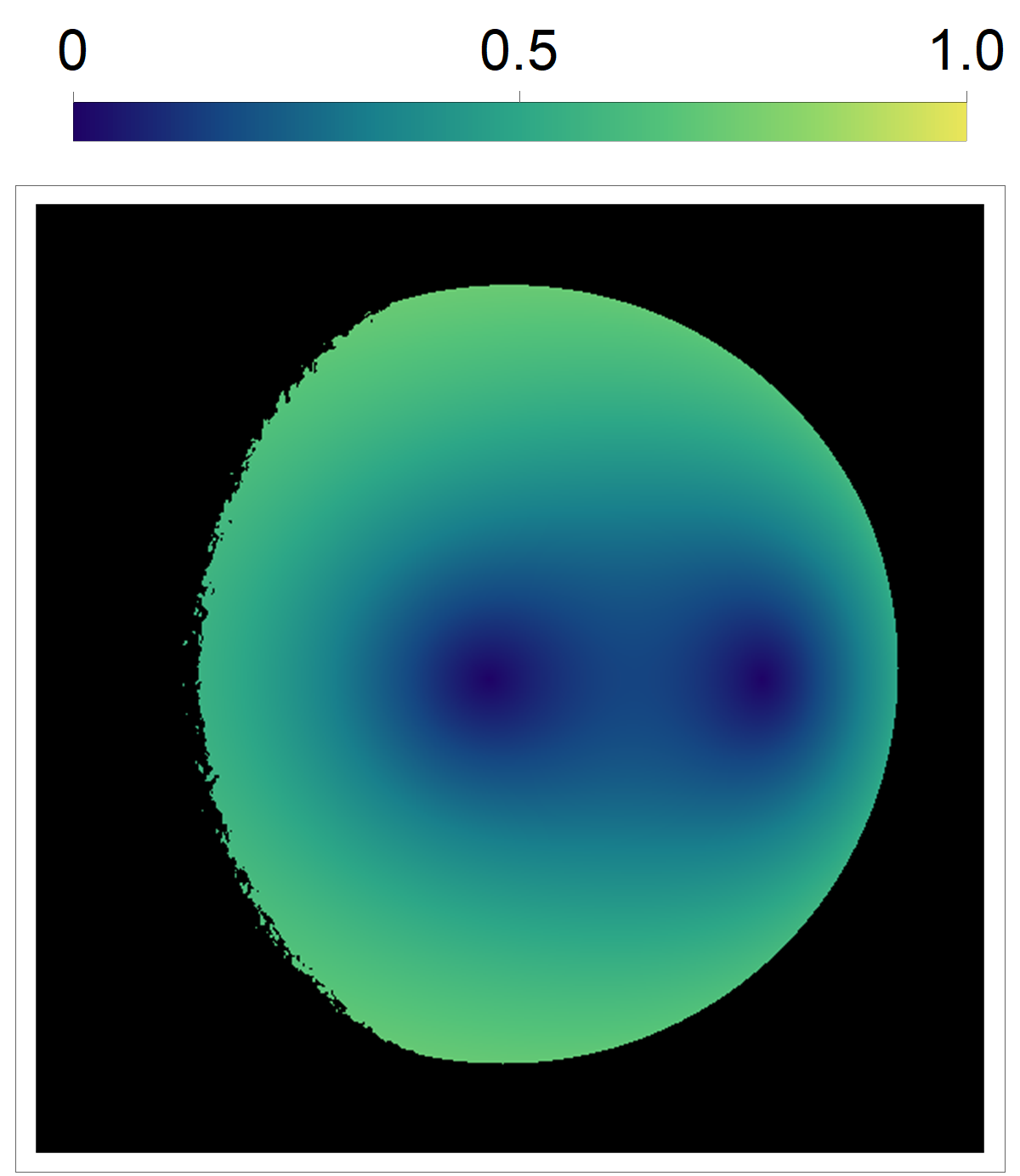}}\label{fig:LUCID_diatMag_662}\\
        \subfloat[True $|\mathbf{D}|$ 451 nm]{\includegraphics[trim={0 0 0 50},clip,width=.34\columnwidth]{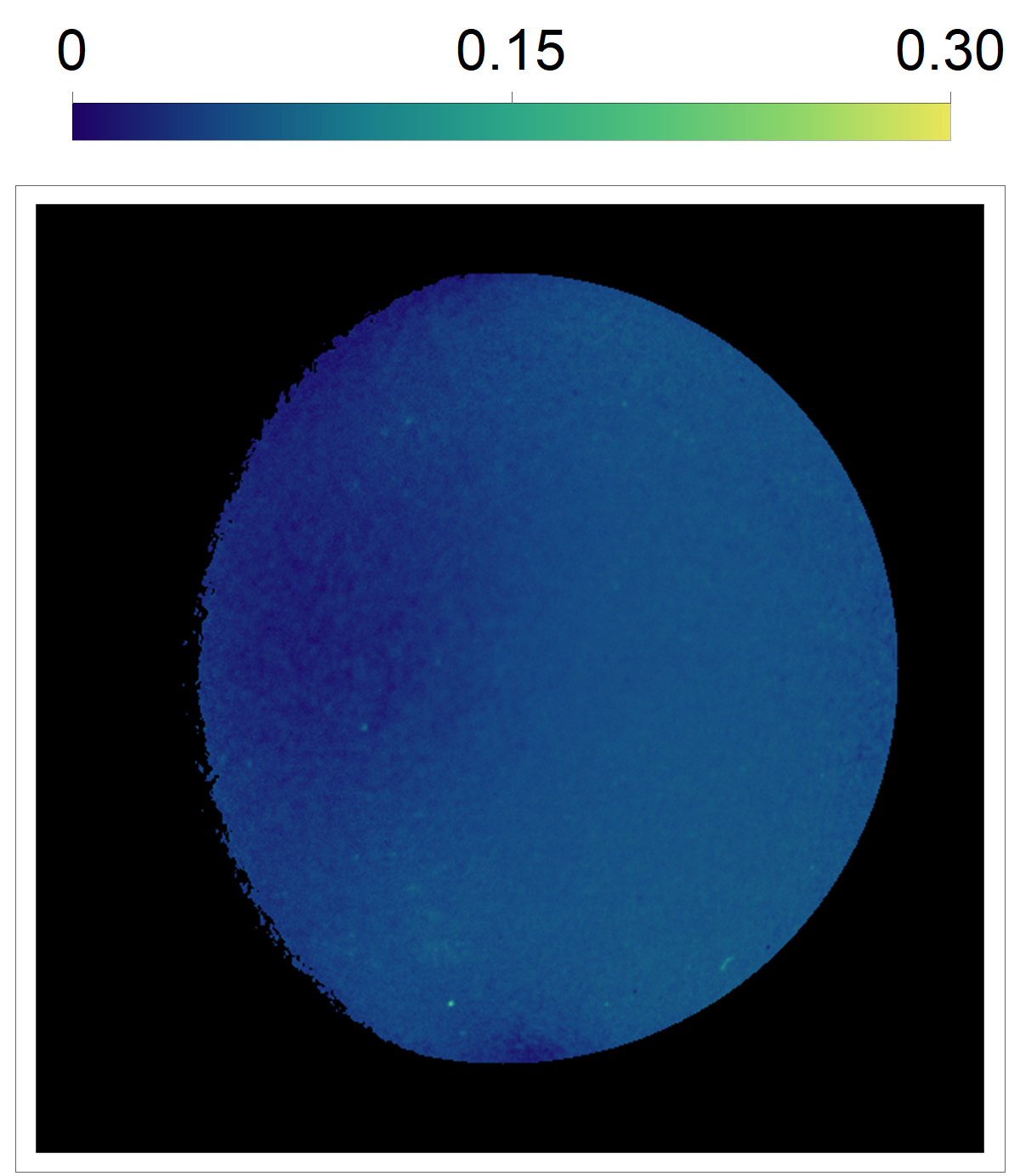}}\label{fig:RGB950_diatMag_451}
        \hspace{.7cm}
        \subfloat[True $|\widetilde{\mathbf{D}}|$ 662 nm]{\includegraphics[trim={0 0 0 50},clip,width=.34\columnwidth]{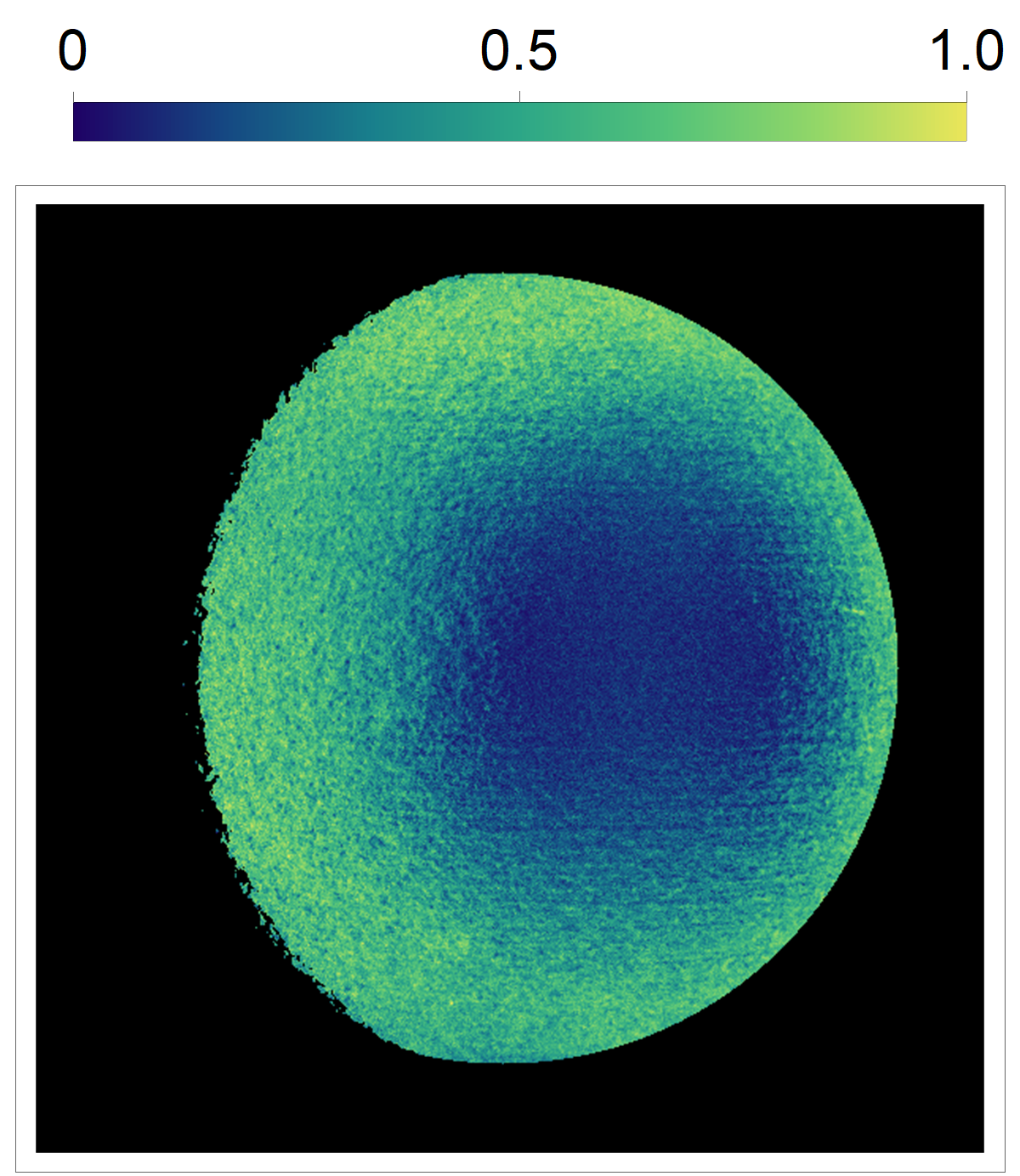}}\label{fig:rgb950_diatMag_662}\\
        \centering
    \makebox[5pt]{{\includegraphics[trim={0 0 0 -5},clip,width=0.35\columnwidth]{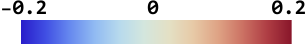}}}
    \hspace{3.5cm}
    \makebox[5pt]{{\includegraphics[trim={0 0 0 -5},clip,width=0.35\columnwidth]{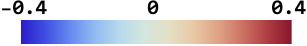}}}\vspace{-.2cm}\\
        \subfloat[$|\mathbf{D}|$ deviation 451 nm]{\includegraphics[trim={0 0 0 50},clip,width=.35\columnwidth]{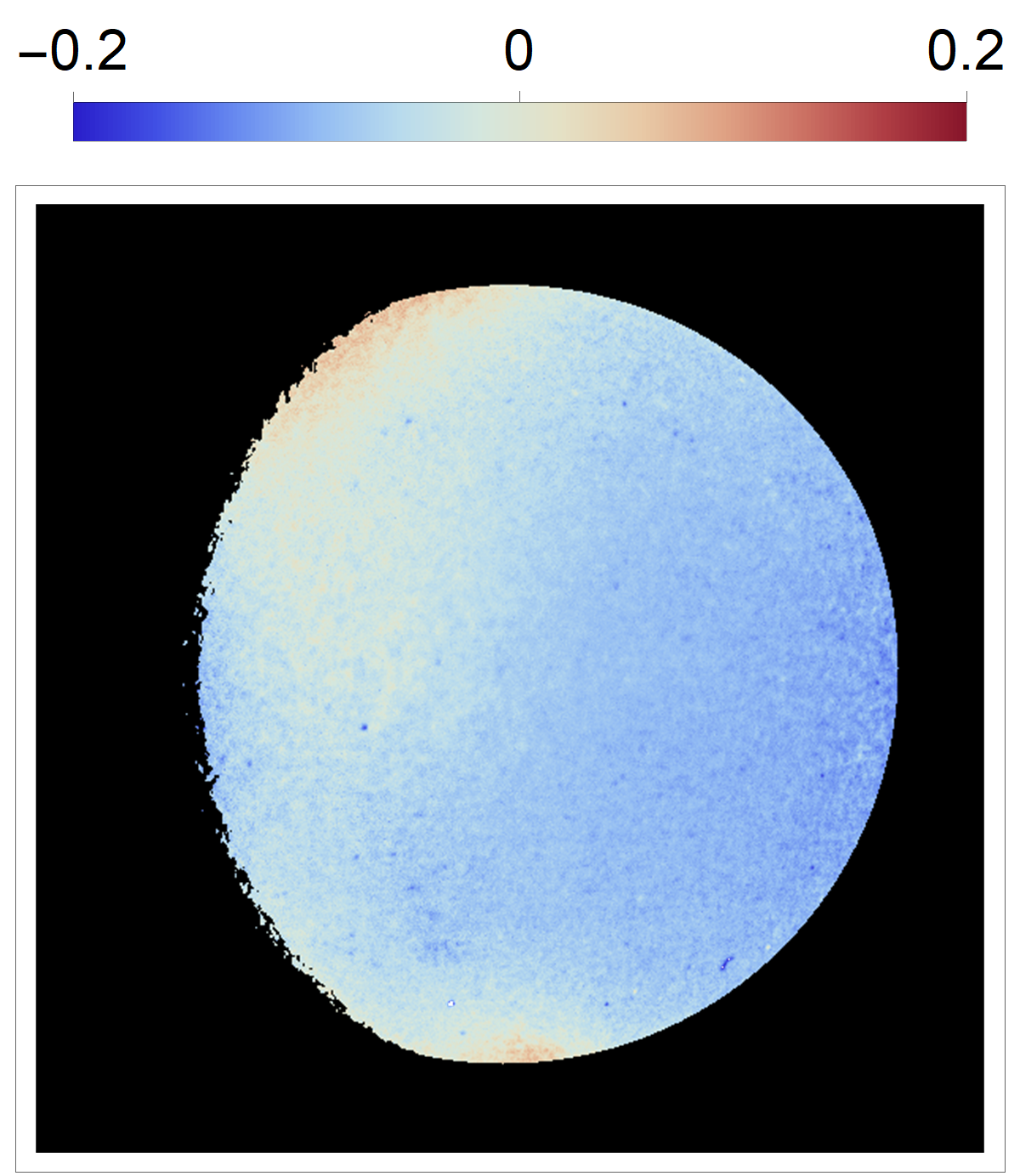}}\label{fig:error_diatMag_451}
        \hspace{.7cm}
        \subfloat[$|\widetilde{\mathbf{D}}|$ deviation 662 nm]{\includegraphics[trim={0 0 0 50},clip,width=.35\columnwidth]{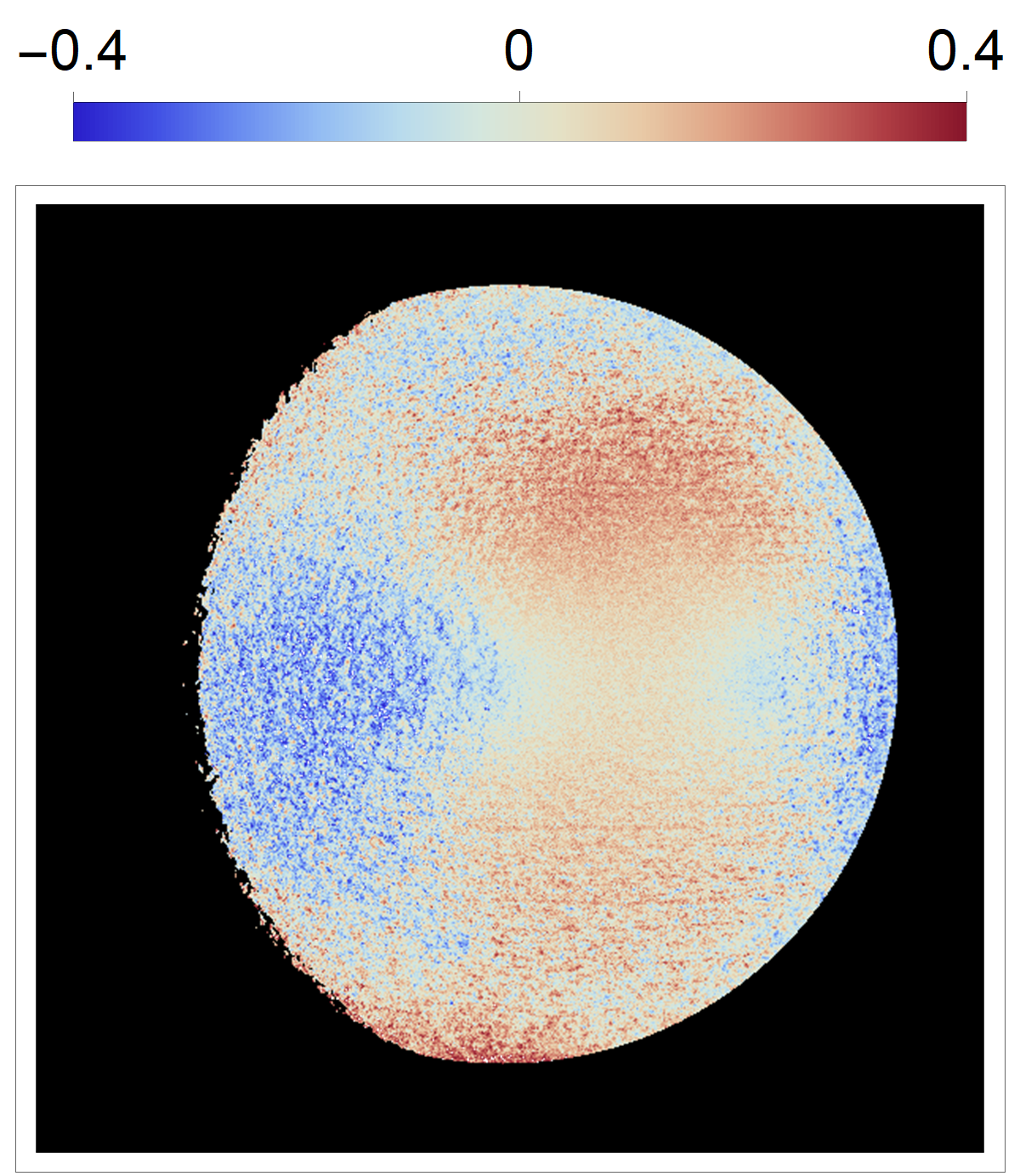}}\label{fig:error_diatMag_662}
            \caption{Diattenuation magnitude $|\mathbf{D}|$ images for the red sphere at $\Omega=35^\circ$. The ground truth from the DRR polarimeter measurements in (a) and the extrapolation from linear Stokes data in (b) at 451 nm are the low albedo case. The ground truth from the DRR measurements in (c) and the extrapolation from linear Stokes data in (d) at 662 nm are the high albedo case. The agreement is summarized with respect to $\Omega$ in Fig.~\ref{fig:diat_error}~(b). The two minima in (c) and (d), referred to as ``neutral points," indicate the presence of at least two polarimetric processes which have comparable diattenuation magnitude and opposing orientations.  }\label{fig:diatMag}
\end{figure}

The measured and modeled $\widehat{\mathbf{m}}_0$ are compared by their diattenuation magnitudes $|\mathbf{D}|$ and orientations $\psi$. Diattenuation is represented in the top row of a MM. 
In Fig.~\ref{fig:aolp} $\psi$ and in Fig.~\ref{fig:diatMag} $|\mathbf{D}|$ are compared at $\Omega=35^\circ$ based on the Rusinkiewicz coordinates from the DRR measurements rather than with the LUT method used in Sect.~\ref{sect:sphere} because the models can be evaluated at arbitrary scattering geometries. In Fig.~\ref{fig:aolp} (a) and (c), $\psi$ is primarily vertical due to the low albedo case of 451 nm illumination being dominated by Frensel reflection. In Fig.~\ref{fig:aolp} (b) and (d), $\psi$ is vertical in the center where Fresnel reflection dominates the near-specular geometries but the diffuse term dominates away from specular, which is consistent with high albedo behavior. 

\begin{figure}
    \centering
    \subfloat[Diattenuation orientation error]{\includegraphics[width=.9\columnwidth]{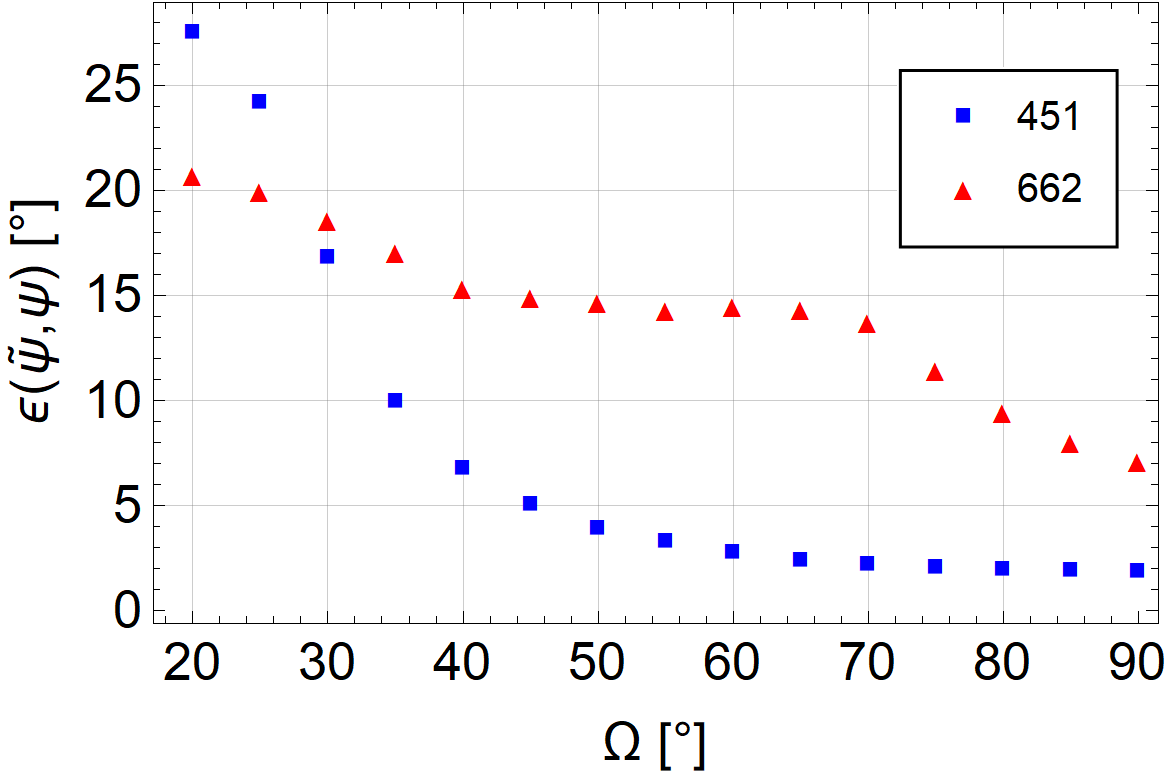}}\label{fig:diat_ori_error}\\
    \subfloat[Diattenuation magnitude error]{\includegraphics[width=.9\columnwidth]{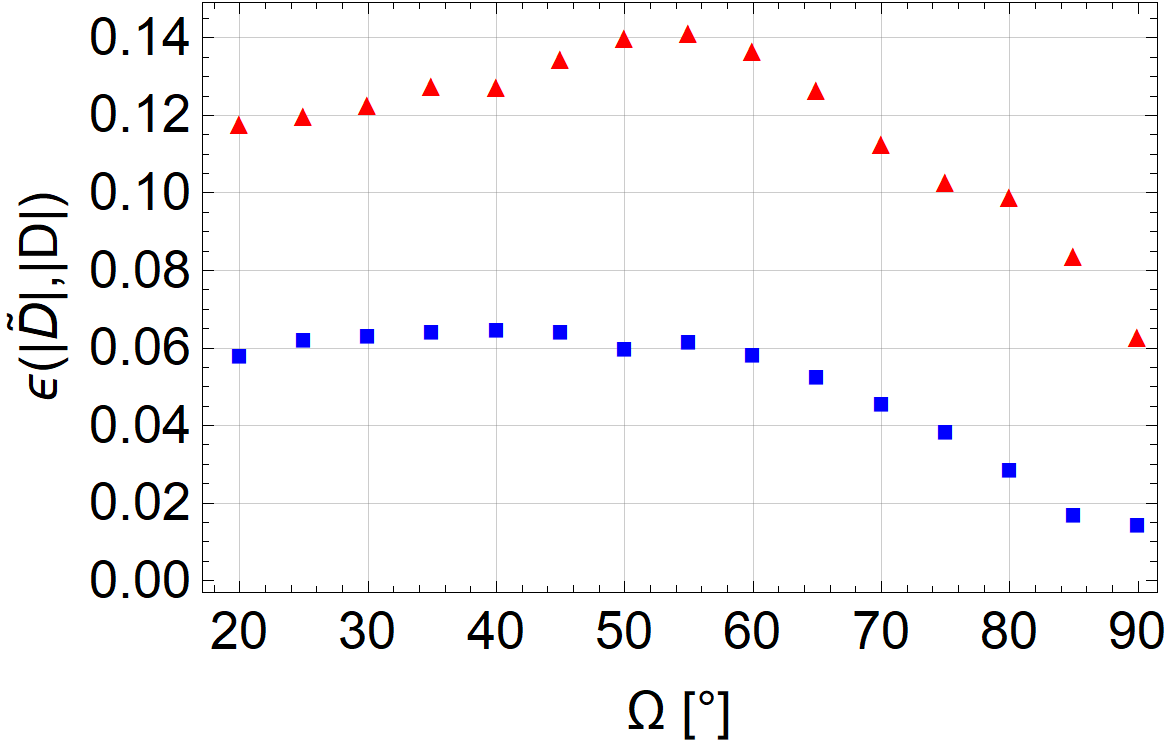}}\label{fig:diat_mag_error}
    \caption{Error (a) in the diattenuation orientation $\widetilde{\psi}$ according to Eq.~\ref{eq:diatOriErr} and (b) in the diattenuation magnitude $|\widetilde{\mathbf{D}}|$ according to Eq.~\ref{eq:diatMagErr} of the $\widehat{\mathbf{m}}_0$ models at 451 nm (blue squares) and 662 nm (red triangles).  RMSD for fifteen acquisition geometries $\Omega$. The orientation error for 451 nm hovers around $2^\circ$ for large $\Omega$. The more complicated orientation pattern for diffuse reflection, which is more prominent at 662 nm, is matched less well as shown in the generally larger error for this high albedo case. The error in diattenuation magnitude is strictly lower for 451 nm than for 662 nm for all $\Omega$. Again, this is likely due to the simpler first-surface model dominating at 451 nm whereas the more complex diffuse model has a stronger contribution at 662 nm.
    }\label{fig:diat_error}
\end{figure}

The agreement in $\psi$ and $|\mathbf{D}|$ are summarized numerically in Fig.~\ref{fig:diat_error} according to their root-mean-squared deviation (RMSD) values over the sphere at each acquisition geometry $\Omega$ as in
\begin{equation}
    \begin{split}
        \epsilon(\widetilde{\psi},\psi) &=\sqrt{\frac{1}{K}\sum_{n=1}^{K}||(\widetilde{\psi}_{k}-\psi_{k})||^2}\\
        &=\sqrt{\frac{1}{K}\sum^{K}_{n=1}\left[\frac{1}{2}\text{acos}\left(\widetilde{\mathbf{d}}_k\cdot\mathbf{d}_k\right)\right]^2},
    \end{split}
    \label{eq:diatOriErr}
\end{equation}
and 
\begin{equation}
    \epsilon(|\widetilde{\mathbf{D}}|,|\mathbf{D}|) =\sqrt{\frac{1}{K}\sum_{n=1}^{K}\left|\left||\widetilde{\mathbf{D}}|_k-|\mathbf{D}|_k\right|\right|^2}
\label{eq:diatMagErr}
\end{equation}
where $K$ is the total number pixels, and $k$ is the pixel index. The dot product of the normalized linear diattenuation parameters, $\mathbf{d}=[D_1,D_2]/|\mathbf{D}|$, is used to avoid angular phase wrapping effects caused by the fact that $\psi=0^\circ$ and $\psi=180^\circ$ describe the same orientation.

In Fig.~\ref{fig:diat_error}~(a), the error in diattenuation orientation trends downward with $\Omega$. The MJM term is dominated by first-surface reflection which has a simpler, easier to model orientation (primarily vertical). This is especially true for 451 nm, where first-surface reflection is overall more dominant due to being low albedo. The error for 451 nm hovers around $2^\circ$ for large $\Omega$. The more complicated orientation pattern for diffuse reflection is matched less well as shown in the generally larger error for 662 nm. In Fig.~\ref{fig:diat_error}~(b), the error in diattenuation magnitude is strictly lower for 451 nm than for 662 nm for all $\Omega$. This is likely due to the simpler first-surface model dominating at 451 nm whereas the more complex diffuse model has a stronger contribution at 662 nm. 
 
\subsection{$\widetilde{\xi}_0$ Estimation from Stokes for a Sphere}\label{sect:sphere}

To measure polarized behavior at many scattering geometries in fewer acquisitions, the measured object was a red 3D printed sphere of 1-inch diameter. The sphere was positioned at the center of rotation of a goniometric arm which was used to perform polarimetric measurements at fifteen angles between the camera and the light source denoted with $\Omega$. Polarimetric measurements were performed at 662 nm and 451 nm. These two wavelengths represent high and low albedo cases, respectively, for the red sphere. Per Umov's effect, these correspond to an expectation of high depolarization (low $\xi_0$) and low depolarization (high $\xi_0$), respectively.

The DRR MM polarimeter and linear Stokes camera were both used with  8 mm focal length lenses but have different detector sizes and resolutions. Additionally, the reliably positioning the two cameras to have the same view proved challenging. A pixel-to-pixel comparison between DRR and Stokes camera images would therefore mean comparing polarimetric behavior at different scattering geometries. Instead, the estimated $\widetilde{\xi}_{0}$ from the linear Stokes camera is compared to the values from the DRR polarimeter by tabulating a lookup table (LUT) of $\xi_0$ values with respect to scattering geometry for each polarimeter. These LUTs are binned according to three angles in the Rusinkiewicz coordinate system, $\theta_h$, $\theta_d$, and $\phi_d$ with 91, 91, and 361 bins assigned respectively. The Rusinkiewicz coordinate system is explained further in Sect.~\ref{sect:rusink}. $\theta_d$ and $\theta_h$ are binned in steps of 1$^\circ$ from $0^\circ$ to $90^\circ$, and $\phi_d$ is binned in steps of 1$^\circ$ from $-180^\circ$ to $180^\circ$. The LUT is then evaluated at the scattering geometries of a virtual sphere matching the linear Stokes camera measurement to compare the true and estimated $\xi_0$ at identical points. 

\begin{figure}[!ht]
\centering
    \makebox[5pt]{{\includegraphics[trim={0 0 0 -5},clip,width=0.35\columnwidth]{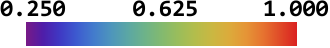}}}
    \hspace{3.7cm}
    \makebox[5pt]{{\includegraphics[trim={0 0 0 -5},clip,width=0.35\columnwidth]{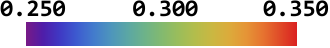}}}\vspace{-.2cm}    \\
    \centering
        \subfloat[Estimated $\widetilde{\xi}_0$ 451 nm]{\includegraphics[trim={0 0 0 50},clip,width=.34\columnwidth]{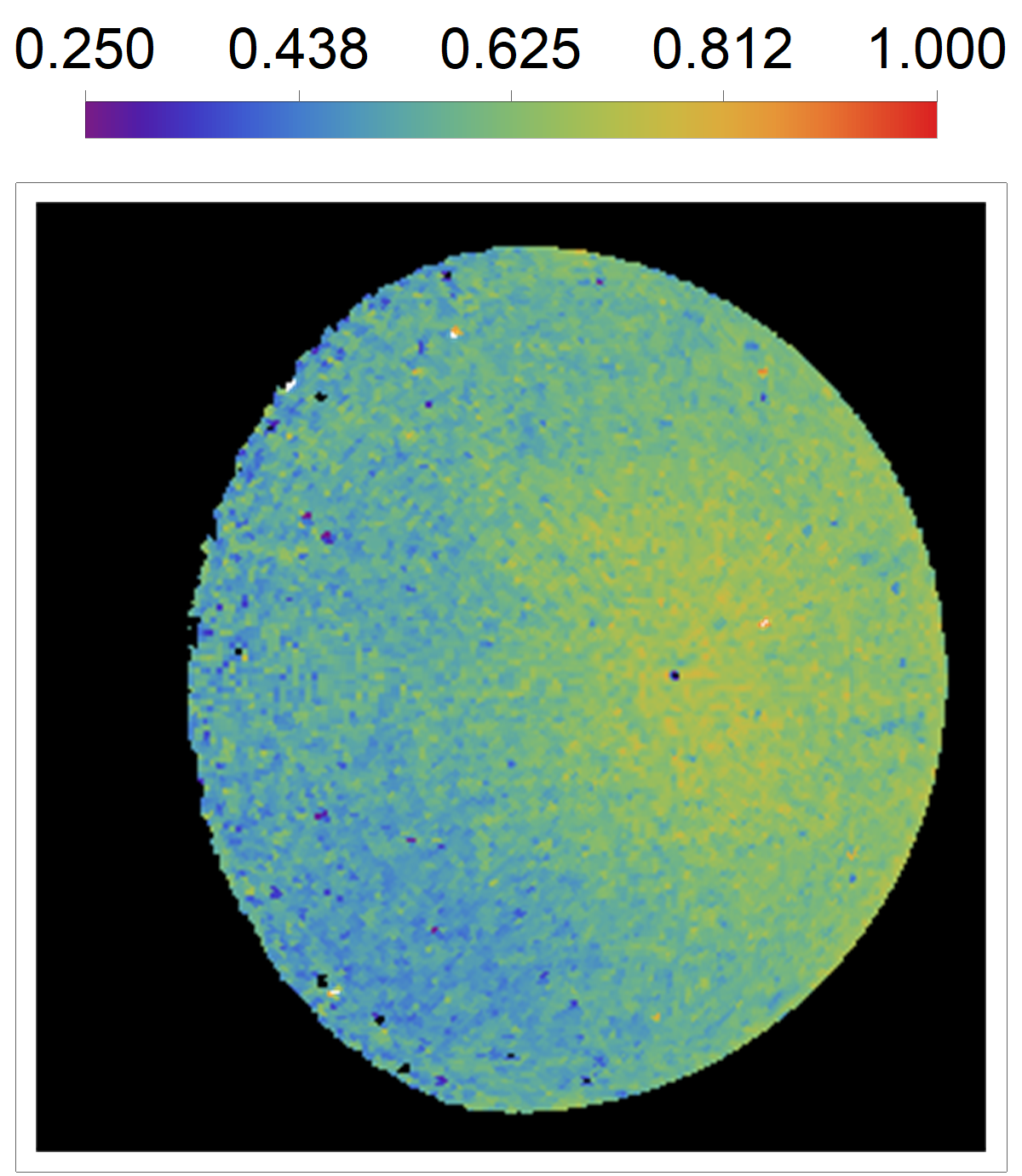}}\label{fig:est_sphere_xi0_451}
        \hspace{.7cm}
        \subfloat[Estimated $\widetilde{\xi}_0$ 662 nm]{\includegraphics[trim={0 0 0 50},clip,width=.34\columnwidth]{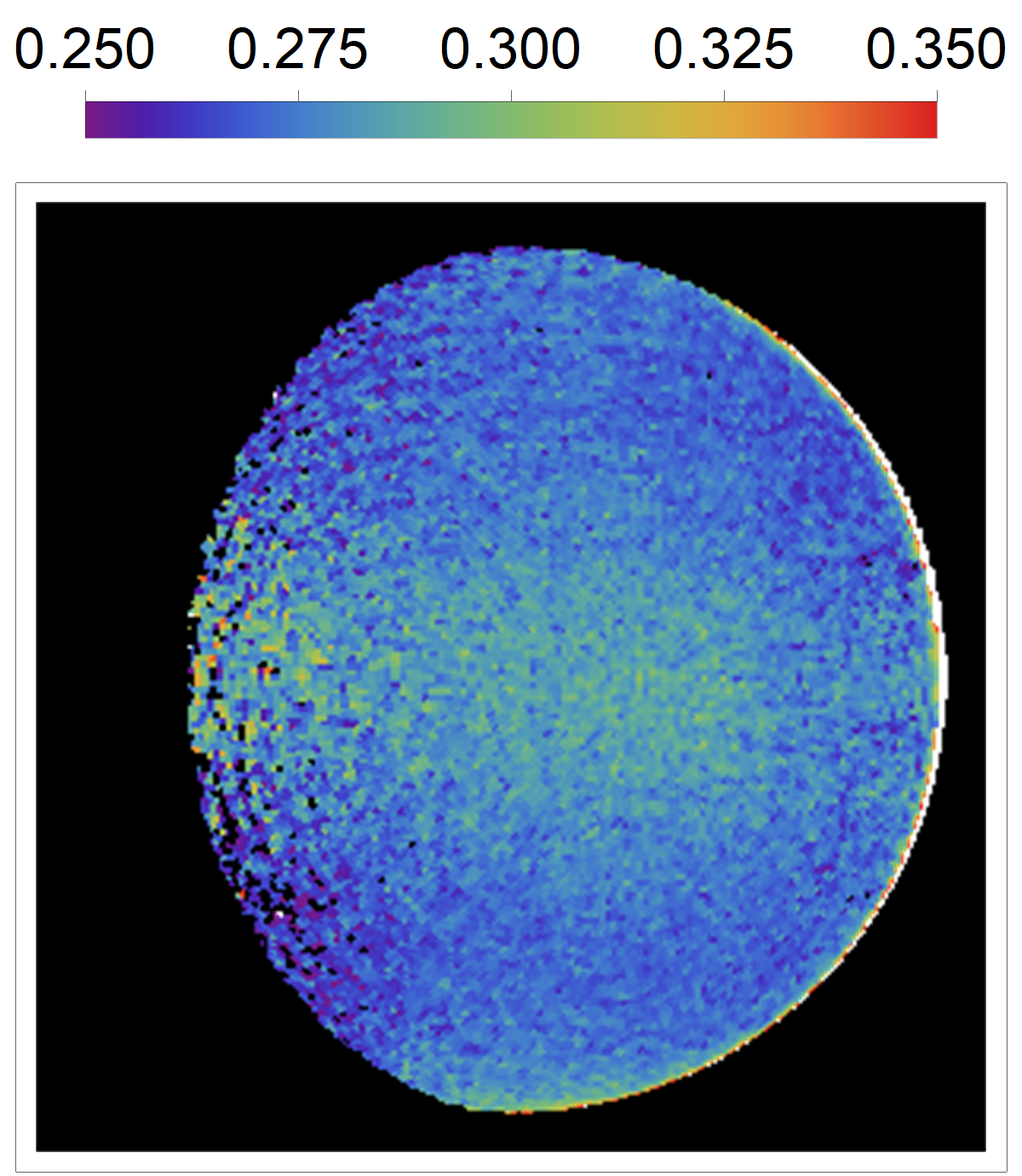}}\label{fig:est_sphere_xi0_662}\\    
        \subfloat[True $\xi_0$ 451 nm]{\includegraphics[trim={0 0 0 50},clip,width=.34\columnwidth]{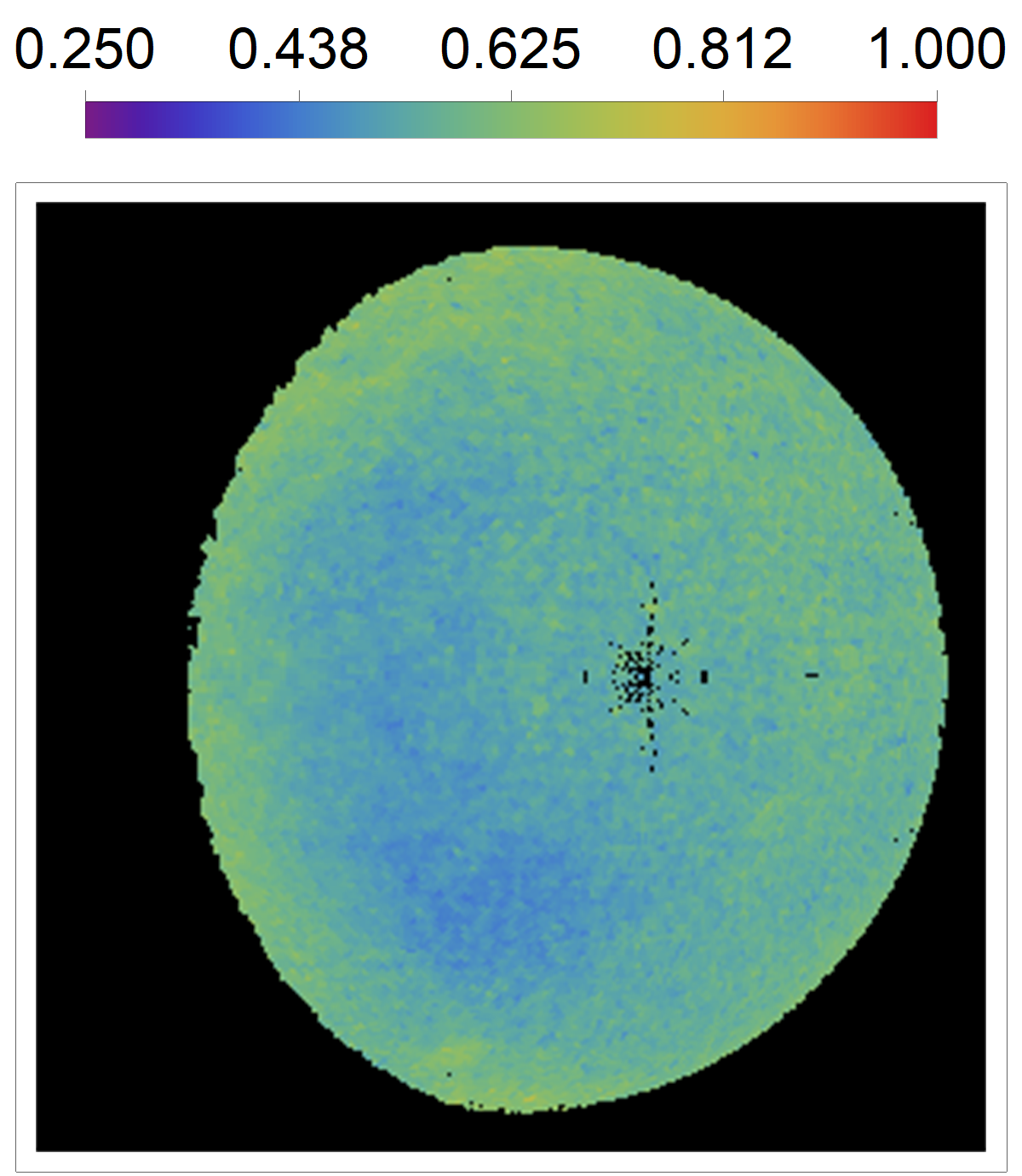}}\label{fig:true_sphere_xi0_662}
        \hspace{.7cm}
        \subfloat[True $\xi_0$ 662 nm]{\includegraphics[trim={0 0 0 50},clip,width=.34\columnwidth]{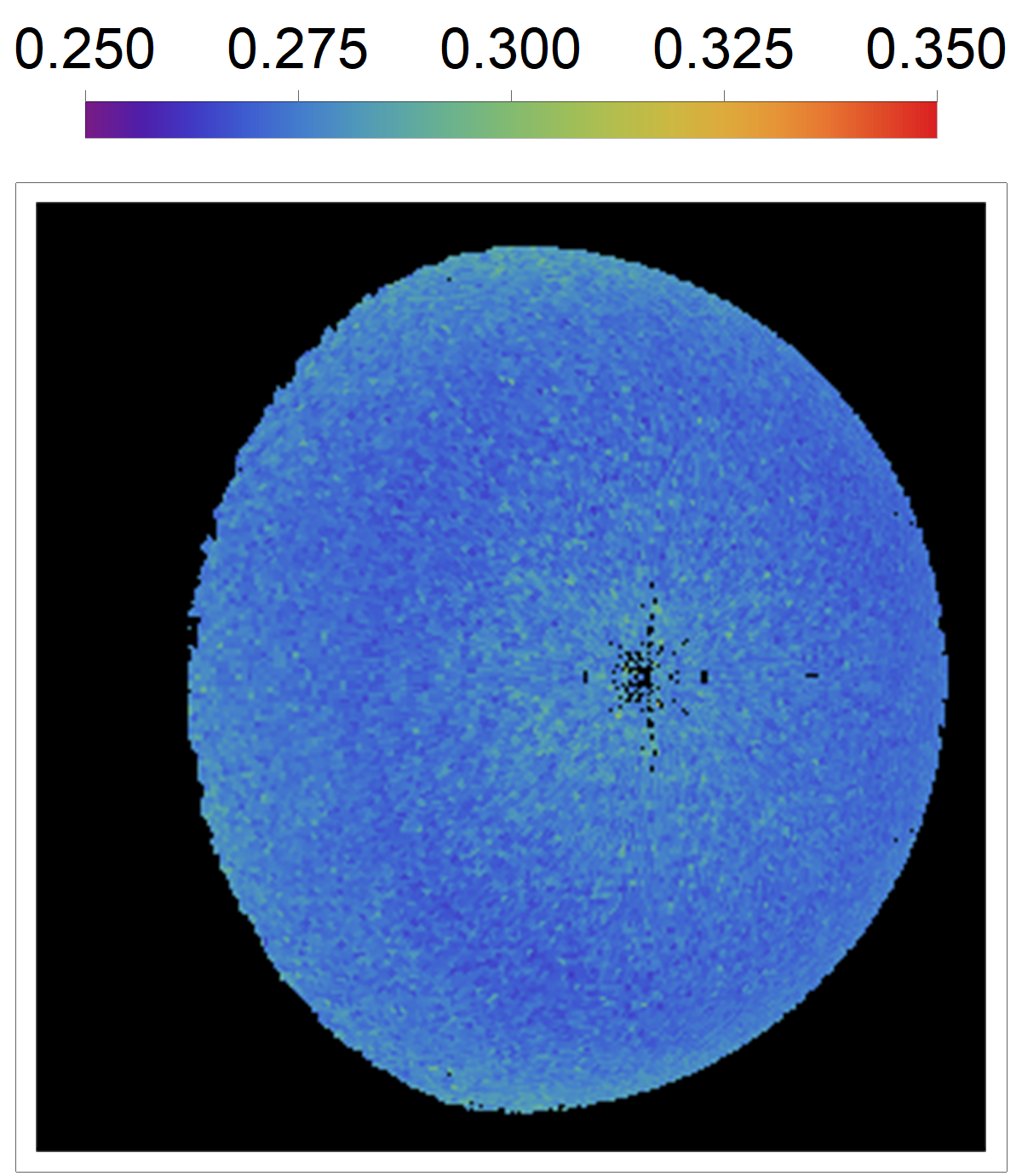}}\label{fig:true_sphere_xi0_451}\\
        \centering
    \makebox[5pt]{{\includegraphics[trim={0 0 0 -5},clip,width=0.35\columnwidth]{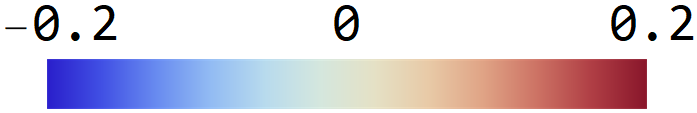}}}
    \hspace{3.7cm}
    \makebox[5pt]{{\includegraphics[trim={0 0 0 -5},clip,width=0.35\columnwidth]{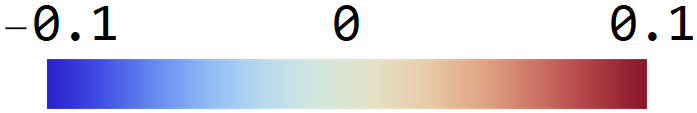}}}\vspace{-.2cm}    \\
        \subfloat[ $\xi_0$ deviation 451 nm]{\includegraphics[trim={0 0 0 51},clip,width=.34\columnwidth]{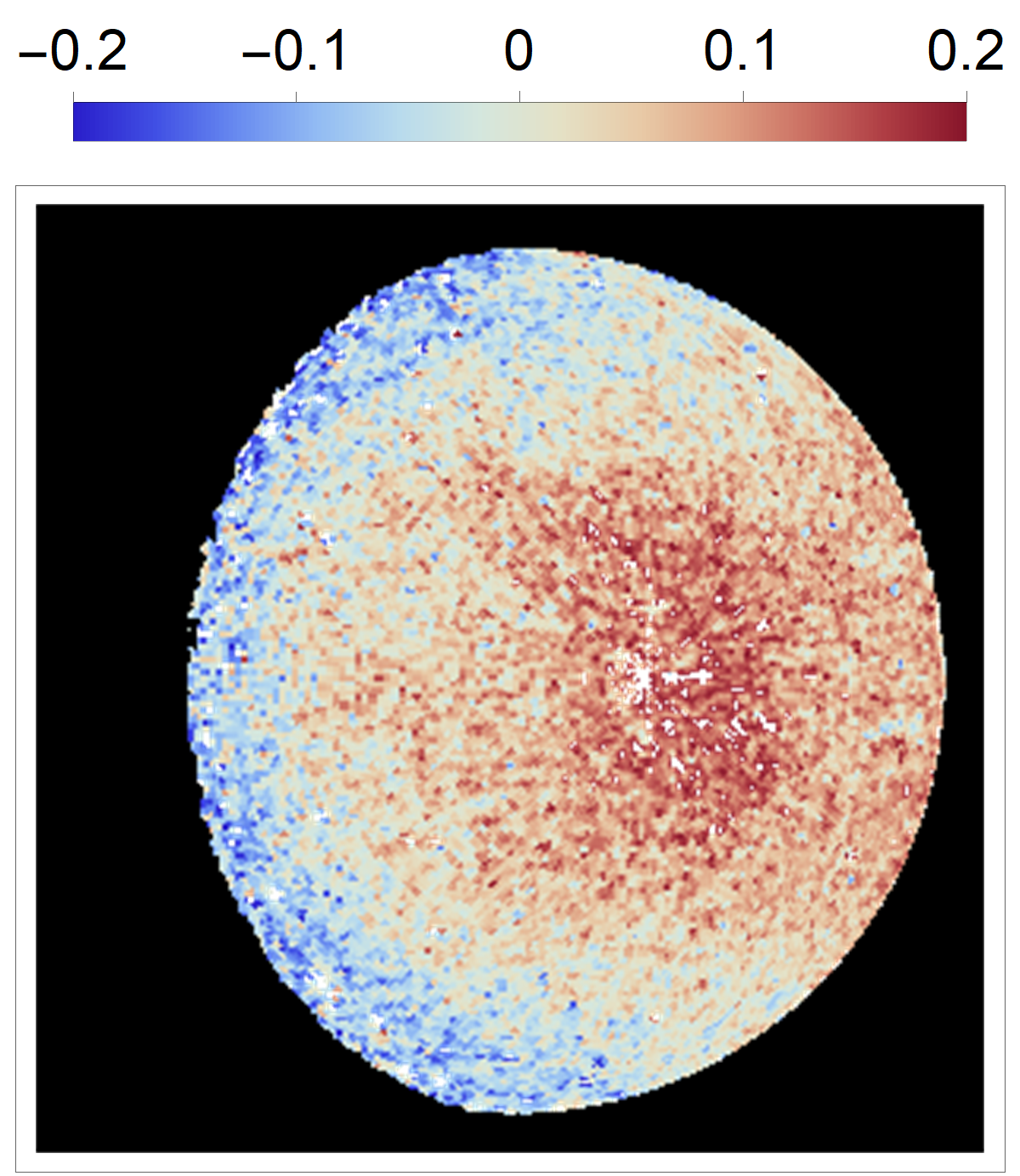}}\label{fig:error_451}  
        \hspace{.7cm}
        \subfloat[ $\xi_0$ deviation 662 nm]{\includegraphics[trim={0 0 0 50},clip,width=.34\columnwidth]{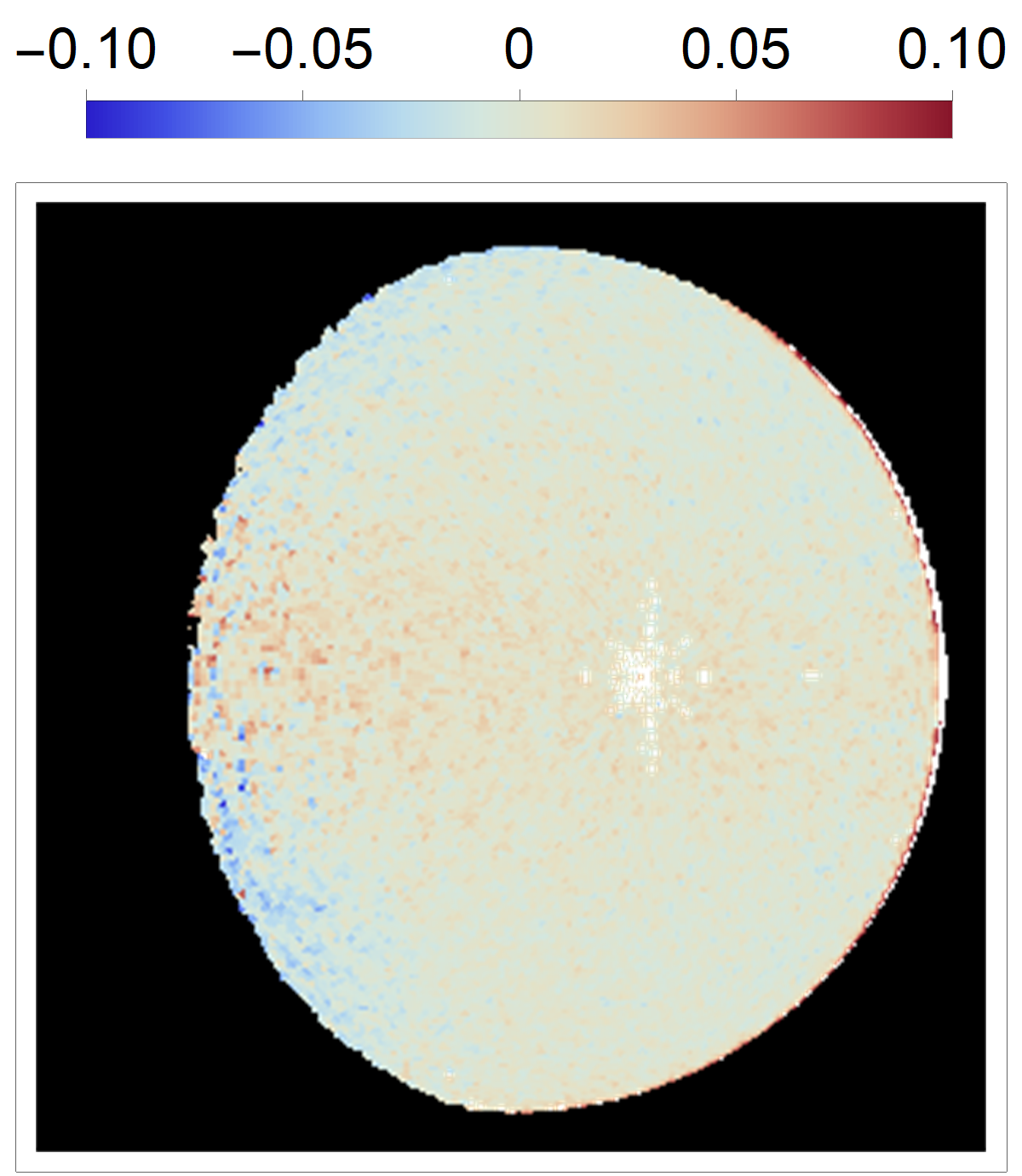}}\label{fig:error_662}
        \caption{Results for the estimation of $\widetilde{\xi}_0$ for angle between the camera and light source $\Omega=35^\circ$. The estimated values from the linear Stokes camera in (a) and the ground truth values from the DRR polarimeter in (b) at 451 nm are compared in (e) by taking their difference. The estimated values from the linear Stokes camera  in (c) and the  ground truth values from the DRR polarimeter in (d) at 662 nm are compared in (f) by taking their difference. 662 nm is high albedo for the red sphere, so depolarization is high and $\xi_0$ is low. 451 nm is low albedo, so $\xi_0$ is larger. In (e) and (f), positive values indicate an overestimation of the contribution of $\widehat{\mathbf{m}}_0$ while negative values indicate an overestimation of the contribution of $\mathbf{m}_{ID}$. Because the eigenvalues are normalized to sum to unity, the errors shown in (e) and (f) can be thought of as related to the fraction of the light that is attributed to the wrong MM component. }\label{fig:xi0_sphere}
\end{figure}

The estimations of $\widetilde{\xi}_0$ from the linear Stokes camera over the sphere at $\Omega=35^\circ$ are shown in Fig.~\ref{fig:xi0_sphere} (a) and (b), the ground truth values measured in the DRR polarimeter are shown in Fig.~\ref{fig:xi0_sphere} (c) and (d), and the difference between estimate and ground truth are shown in Fig.~\ref{fig:xi0_sphere} (e) and (f). The white region in the center consists of specific Rusinkiewicz coordinates that the instruments do not have in common. 662 nm is high albedo for the red sphere, so the expectation according to Umov's effect is that depolarization is high and $\xi_0$ is low. Conversely, 451 nm is low albedo so $\xi_0$ is expected to be larger. Both of these trends are observed in the ground truth measurements as well as in the estimations. In (e) and (f), positive values indicate an overestimation of the contribution of $\widehat{\mathbf{m}}_0$ while negative values indicate an overestimation of the contribution of $\mathbf{m}_{ID}$. Because the eigenvalues are normalized to sum to unity, the errors shown in (e) and (f) can be thought of as related to the fraction of the light that is attributed to the wrong MM component. 

\begin{figure}
    \centering%
    \includegraphics[width=.9\columnwidth]{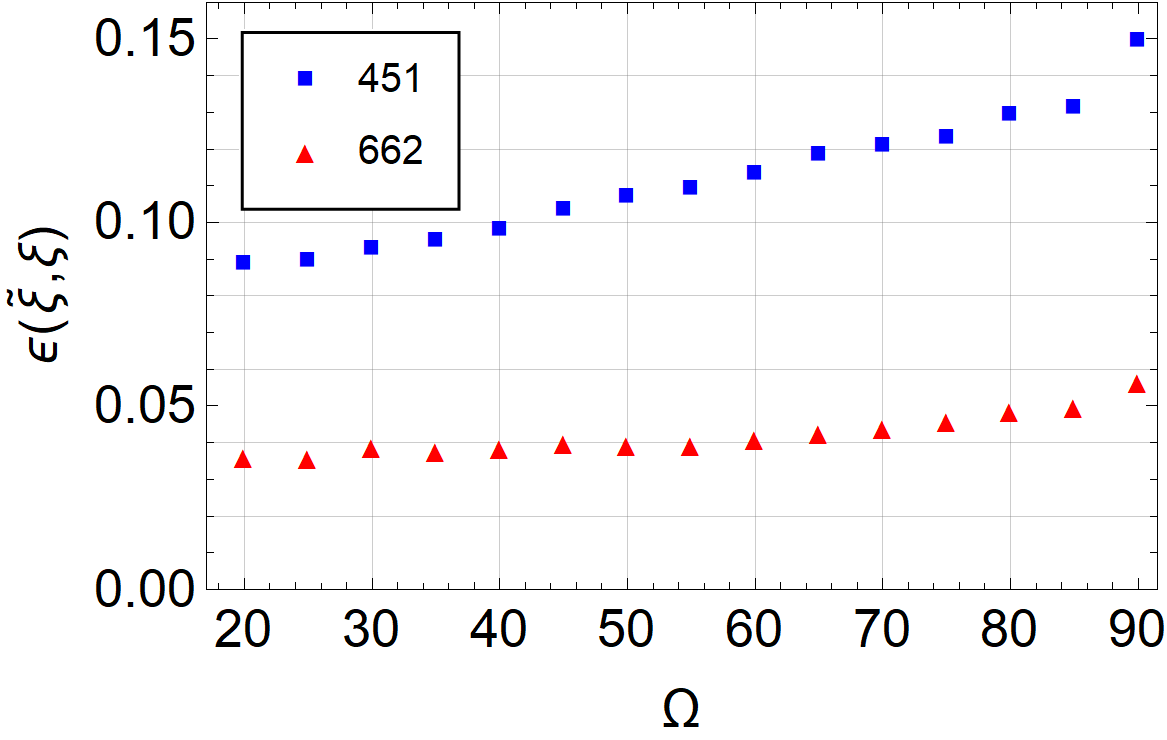}
    \caption{RMSD in $\widetilde{\xi_0}$ estimate from Stokes imaging as compared to ground truth from DRR MM imaging according to Eq.~\ref{eq:xi0Err} at 451 nm (blue squares) and 662 nm (red triangles) at the fifteen acquisition geometries $\Omega$. RMSD values at 451 nm, the low albedo case, are strictly greater than those for 662 nm. This is potentially explained by the material being less depolarizing according to Umov's effect than at 662 nm and therefore more sensitive to disagreement between the ground truth and modeled $\widehat{\mathbf{m}}_0$. Additionally, the eigenspectrum at 451 nm is further from TD than for 662 nm, so a TD model may be less appropriate. The lower throughput due to the low albedo could exacerbate the effects of noise and degrade the estimation due to the smaller dynamic range of the linear Stokes camera.}
    \label{fig:xi0_error}
\end{figure}

The error averaged over the image of the sphere is summarized quantitatively by the RMSD
\begin{equation}
    \epsilon(\widetilde{\xi}_0,\xi_0) =\sqrt{\frac{1}{K}\sum_{k=1}^{K}||\widetilde{\xi}_{0,k}-\xi_{0,k}||^2}, 
\label{eq:xi0Err}
\end{equation}
where $K$ is the total number pixels, and $k$ is the pixel index. Figure~\ref{fig:xi0_error} shows the trend of $\epsilon(\widetilde{\xi}_0,\xi_0)$ as a function of the angle between the camera and light source $\Omega$. The error at 451 nm is larger than at 662 nm for every $\Omega$. According to Umov's effect, the depolarization at 451 nm for a red object should be lower, suggesting that the estimation is more sensitive to disagreement between the ground truth and model $\widehat{\mathbf{m}}_0$. Additionally, the relatively lower overall reflectance at 451 nm also makes the linear Stokes camera measurements more susceptible to noise.

\subsection{Mueller Extrapolation from Stokes for a Stanford Bunny}
To extend the demonstration of the efficient TD-MM model to non-spherical objects, the MM image of a Stanford bunny was extrapolated from linear Stokes camera measurements. This Stanford bunny was printed from the same material as the sphere in Sect.~\ref{sect:sphere}, so the same pBRDF models can be used. These models are detailed in Sect.~\ref{sect:model}. 

Linear Stokes measurements are performed using the COTS camera and one polarized illumination state from the DRR light source as in Sect.~\ref{sect:sphere}. The $\widehat{\mathbf{m}}_0$ models are evaluated at the Rusinkiewicz angles calculated from the .stl file used to 3D print the Stanford bunny. The $\widehat{\mathbf{m}}_0$ model and Stokes data are used to estimate $\widetilde{\xi}_0$ following Sect.~\ref{sect:method}. MM extrapolation is performed by plugging the estimated $\widetilde{\xi}_0$ back into Eq.~\ref{eq:1par_a}.

\begin{figure*}[!ht]
\centering
    \makebox[5pt]{{\includegraphics[trim={0 0 0 -5},clip,width=0.25\textwidth]{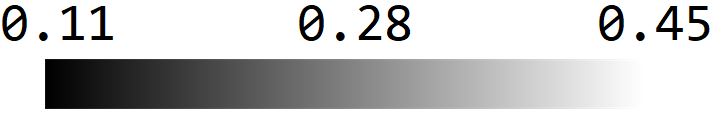}}}
    \hspace{5.6cm}
    \makebox[5pt]{{\includegraphics[trim={0 0 0 -5},clip,width=0.25\textwidth]{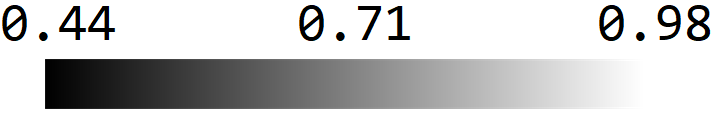}}}
    \hspace{5.6cm}
    \makebox[5pt]{{\includegraphics[trim={0 0 0 -5},clip,width=0.25\textwidth]{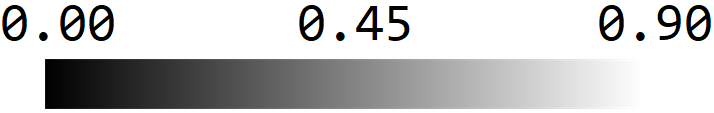}}}\vspace{-.2cm}    \\
    \centering
        \subfloat[Crossed linear polarizers 451 nm]{\includegraphics[trim={0 0 0 0},clip,width=.32\textwidth]{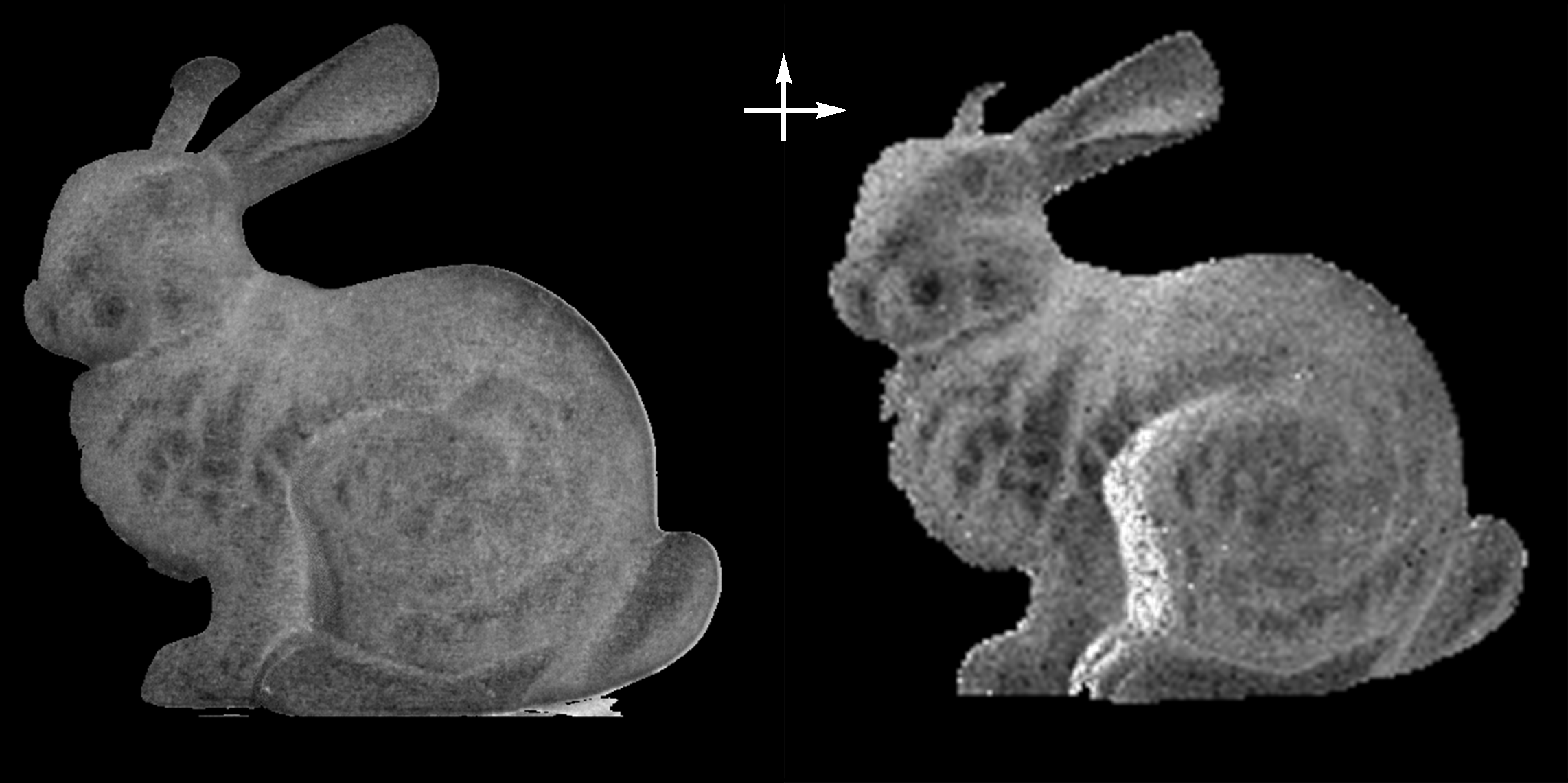}}\label{fig:crossed451}
        \subfloat[Aligned linear polarizers 451 nm]{\includegraphics[trim={0 0 0 0},clip,width=.32\textwidth]{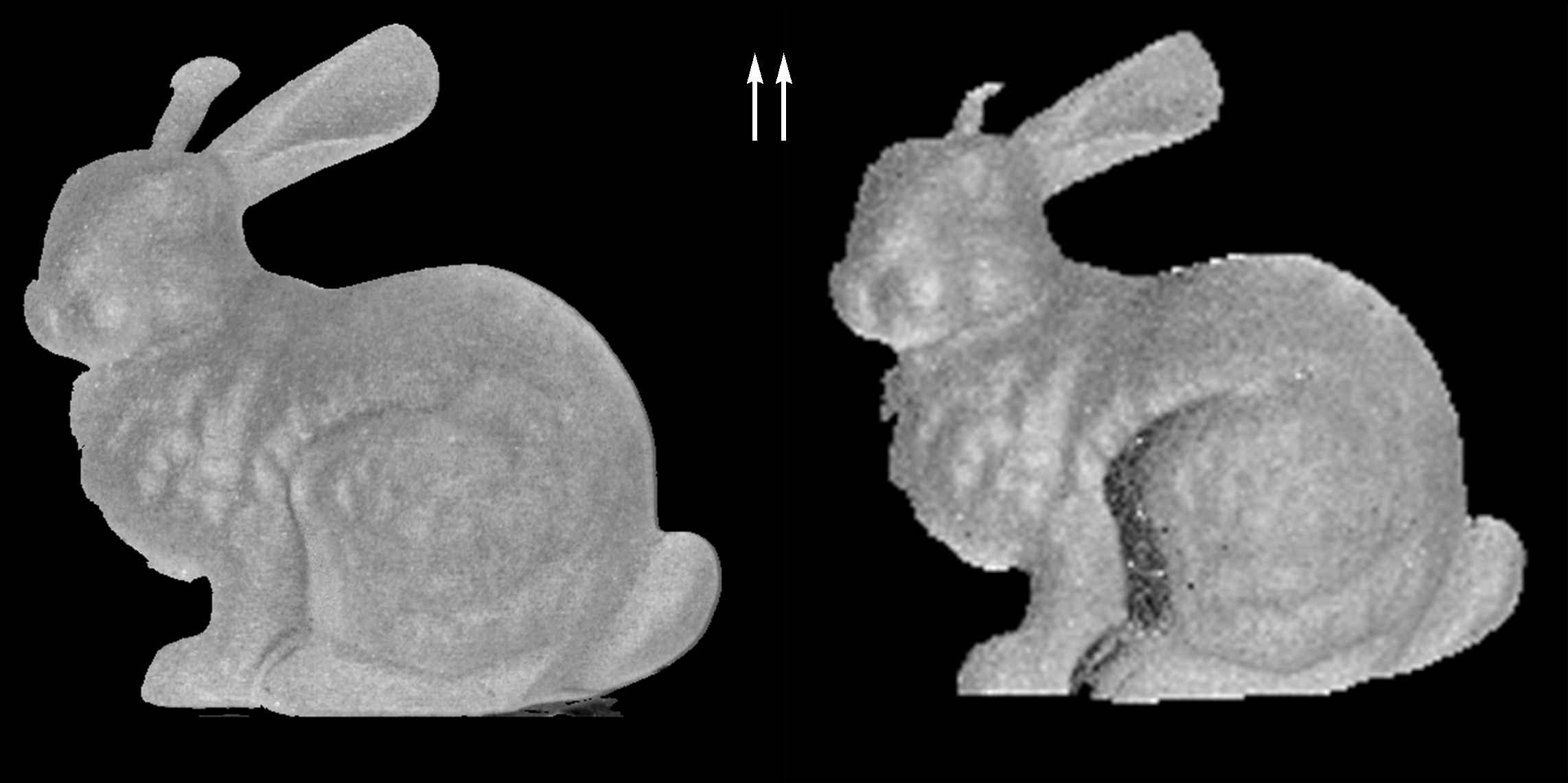}}\label{fig:aligned451}
        \subfloat[Crossed circular polarizers 451 nm]{\includegraphics[trim={0 0 0 0},clip,width=.32\textwidth]{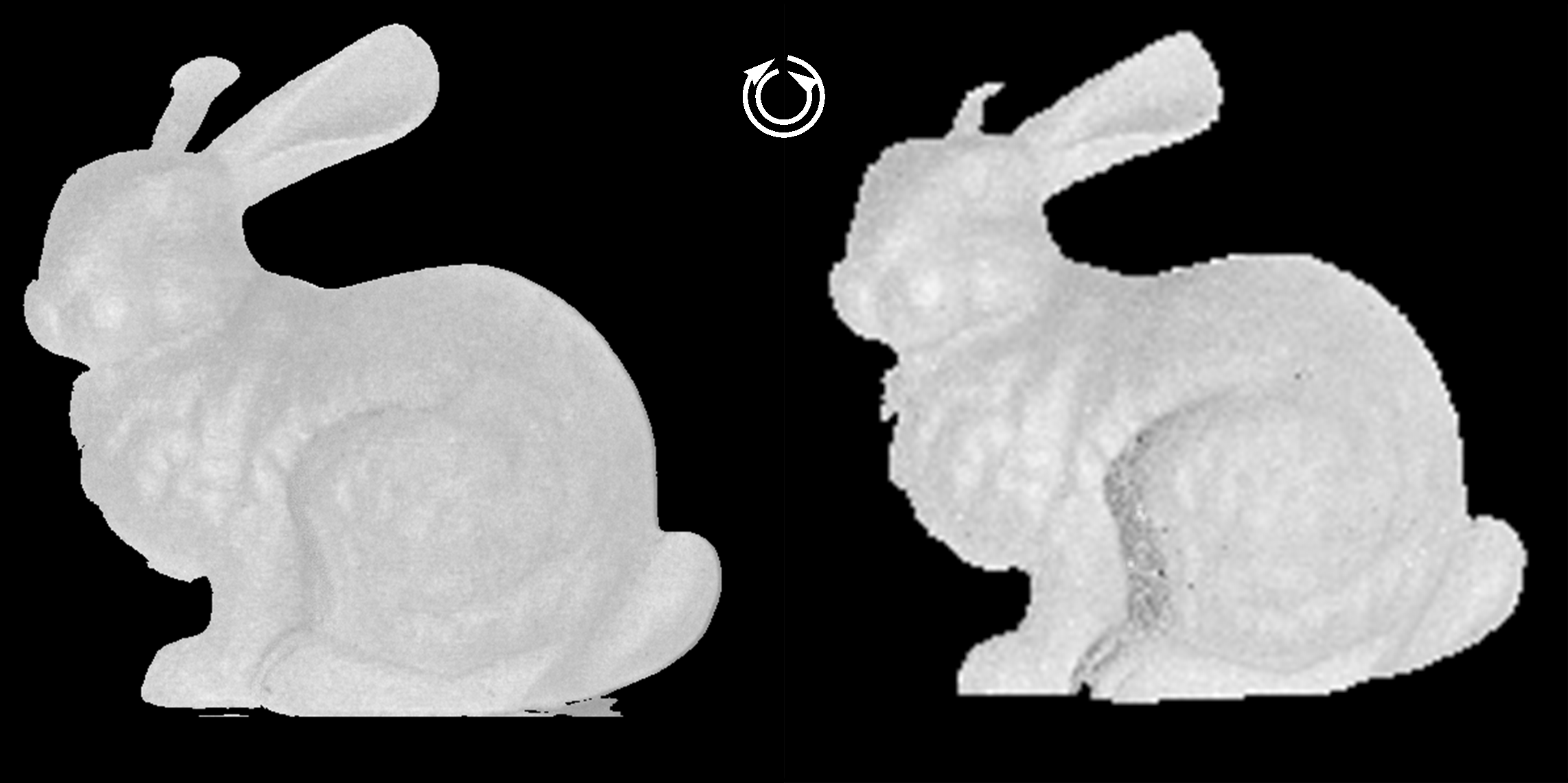}}\label{fig:circular451}\\
        \vspace{0.2cm}
        \centering
    \makebox[5pt]{{\includegraphics[trim={0 0 0 -5},clip,width=0.25\textwidth]{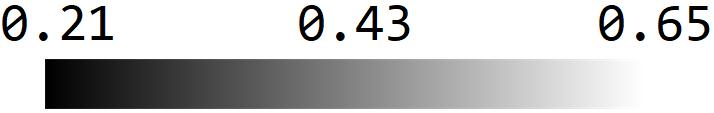}}}
    \hspace{5.6cm}
    \makebox[5pt]{{\includegraphics[trim={0 0 0 -5},clip,width=0.25\textwidth]{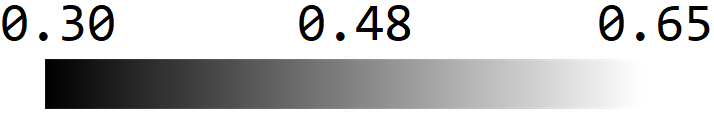}}}
    \hspace{5.6cm}
    \makebox[5pt]{{\includegraphics[trim={0 0 0 -5},clip,width=0.25\textwidth]{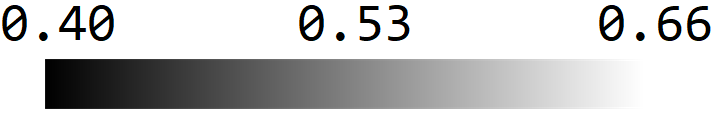}}}\vspace{-.2cm}    \\
        \subfloat[Crossed linear polarizers 662 nm]{\includegraphics[trim={0 0 0 0},clip,width=.32\textwidth]{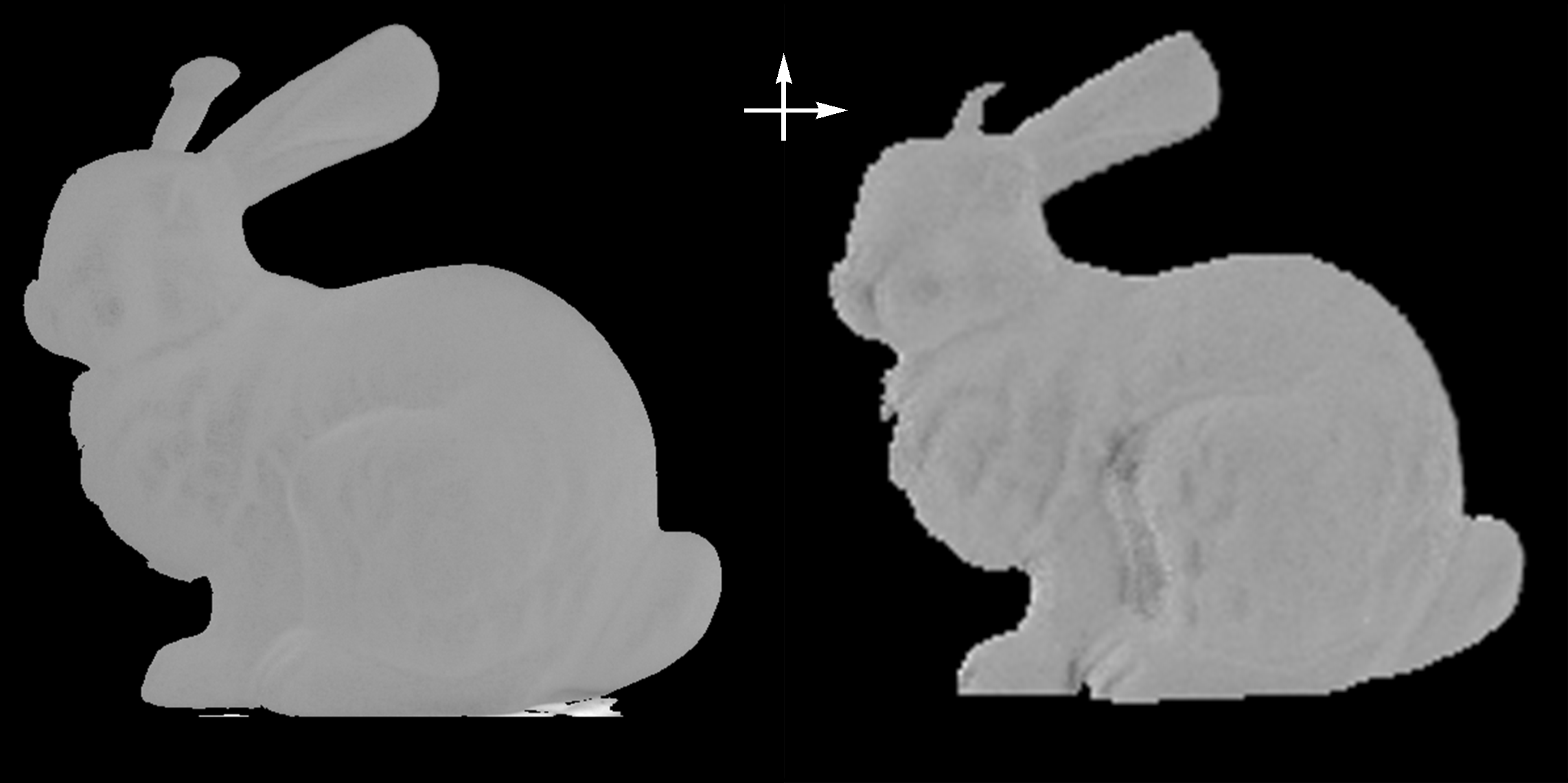}}\label{fig:crossed662}
        \subfloat[Aligned linear polarizers 662 nm]{\includegraphics[trim={0 0 0 0},clip,width=.32\textwidth]{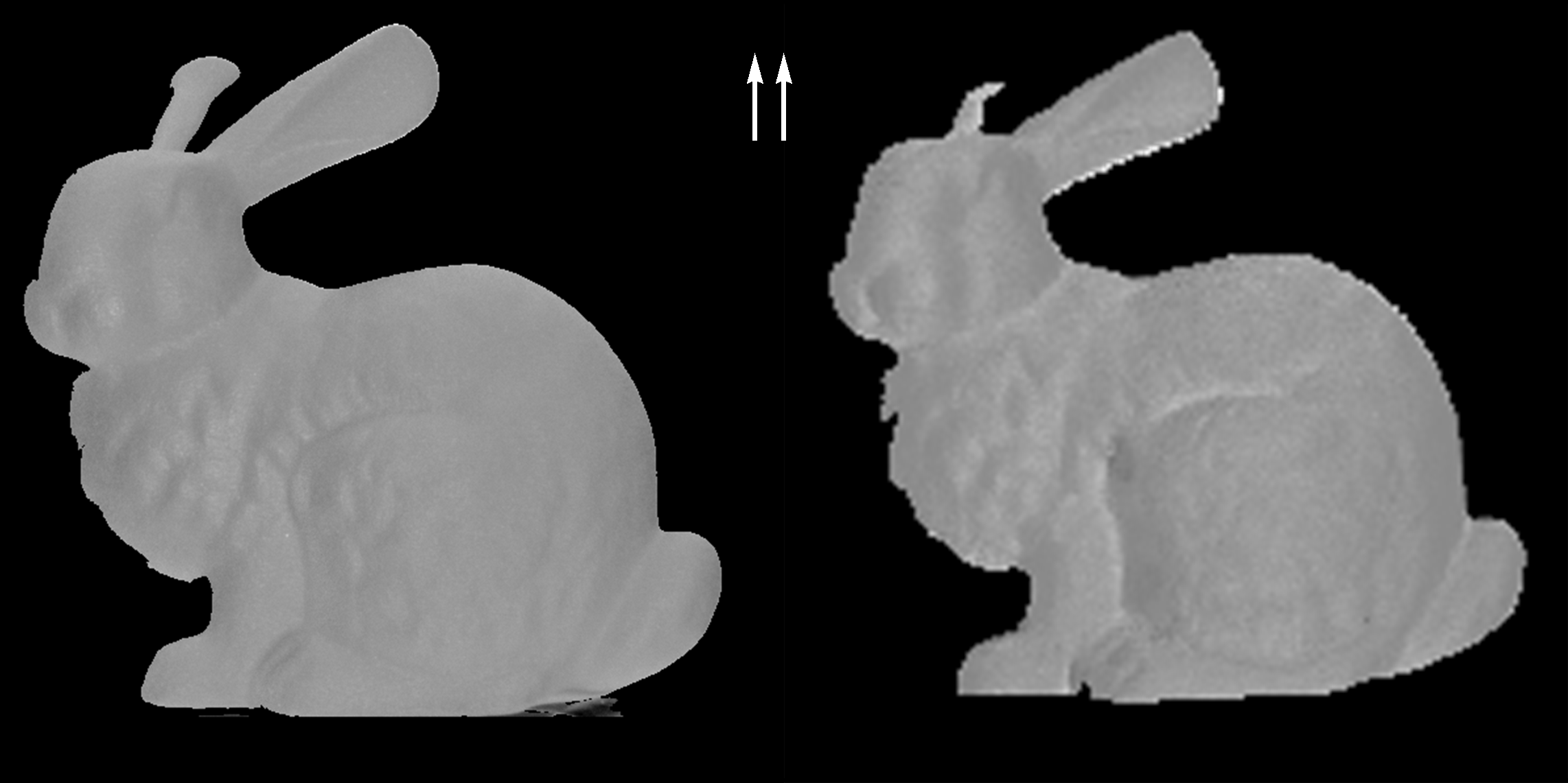}}\label{fig:aligned662}
        \subfloat[Crossed circular polarizers 662 nm]{\includegraphics[trim={0 0 0 0},clip,width=.32\textwidth]{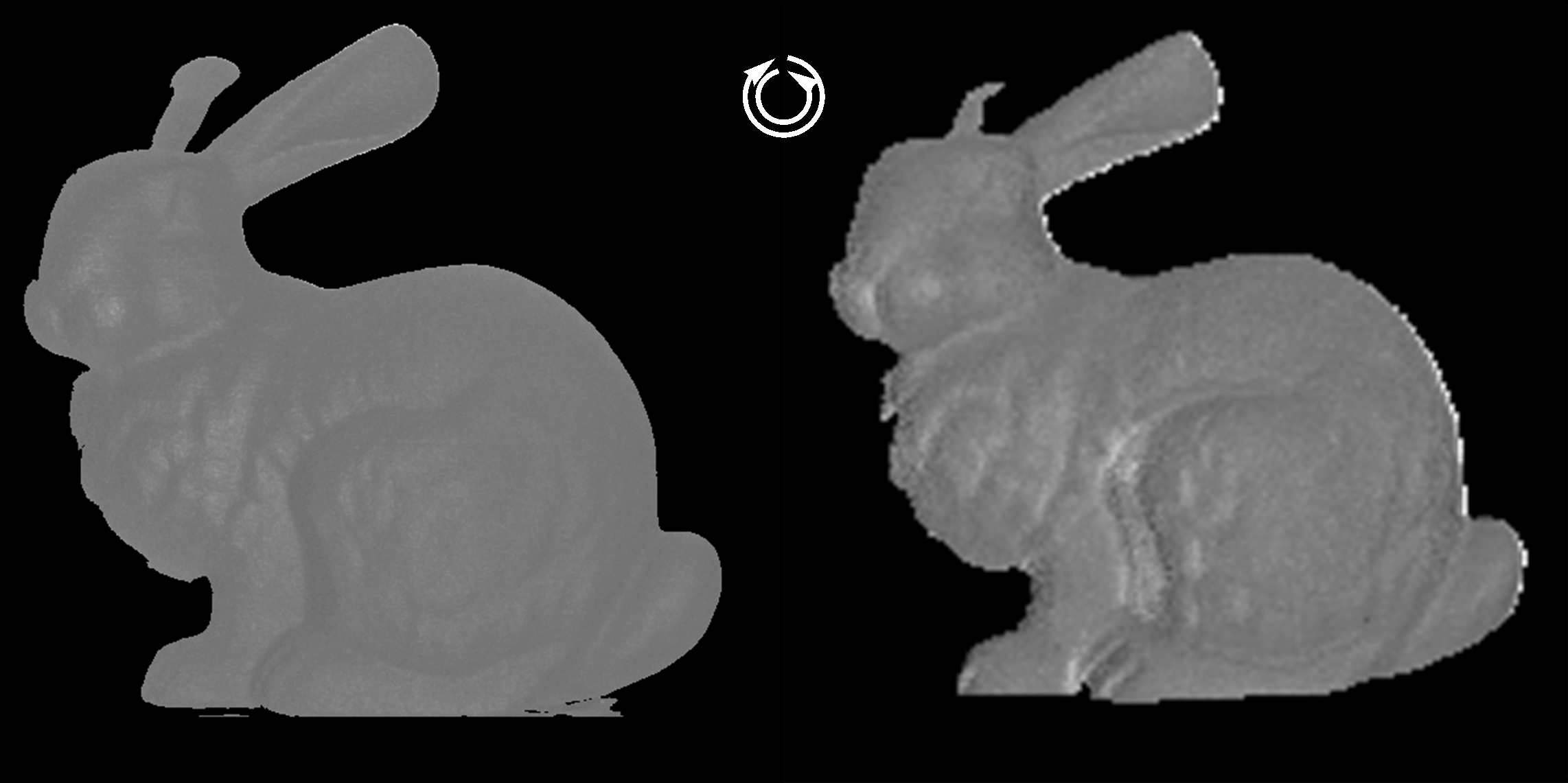}}\label{fig:circular662}
        \caption{Each image pair in (a-f) is arranged so that the simulated measurements are on the left the normalized MM image of the Stanford bunny from the DRR polarimeter and (right) the normalized extrapolated MM image from linear Stokes data at 451 nm. In (a) vertically polarized illumination and a horizontal analyzer, in (b) vertically polarized illumination and a vertical polarizer, and (c) between left circular polarization illumination and a right circular polarizer denoted between each image pair. In (d-f) the same experiments is repeated at 662 nm. The MM images were normalized, so variation in the intensity transmitted through the second polarizer is due to diattenuation, depolarization, and geometric transformation of the illumination polarization. If the estimate of $\xi_0$ were too close to 0.25 (\emph{i.e.} extrapolating to the ideal depolarizer), then there would be no variation and all three polarizer conditions would look flat. Overall, there is good visual agreement, with regions of high and low intensity in the DRR results tending to correspond to high and low intensity in the Stokes camera extrapolated results. It is interesting to note that, despite the MJM models having only residual circular polarization properties from the complex refractive indices, the crossed circular polarizer images show some agreement. This is because of the variation in the depolarization due to the $\xi_0$ estimate. There is inherent disagreement between the measurement simulated from DRR data and from the linear Stokes data due to the slightly different viewing geometry, resolution, and camera properties between the two polarimeters. Qualitatively, the agreement between DRR and Stokes results is lowest on the side of the leg. This is the region where the TD assumption is weakest, as shown in Fig.~\ref{fig:bunny_xi}. Additionally, this region of high curvature is where misregistration between the modeled geometry shown in Fig.~\ref{fig:measAngles}~(b,d,f) and the geometry measured with the Stokes camera would cause large deviation in the modeled and measured $\widehat{\mathbf{m}}_0$, reducing the quality of the extrapolation. }\label{fig:bunnyFluxSim}
\end{figure*}

Figure~\ref{fig:bunnyFluxSim} shows a visual comparison of the Stanford bunny between various polarizers using the DRR measurements (left) and extrapolated MM image from linear Stokes data (right). Each pair of images is on the same color scale. Crossed linear, aligned linear, and crossed circular measurements are simulated. The MM images are normalized, so variation in the throughput of the second polarizer is due to the diattenuation, depolarization, and geometric transformation of the incident polarizer. The overall appearance between the results from DRR measurement and Stokes extrapolation are in agreement, particularly with respect to wavelength. At 451 nm, shown in Fig.~\ref{fig:bunnyFluxSim}~(a-c), the smaller amount of depolarization expected for a low albedo material appears as greater polarimetric modulation as compared to 662 nm, shown in Fig.~\ref{fig:bunnyFluxSim}~(d-f). The overall trend with geometry tends to match as well: regions of the bunny which appear brighter or darker are in the results from DRR measurement correspond to brighter or darker regions in the results extrapolation from Stokes data. Notably, the region with the most significant difference is the left side of the bunny's leg. This is the region where the TD assumption is weakest due to a higher angle of scattering, as shown in Fig.~\ref{fig:bunny_xi}. Additionally, this region of high curvature is where misregistration between the modeled geometry shown in Fig.~\ref{fig:measAngles}~(b,d,f) and the geometry measured with the Stokes camera would cause large deviation in the modeled and measured $\widehat{\mathbf{m}}_0$, reducing the quality of the extrapolation. There is unavoidable disagreement between the DRR simulations and the linear Stokes images due to even slight differences in viewing geometry, resolution, and camera properties of the two polarimeters. An example of such disagreement due to different camera systems can clearly be seen on the left ear which has a significantly different shadow between DRR and Stokes results. 

\section{Conclusions and Discussion}
In contrast with precision optical components, depolarization is the dominant polarimetric property of everyday materials. The TD-MM pBRDF model lends itself to efficient descriptions for such depolarization-dominant materials. This work presents a closed-form Mueller-matrix valued pBRDF that is efficiently represented by a reduced parameter set: four wavelength-dependent material constants that define the polarization properties and a parameter for depolarization that depends on both geometry and wavelength. The material constants determine how the six degrees of freedom for the diattenuation and retardance vary with geometry. This reduced parameterization assumes a TD coherency eigenspectrum and specifies two distinct non-depolarizing MJMs for first-surface and diffuse polarized light scattering. A microfacet Fresnel MJM is a popular polarized-light scattering model that is simple enough to describe first-surface scattering in many cases. However, polarization from diffuse scattering becomes non-negligible when a wider range of scattering geometries are considered. 

In this work, the TD-MM pBRDF model is demonstrated over a large range of scattering geometries by combining polarized first-surface and diffuse contributions into a single MJM model. The first-surface MJM is Fresnel reflection based on microfacet theory, but purely radiometric effects such as shadowing and masking are ignored. The polarimetric properties of microfacet Fresnel reflections are invariant to the surface normal. However, the diffuse MJM is a diattenuator with orientation that strongly depends on the surface normal in a pattern centered where $\theta_h=0^\circ$ (\emph{i.e.} specular reflection). This orientation is parallel to the plane of incidence with the caveat that 45$^\circ$ and 135$^\circ$ are flipped due to reflection. The magnitude of the diffuse contribution is determined by material constants. 

The four material constants depend chiefly on the spectrally-dependent albedo of the material. The depolarization parameter depends on both the albedo and scattering geometry. The error in the TD assumption is inversely proportional to depolarization. For example, the red object used in this work is higher albedo at 662 nm compared to 451nm and the TD assumption is more appropriate for the high albedo case, as shown in Fig.~\ref{fig:bunny_xi}. 

Multi-angle MM measurements at two illumination wavelengths (for both high and low albedo conditions) of a red 3D printed material are compared to our pBRDF model. The RMSD in diattenuation orientation for the modeled versus measured dominant MJM (see Fig.~\ref{fig:diat_error}~(a)) averaged over acquisition geometry was 7.49$^\circ$ and 14.29$^\circ$ at 451 nm (low albedo) and 662 nm (high albedo), respectively. The RMSD in diattenuation magnitude (see Fig.~\ref{fig:diat_error}~(b)) averaged over acquisition geometry was 4.96\% at 451 nm and 11.73\% at 662 nm. Next, as an example application, the MJM term of our pBRDF model is evaluated at assumed values of the material constants, see Table~\ref{tab:matProps}, and the geometry-dependent depolarization parameter and the average reflectance are estimated from only linear Stokes images. Despite the MJM model diattenuation properties matching the measurement better at 451 nm than at 662 nm, using the MJM models to estimate $\xi_0$ resulted in RMSD values (see Fig.~\ref{fig:xi0_error}) averaged over acquisition geometry of 11.11\% at 451 nm and 4.24\% at 662 nm. This performance difference can be understood by noting that the depolarization magnitude at 662 nm is higher and therefore the estimation is less sensitive to MJM model accuracy. 

To demonstrate the generalization of the TD-MM model to more complicated geometry, MM images of a 3D printed Stanford bunny were extrapolated from linear Stokes images. To compare extrapolated MM images to ground truth from DRR polarimetry, Fig.~\ref{fig:bunnyFluxSim} shows simulated measurements of the bunny between various polarizers. Overall agreement is good, with regions of high and low transmitted intensity in the extrapolated results corresponding to those of the DRR results. The largest deviations occur on the left side of the hind leg where Fig.~\ref{fig:bunny_xi} shows the TD eigenspectrum assumption is weakest as well as being the region where registration between the geometry modeled (shown in Fig.~\ref{fig:measAngles} (b,d,f)) and the actual measured geometry is most sensitive. 

Quantitative results are presented for the sphere and visual agreement is assessed for the Stanford bunny, but the success of the model is application-dependent. 

The quantitative results for the sphere and the visual assessment performed on the Stanford bunny highlight an important point: the accuracy of a polarimetric model does not need to exceed the accuracy of the polarimeter in use. For example, the polarimetric accuracy of a model required to approach photorealism in a polarization-aware physics-based rendering is lower than the accuracy required in silicon wafer metrology. In the former application, a COTS Stokes polarimeter might be used to characterize a material based on an efficient, reduced parameterization with sufficient accuracy. Facilitating the characterization of many materials according to a simple model in a reduced length of time may present a significant benefit over the cost of more rigorous characterization.

The TD-MM pBRDF model is intended to generalize to materials other than the red 3D printing material used for demonstration purposes in this work. By combining first-surface and diffuse polarized reflection in amounts that are material-dependent the model is designed to capture the most relevant features of polarized light scattering. To apply the model in Eq.~\ref{eq:mixed} to a given material, the albedo-dependent material constants $n_\lambda$, $a_\lambda$, and $b_\lambda$ must be determined. One method to estimate these values is MM imaging at a range of scattering geometries and least-squares fitting to the dominant process. This has the advantage of also providing a ground truth of the eigenspectrum to gauge the appropriateness of the  TD assumption. Another approach, based on the \emph{a priori} TD assumption, is to maximize the linear correlation coefficient between polarized measurements and measurements simulated from the MJM model defined by the material constants. Modulation in the polarimetric measurements of a TD MM, see Eq.~\ref{eq:basis}, is determined only by the $\widehat{\mathbf{m}}_0$ term, since the ideal depolarizer $\mathbf{m}_{ID}$ is an offset. Therefore, polarimetric measurements of the full depolarizing MM at a given scattering geometry are linearly related to polarimetric measurements of the dominant MJM. The slope depends on $\xi_0$ and $M_{00}$, which in turn each depend on scattering geometry, but the material constants $n_\lambda$, $a_\lambda$, and $b_\lambda$ are geometry-independent. Therefore, $n_\lambda$, $a_\lambda$, and $b_\lambda$ could be globally tuned to optimize the linear fit across the pixel-dependent scattering geometry.

\bibliographystyle{IEEEtran}
\bibliography{sample}


\section{Biography Section}

\begin{IEEEbiography}
[{\includegraphics[width=1in,height=1.25in,clip,keepaspectratio]{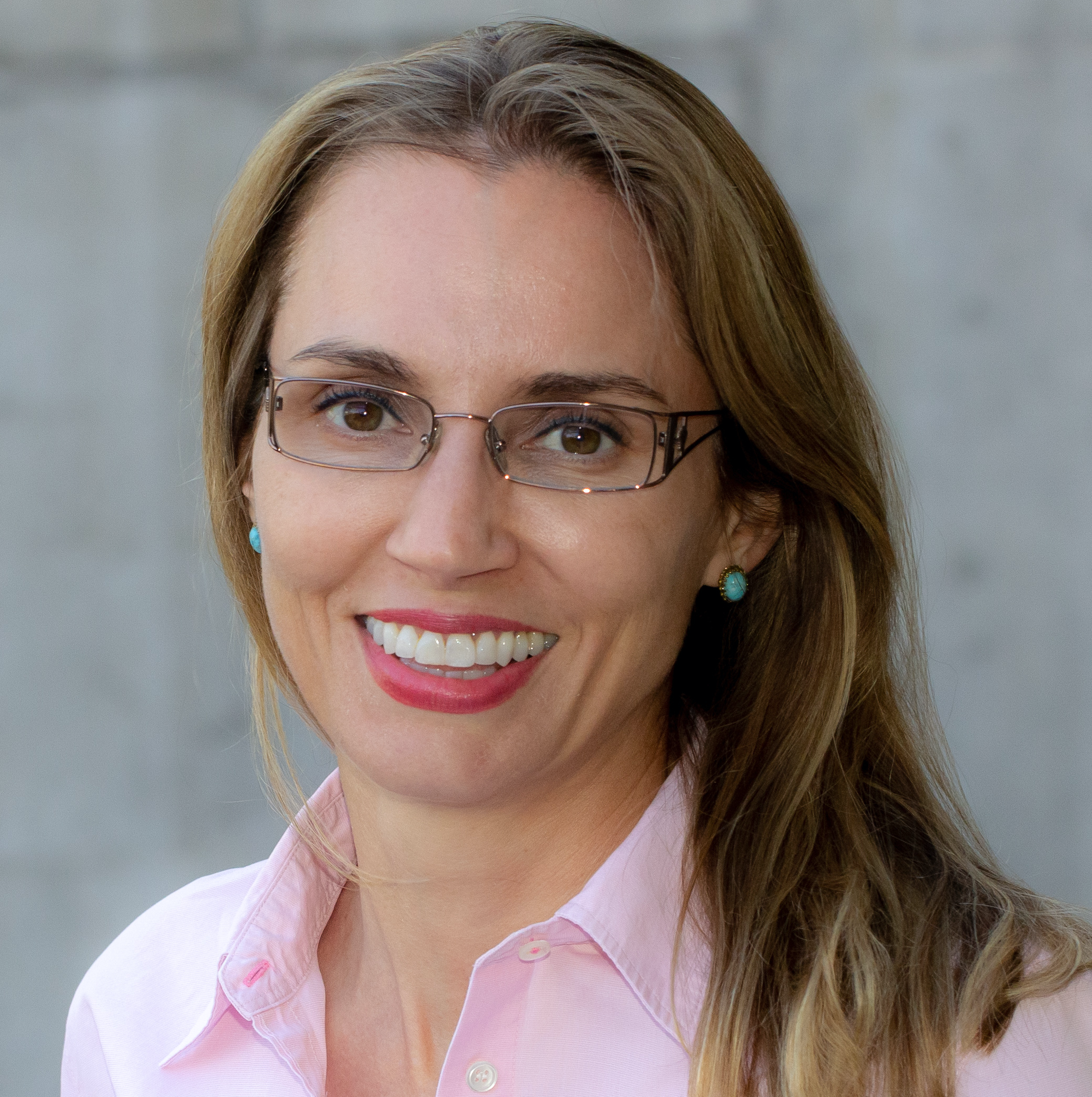}}]{Meredith Kupinski} received the M.S. and Ph.D. degrees in optical science from The University of Arizona, Tucson, AZ, USA, in 2003 and 2008, respectively. She is currently an Assistant Professor with the Wyant College of Optical Sciences, The University of Arizona. Her research interests include task-relevant metrics for polarimeter design, stochastic systems analysis and information theory.
\end{IEEEbiography}

\begin{IEEEbiography}
[{\includegraphics[trim= 150 10 60 10, clip, width=1in,height=1.25in,keepaspectratio]{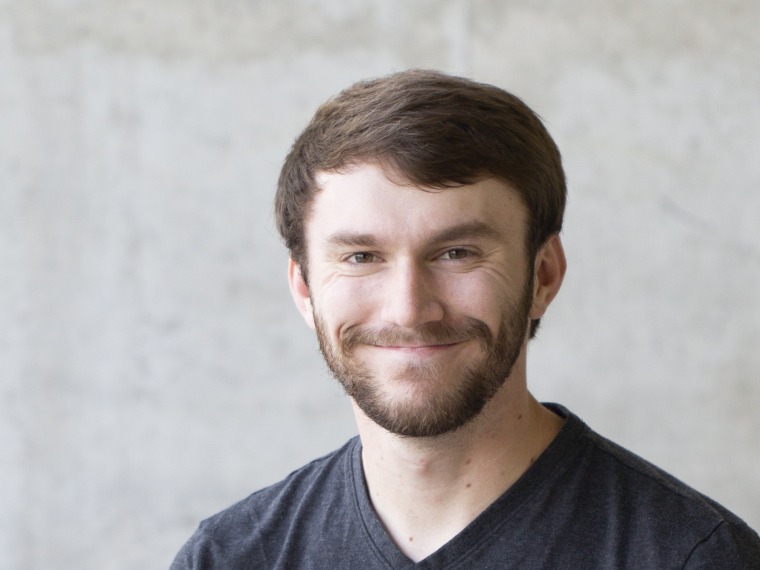}}]{Quinn Jarecki} received the B.S. and B.M. degrees in optical science and music performance, respectively, from The University of Arizona, Tucson, AZ, USA, in 2015. He is currently a Ph.D candidate at the Wyant College of Optical Sciences, The University of Arizona. His research interests include full and partial Mueller polarimetric imaging and polarized light scattering.

\end{IEEEbiography}

\vspace{11pt}

\vfill

\end{document}